\documentclass[12pt]{article}
\usepackage{amsmath} 
\usepackage{amssymb} 
\usepackage{amssymb}
\usepackage{amsmath}
\usepackage{indentfirst}
\usepackage{amsfonts,amssymb}
\usepackage{mathrsfs}
\usepackage{CJK,CJKnumb}
\usepackage{color}
\usepackage{amsthm}
\usepackage{graphicx}
\usepackage{enumerate}
\usepackage{epstopdf}
\usepackage{multirow}
\usepackage{float}
\usepackage{bm}

\theoremstyle{definition}
\newtheorem{Definition}{Definition}[section]
\newtheorem{Theorem}{Theorem}[section]
\newtheorem{Lemma}{Lemma}[section]
\newtheorem{Corollary}{Corollary}[section]

\newtheorem{Remark}{Remark}[section]

\numberwithin{equation}{section}

\def\sF{{\mathscr F}}

\def\sB{{\mathscr B}}
\def\sX{{\mathscr X}}

\def\sL{{\mathscr L}}

\begin{document}
	
	\title{  {On data-driven robust distortion risk measures \\ for non-negative risks with partial information }
		\thanks{ Supported in part by the National Natural Science Foundation of China (No: 12271415)
			and
			the Fundamental Research Funds for the Central Universities (104972026KFYjc0094)} \\
		\footnotetext{ Email addresses: hxymath@whu.edu.cn (X. Han), yjhu.math@whu.edu.cn (Y. Hu),
			rwang@whu.edu.cn (R. Wang), lxwei@whut.edu.cn (L. Wei).  \\
			\indent \ $^\ast$Corresponding author: Xiangyu Han
		}
	}

	\vspace{0.5cm}
	
	\author{Xiangyu Han$^{1, \ast}$, \quad Yijun Hu$^{1}$, \quad Ran Wang$^{1}$, \quad Linxiao Wei$^{2}$ \\
		\\
		$^1$ {School of Mathematics and Statistics}\\
		{Wuhan University} \\
		{Wuhan, Hubei 430072}\\
		{People's Republic of China}\\
		\\
		$^2$ {School of Mathematics and Statistics}\\
		{Wuhan University of Technology} \\
		{Wuhan, Hubei 430070}\\
		{People's Republic of China}
	}
	
	\vspace{0.2cm}
	
	\date{\today}
	
	\maketitle
	
	\noindent{\bf Abstract:} \quad
	In this paper, by proposing two new kinds of distributional uncertainty sets, we explore robustness of distortion risk measures
	against distributional uncertainty.
	To be precise, we first consider a distributional uncertainty set which is characterized solely by a ball determined by general Wasserstein
	distance centered at certain empirical distribution function,
	and then further consider additional constraints of known first moment and any other higher moment of the underlying loss distribution function.
	Under the assumption that the distortion function is strictly concave and twice differentiable, and that the underlying loss random variable
	is non-negative and bounded, we derive closed-form expressions for the distribution functions which maximize a given distortion risk measure over the
	distributional uncertainty sets respectively. Moreover, we continue to study the general case of a concave distortion function
	and unbounded loss random variables.
	Comparisons with existing studies are also made. Finally, we provide a numerical study to illustrate the proposed models and results.
	Our work provides a novel generalization of several known achievements in the literature.

	\vspace{0.2cm}
	
	
	\vspace{0.2cm}
	

	\newpage
	
	\section{Introduction}\label{sec:1}
	
	Risk measures are prevalently used in various contexts in both insurance and finance, such as insurance pricing and regulatory capital calculation, etc.
	Classical risk measures are defined on univariate risks, i.e. on random variables defined on certain probability space, for example,
	see Artzner et al. (1999), F\"{o}llmer and Schied (2002), Frittelli and Rosazza Gianin (2002) and Wang et al. (1997).
	In practice, these classical risk measures usually require the acquirement of accurate specification of distribution functions of the random variables.
	For instance, this is the case of value-at-risk (VaR) or expected shortfall (ES), which are standard risk measures popularly used in
	insurance and finance. ES is sometimes known as tail value-at-risk (TVaR) or conditional value-at-risk (CVaR).
	From the practical point of view, it is usually difficult to accurately capture the \textit{true} specifications of
	distribution functions of the random variables, because the distribution functions usually have to be estimated from data (i.e. samples)
	or statistical simulation.
	Alternatively, it is often inadequate to theoretically assume that the distribution functions of random variables are accurately known.
	Therefore, the so-called distributional uncertainty problem naturally arises. Distributional uncertainty issue is especially crucial for
	law-invariant risk measures, because the value of law-invariant risk measure of a random variable depends only
	on the specification of its distribution function. Distortion risk measures (DRMs) are typical law-invariant ones,
	which include VaR, ES and range value-at-risk (RVaR) as special cases.
	For more details about risk measures, we refer to F\"{o}llmer and Schied (2016).
	
	\vspace{0.2cm}
	
	To deal with distributional uncertainty, it is natural to establish robust risk measures. Instead of evaluating the risk measure for a
	\textit{single} specification of the distribution function of a random variable, a robust risk measure evaluates the worst-case (or best-case)
	risk measure for distribution function within certain set consisting of particular distribution functions,
	while such a set is usually referred to as a distributional uncertainty set.
	In other words, a robust risk measure is to determine the maximum (or minimum) value of a given risk measure for distribution function
	over certain distributional uncertainty set, and to identify the corresponding maximizing (or minimizing) distribution function,

	\vspace{0.2cm}
	
	To enhance the robustness of involved risk measures against distributional uncertainty, the choice of an appropriate distributional uncertainty set
	is of central importance.
	In the literature, there have been diverse models about evaluating risk measures for (loss) distribution functions with partial information,
	such as known moments, where the distributional uncertainty set is determined by moment conditions.
	For instance, H\"{u}rlimann (2002) studied the worst-case VaR and CVaR, where the distributional uncertainty sets were determined by
	known fist $n$  moments. Zhu and Fukushima (2009) studied the worst-case CVaR under three kinds of
	distributional uncertainty sets:  mixture distributional uncertainty, box uncertainty and ellipsoidal uncertainty.
	Cornilly et al. (2018) studied upper bounds for strictly concave distortion risk measures, where the distributional uncertainty sets
	were characterized by known any $n$ moments. Li (2018) studied the worst-case spectral and law-invariant coherent risk measures,
	where the distributional uncertainty sets were determined by known first and second moments.
	Li et al. (2018) studied the worst-case RVaR, where four kinds of distributional uncertainty sets were considered: (1) known mean and
	variance; (2) known mean, variance and symmetric random variables; (3) known mean, variance and unimodal distribution functions;
	and (4) known mean, variance and  unimodal-symmetric random variables.
	Recently, Bernard et al. (2020, 2023) studied the worst-case VaR and RVaR, where the distributional uncertainty set
	consisted of distribution functions with either fixed mean and variance being less than certain constant or fixed mean, variance being less than
	certain constant and unimodal distribution functions.
	Shao and Zhang (2023) studied the worst-case DRMs, where the distributional uncertainty set consisted of distribution functions
	with either known mean and variance or known mean, variance and symmetric random variables.
	Shao and Zhang (2024) studied the worse-case DRMs, where the distributional uncertainty set consisted of distribution functions
	with either known mean and $k$th centered moment or known mean, $k$th centered moment and symmetric random variables.
	Zuo and Yin (2025) studied the worst-case distortion risk metrics, where the distributional uncertainty sets consisted of distribution functions
	with known mean and variance.
	For more studies about applications of moment constraints in related optimization problems,
	see also Ghaoui et al. (2003), Pflug and Wozabal (2007), Popescu (2007), Bertsimas et al. (2010), Bernard et al. (2018), Cai et al. (2025),
	Liu et al. (2025), and the references therein.
	Motivated by above observation, in the present paper, we will assume that the first moment and another higher order moment are known
	when we construct distributional uncertainty sets.
	
	\vspace{0.2cm}
	
	Besides considering moment conditions in determining a distributional uncertainty set,
	the Wasserstein distance, particularly of order 2, has also emerged as a powerful tool for modeling distributional uncertainty in risk assessment context.
	Unlike other divergence-based approaches, such as the Kullback-Leibler divergence, the Wasserstein distance can handle distributions with differing supports, making it highly adaptable in situations where the worst-case distribution may not share the same support as the reference distribution.
	This flexibility is especially useful when dealing with discrete and continuous distributions together, such as the cases involving DRMs.
	By lifting the absolute continuity restriction, the Wasserstein distance allows for a broader uncertainty set that accommodates both discrete and
	continuous distributions.
	In addition, its non-parametric feature makes it immune to specific distribution types, helping alleviate the risk of model misspecification.
	For instance, recently, Bernard et al. (2024) studied the worst- and best-case DRMs, where the distributional
	uncertainty set was assumed to be either a ball determined by 2-order Wasserstein distance or furthermore with known mean and variance as well.
	Quite recently, Liu et al. (2025) derived sharp upper and lower bounds for distortion risk metrics under distributional uncertainty,
	where the distributional uncertainty sets were characterized by four key features of the underlying distribution function:
	mean, variance, unimodality and Wasserstein distance of order 2 to a given reference distribution function. Wasserstein distance is also employed in the study of distributionally robust insurance; for instance, see Boonen and Jiang (2025a, 2025b) and Cai et al. (2024).
	For more studies about application of Wasserstein distance in other related optimization problems, see also Mohajerin Esfahani and Kuhn (2018),
	Blanchet et al. (2022), Pichler and Xu (2022), Gao and Kleywegt (2023), Tang and Yang (2023), Cai et al. (2024), and the references therein.
	In the present paper, we will also take into account the general Wasserstein distance,
	when we construct distributional uncertainty sets. Moreover, following the mainstream in related research, we choose the order $p$ of
	the Wasserstein distance to be the same as the higher order $p$ of the moment.
	
	\vspace{0.2cm}
	
	From the statistical point of view, a random variable can be characterized by its samples. Alternatively, the distribution function of
	a random variable can be characterized by its empirical distribution function. Hence, it seems both natural and reasonable to take into account
	the empirical distribution function when constructing a distributional uncertainty set.
	For instance, recently, Tang and Yang (2023) studied the worst-case moments, where the \textit{true} probability distribution of a non-negative
	risk variable was assumed to lie within a ball specified through a general Wasserstein distance centered at its empirical distribution.
	For more related studies, see also Mohajerin Esfahani and Kuhn (2018), Blanchet et al. (2022), and the references therein.
	Following the mainstream in related research, in this paper, we will also incorporate the empirical distribution function into
	the construction of distributional uncertainty sets.

	\vspace{0.2cm}
	
	In the present paper, by introducing two new kinds of distributional uncertainty sets, we study the worst-case DRMs
	based on data (samples). Precisely, we first consider a distributional uncertainty set which is solely characterized by a ball determined by
	general $p$-order Wasserstein distance around certain empirical distribution function,
	and then further consider additional constraints of known mean and $p$th moment of the underlying (loss) distribution function.
	Such two distributional uncertainty sets are not well studied in the literature.
	Under the assumption that the distortion function is strictly concave and twice differentiable, and that the underlying loss random variable
	is nonnegative and bounded, we derive closed-form expression for the distribution function which maximizes a given distortion risk measure over the
	distributional uncertainty sets, respectively. Moreover, we continue to study the general case of a concave distortion function
	and unbounded loss random variables.
	Comparisons with existing studies are also made. Finally, we provide a numerical study to illustrate the proposed models and results.
	This study provides a novel generalization of several known achievements in the literature.

	\vspace{0.2cm}
	
	It should be noted that the models that arise here are far from simple hybrids of models involving either
	Wasserstein distance constraints or moment constraints alone. They are also not trivial generalizations of known models in the literature.
	Indeed, it turns out not only that they are delicate problems to develop an appropriate method for deriving the distribution functions maximizing
	the DRMs, but also that the maximizing distribution functions to such optimization models are also more complicated; see especially the preparation
	Lemmas 3.2-3.8, Theorems 3.1 and 3.2 and Corollary 3.1 below.
	
	\vspace{0.2cm}
	
	It might be helpful for us to briefly comment on the main contributions of the present paper.
	First, in order to explore the robustness of DRMs with concave distortion functions to distributional uncertainty,
	we propose two new kinds of distributional uncertainty sets, which are not well studied in the literature. Closed-form expressions for the
	distribution functions maximizing the DRMs are derived. This paper significantly generalizes the recent relevant works of Bernard et al. (2024)
	by considering general $p$-order Wasserstein distance, known mean and $p$th moment of the underlying distribution function; see Theorems 3.1 and 3.2,
	the discussions after Remark 3.3 and Remark 3.8 below. From the theoretical point of view, we believe that it is also reasonable to
	take into account the general $p$-order Wasseratein distance and $p$th moment besides mean when thinking about distributional uncertainty set,
	and this consideration is the starting point of the present study;
	for instance, for the later case (i.e. known mean and another higher order moment), see Cornilly et al. (2018, Theorem 2.2).
	Second, compared with the relevant works of Bernard et al. (2024) and Cornilly et al. (2018), the generalizations achieved in the present paper
	are non-trivial, because new arguments need to be developed to derive the closed-form expressions for the maximizing distribution functions.
	Inspired by Rustagi (1957), Cornilly et al. (2018) and Lagrangian multiplier method, we manage to design a new approach to construct the closed-form
	expressions for the maximizing distribution functions.
	These newly developed arguments are far more delicate and complicated; see the preparation Lemmas 3.2-3.8, Theorems 3.1 and 3.2 and Corollary 3.1 below.
	We believe that these newly developed arguments are also interesting in their own right.
	Third, we directly connect our models with data (samples), so that the expected maximizing distribution functions emerge more trackable
	from the statistical point of view.
	
	\vspace{0.2cm}
	
	It should also be mentioned that our models are closely related to the ones studied by Bernard et al. (2024). Recently, Bernard et al. (2024) studied
	robust DRMs. They derived both maximizing distribution functions and the maximum values of the DRMs for distribution functions specified within
	certain ball, which is determined by the $2$-order Wasserstein distance and around a given reference distribution function.
	Besides considering the $2$-order Wasserstein distance constraint solely,
	they further considered the case of additional moment constraints of fixed mean and variance of the underlying distribution functions.
	Nevertheless, there are significant differences between this paper and  Bernard et al. (2024), which we explain briefly here.
	First, the methods used in this paper are different from that of Bernard et al. (2024).
	Unlike the methods employed by Bernard et al. (2024), in the present paper,  we employ Lagrangian multiplier method, since the involvement of general $p$-order Wasserstein distance makes the models we consider much more complicated.
	Specifically, inspired by Rustagi (1975), Cornilly et al. (2018) and by Lagrangian multiplier method,
	we manage to design a new approach to derive the distribution functions maximizing the DRMs.
	We are not certain whether the approaches of Bernard et al. (2024) could be applied to the present models.
	Second, about the moment constraints, we employ the mean and another higher order moment rather than the first two moments.
	Besides the case of first two moments constraints, we believe that it is also worth studying the case where the mean and another
	higher order moment are involved; for instance, see Cornilly et al. (2018, Theorem 2.2), where the mean and another higher order moment
	were assumed to be known.
	Third, instead of using a general reference distribution function, we directly employ the empirical distribution function to connect our models
	with data (samples), so that the expected results emerge more trackable from the statistical point of view.
	When the order of Wasserstein distance is chosen to be $2$, then our main results Theorems 3.1 and 3.2 and Corollary 3.1
	are respectively in accordance with Theorems 3.1 and 4.8 of Bernard et al. (2024); see the discussions after Remark 3.3 and Remark 3.8 below.
	Taking the above considerations into account, the present study can be viewed as a meaningful complement to that of Bernard et al. (2024).
	
	\vspace{0.2cm}
	
	The rest of this paper is organized as follows.  In Section 2,  we prepare the preliminaries including the description of optimization problems
	we are interested in. Section 3 is devoted to the presentation of the main results of this paper about the worst-case DRMs under
	both distributional uncertainty sets.
	Section 4 provides numerical studies illustrating the proposed models and results obtained.
	Concluding remarks are summarized in Section 5.
	
	\vspace{0.2cm}
	
	In order to make a better flow, we will postpone all proofs of main results and lemmas, presented in Section 3, of this paper to the appendix.

	\section{Preliminaries}
	
	\subsection{Notations}
	
	In insurance, the random loss of an insurance risk faced by an insurer is commonly described by a non-negative random variable defined on
	some probability space, which could be a claim or an aggregate of claims.
	Note that in finance, the random loss of a financial position is commonly described by a random variable defined on certain probability space.
	From the practical point of view, a risk analyst might more concern the losses of financial positions,
	which are also commonly described by  non-negative random variables.
    
	\vspace{0.2cm}

	Let $(\Omega, \sF, \mathbb{P})$ be a fixed probability space.
	In this paper, we consider non-negative random variables  defined on $(\Omega, \sF, \mathbb{P})$.
	Furthermore, we consider non-negative bounded random variables  on $(\Omega, \sF, \mathbb{P})$.
	In this situation, without loss of generality, we can further assume that the random variables take values in $[0,1]$,
	otherwise by a scaling one can make the scaled non-negative random variables take values in $[0,1]$.
	We adopt notation $\mathbb{E}$ to stand for the (conditional) expectation operator with respect to the probability measure $\mathbb{P}$.

	\vspace{0.2cm}

	Denote by $\mathcal{X}$ the set of all non-negative random variables on $(\Omega, \sF, \mathbb{P})$ taking values in $[0,1]$.
	Moreover, we assume that the probability space $(\Omega, \sF, \mathbb{P})$ is rich enough to support all discrete random variables.
	Given a random variable $X$ on $(\Omega, \sF, \mathbb{P})$ and let $ x_1, \ldots, x_N$  be samples from it with size $N \geq 1$,
	we denote by \(\hat{X}\) the discrete random variable on $(\Omega, \sF, \mathbb{P})$, which takes values $ x_1, \ldots, x_N$
	with uniform probability mass function
	$\mathbb{P}(\hat{X} = x_i) = \frac{1}{N}$, $1\leq i \leq N$. Without loss of generality, we assume that $x_1 \leq  \ldots  \leq x_N$.

	\vspace{0.2cm}

	For a random variable $\xi$ on $(\Omega, \sF, \mathbb{P})$,   denote by $\mathbb{P}_\xi$ 
	the probability distribution of \(\xi\) under \(\mathbb{P}\), i.e.,
	\begin{align*}
		\mathbb{P}_\xi(B) := \mathbb{P}(\xi \in B), \quad B \in \sB(\mathbb{R}),
	\end{align*}
	which is a probability measure on $\sB(\mathbb{R})$. The support of $\mathbb{P}_\xi$ is contained in $[0,1]$ whenever $\xi \in \mathcal{X}$.
	For a random variable \(\xi\) on \((\Omega, \sF, \mathbb{P})\), the distribution function $F_\xi$ of $\xi$ under \(\mathbb{P}\) is defined  by
	$F_\xi(x) :=\mathbb{P}(\xi \leq x)= \mathbb{P}_\xi((-\infty, x]) $, $x \in \mathbb{R}$, and its left-continuous inverse function at probability level $q \in (0,1)$ is defined by
	\begin{align*}
		F_\xi^{-1}(q) := \inf \left\{ x \in \mathbb{R} : F_\xi(x) \ge q \right\} = \sup \left\{ x \in \mathbb{R} : F_\xi(x) < q \right\},
	\end{align*}
	with the conventions $F_\xi^{-1}(0):=\operatorname{essinf} \xi$ and $F_\xi^{-1}(1):=\operatorname{esssup} \xi$.
	Conversely, $\mathbb{P}_\xi$ can be recovered from $F_\xi$ as the Lebesgue--Stieltjes measure $L_{F_\xi}$ induced by $F_\xi$; that is,
	$\mathbb{P}_\xi = L_{F_\xi}$ on $\sB(\mathbb{R})$.
	In what follows, we shall freely switch between the probability distribution $\mathbb{P}_\xi$ and its distribution function $F_\xi$ via this one-to-one correspondence.

	\vspace{0.2cm}

	Note that for the given random variable $X$ and the given samples $ x_1, \ldots, x_N$, 
	the distribution function $F_{\hat{X}}$ of $\hat{X}$ under \(\mathbb{P}\)
	is exactly the empirical distribution function of $X$, that is,
	\begin{align}\label{add25073001}
		F_{\hat{X}}(x)=\frac{1}{N}\sum_{i=1}^{N}I_{(-\infty,x]}(x_i), \quad x\in\mathbb{R}.
	\end{align}

A function $f$ on $\mathbb{R}$ is called a distribution function, if it is non-decreasing and right-continuous with $\lim_{x\rightarrow +\infty}f(x)=1$ and $\lim_{x\rightarrow -\infty}f(x)=0.$ 
	A distribution function $F$ on $\mathbb{R}$ is said to be supported on $[0,1]$ if $F(x) = 0$ whenever $x<0$, and $F(x) = 1$ whenever $x\geq 1$.
	We denote by $\mathcal{F}$ the class of all distribution functions supported on $[0,1]$.
     Clearly, for any $\xi \in \mathcal{X}$, its distribution function $F_\xi$ belongs to $\mathcal{F}$.
	For a distribution function $F$ on $\mathbb{R}$, we denote by $L_F$ the Lebesque-Stieltjes measure on $\sB(\mathbb{R})$ induced by $F$.
	Clearly, $L_F$ is a probability measure on $\sB([0,1])$ whenever $F \in \mathcal{F}$, since $L_F$ is supported on $[0,1]$ in this case.
	Given a $F \in \mathcal{F}$, we say that a set $A \in \sB([0,1])$ has $F$-measure zero, if $L_F(A) = 0$.
	A function $g : [0,1] \rightarrow [0,1]$ is called a distortion function, if it is non-decreasing with $g(0)=0$ and $g(1)=1$.
	We introduce more notations. For a non-empty set $A$, $I_A$ stands for the indicator function of $A$.
	For $x \in \mathbb{R}$, $\mbox{sign}(x)$ stands for the sign function, that is, $\mbox{sign}(x)= -1$ if $x<0$, $0$ if $x=0$, $1$ if $x>0$.
	
	\vspace{0.2cm}
	
	Given a distortion function $g$, for any non-negative random variable X with distribution function $F_X$, the distortion risk measure $H_g(X)$ of $X$
	under $\mathbb{P}$ is defined as
	\begin{equation}\label{2001}
		H_g(X) := \int_0^\infty g(1 - F_X(x)) \mathrm d x = \int_0^\infty g(\mathbb{P}(X > x)) \mathrm d x.
	\end{equation}
	Notice that \( H_g(X) \)  depends solely on the distribution function $F_X$, and thus in the sequel, we also write $ H_g(F_X) $ instead of $ H_g(X) $.
	
	\vspace{0.2cm}
	
	Throughout this paper, we assume that $g$ is strictly concave and twice differentiable. Under this assumption,
	we know that \( H_g \) is a coherent risk measure. The importance of DRMs with concave distortion function
	(henceforth called concave DRMs)
	is highlighted by the fact that this class coincides with the class of
	coherent risk measures that are law-invariant and comonotonically additive; for instance, see Kusuoka (2001).

	\subsection{Models set-up}
	
	In risk evaluation and risk management, there is a prevalent issue of distributional uncertainty, which influences the evaluation of risks.
	Specifically, the establishment of upper bounds on risk measures is sometimes quite crucial for informed decision-making.
	The worst-case value of a risk measure refers to its maximal value within a class of potential plausible loss distribution functions.
	In this paper, we explore subsets of $\mathcal{F}$ characterized by a permissible deviation from a benchmark empirical distribution function.
	More precisely, the class of potential plausible loss distribution functions will be chosen as certain balls determined by the Wasserstein distance.
	
	\vspace{0.2cm}
	
	Wasserstein distance is an effective way to describe the distance between two probability measures, or equivalently two distribution functions.
	Let us recall the definition of the Wasserstein distance. 
	For more details about Wasserstein distance, we refer to Villani (2009).
	
	\vspace{0.2cm}
	
	\begin{Definition}\label{def201}\ \
		Let $(\sX, d)$ be a Polish space, and let $p \in [1, \infty)$. For any two probability measures $\mu, \nu$ on $\sX$,
		then the Wasserstein distance of order $p$ between $\mu$ and $\nu$ is defined by
		\begin{align*}
			W_p(\mu, \nu) & := \left( \inf_{\pi \in \Pi(\mu, \nu)} \int_{\sX \times \sX}  d(x, y)^p  \mathrm{d}\pi(x, y) \right)^{\frac{1}{p}}\\
			& \ = \inf \left\{  \left( \mathbb{E}_\pi  \left[ d(X, Y)^p \right] \right)^{\frac{1}{p}} :\ \pi \in \Pi(\mu, \nu)\ \mbox{such that}\
			\sL_\pi(X) = \mu,\ \sL_\pi(Y) = \nu\right\},
		\end{align*}
		where $\Pi(\mu, \nu)$ stands for the set of all probability measures on $\sX \times \sX$ with marginal probability measures $\mu$ and $\nu$,
		$X$ and $Y$ are coordinate projections from $\sX \times \sX$ to $\sX$, $\sL_\pi(X)$ and $\sL_\pi(Y)$ are respectively the laws of $X$ and $Y$
		under $\pi$, and $ \mathbb{E}_\pi (\cdot)$ stands for expectation under $\pi$.
	\end{Definition}
	
	\vspace{0.2cm}
	
	For the case of one-dimension, that is, $\sX := \mathbb{R}$, then for any random variables $X$ and $Y$,
	the Wasserstein distance of order $p$ between $F_Y$ and $F_X$ is given by
	\begin{align*}
		W_{p}(F_Y, F_X) :=  W_p(\mu_{_Y}, \nu_{_X}) = \left( \int_0^1 \left| F_Y^{-1}(u) - F_X^{-1}(u) \right|^p  \mathrm{d}u \right)^{\frac{1}{p}},
	\end{align*}
    where $\mu_{_Y}$ and $ \nu_{_X}$ are the Lebesque-Stieltjes measures induced by the distribution functions
	$F_Y$ and $F_X$, respectively; for instance, see Panaretos and Zemel (2019, page 7).
	
	\vspace{0.2cm}
	
	Next, we consider the case where the \textit{true} distribution function $ F_X $ of a random variable $X$ lies in certain ball determined by
	the Wasserstein distance of order $p$ and centered at its empirical distribution function $F_{\hat{X}}$.
	
	\vspace{0.2cm}
	
	From now on, let a (non-negative) random loss $X \in \mathcal{X}$ be arbitrarily fixed, and let $ x_1 \leq \ldots \leq x_N$, $N \geq 1$,
	be samples from it. We consider two types of distributional uncertainty sets:
	for some $\epsilon>0$, $p \geq 1$, $0 < c_1 < 1$ and $0 < c_p <\frac{1}{p}$,
	
	\begin{align*}
		\mathcal{F}_{\epsilon} :=\{F_{X}\in\mathcal{F} ~ : ~ W_{p}(\mathbb{P}_{X}, \mathbb{P}_{\hat{X}})\leq\epsilon\}
		=\{F_{X}\in\mathcal{F} ~ : ~ W_{p}^{p}(\mathbb{P}_{X}, \mathbb{P}_{\hat{X}})\leq\epsilon^{p}\},
	\end{align*}
	and
	\begin{align*}
		\mathcal{F}_{\epsilon} & (c_{1},c_{p})
		:= ~\{F_{X}\in\mathcal{F}_\epsilon :\ \int_{0}^{1}x~\mathrm dF_{X}(x)=c_{1}, \int_{0}^{1}x^{p}~\mathrm dF_{X}(x)=c_{p}\},
	\end{align*}

	where $F_{\hat{X}}$ is the empirical distribution function of $X$ defined by (\ref{add25073001}).
	
	\vspace{0.2cm}

	To reformulate the optimization problem over the Wasserstein ball $\mathcal{F}_{\epsilon}$, we introduce 
	the space of couplings (i.e., joint distributions) of $X$ and its empirical counterpart $\hat{X}$.
	Let $\mathcal{P}([0,1]^2)$ be the set of all probability measures on $([0,1]^2,~\sB([0,1]^2))$, and define
	\begin{align*}
		\Pi(\mathbb{P}_X, \mathbb{P}_{\hat{X}}) :=\left\{\pi\in\mathcal{P}([0,1]^2) ~ : ~\pi_1=\mathbb{P}_{X}, ~ \pi_2 =\mathbb{P}_{\hat{X}}\right\},
	\end{align*}
    where $\pi_1$ and $\pi_2$ are the two marginal probability measures of $\pi$, respectively.
	On the measurable space $([0,1]^2,~\sB([0,1]^2))$, the coordinate projections \(Z\) and \(\hat{Z}\)
	are respectively given by
	$$
	Z(x, \hat{x})=x,  \quad (x,\hat{x})  \in  [0,1] \times [0,1],
	$$

	$$
	\hat{Z}(x, \hat{x})=\hat{x}, \quad  (x,\hat{x}) \in [0,1] \times [0,1].
	$$
    By Definition \ref{def201}, the $p$-th power of the Wasserstein distance of order $p$ is expressed by
	\begin{align}\label{2026051401}
		W_p^p(\mathbb{P}_X, \mathbb{P}_{\hat{X}}) =  \inf_{\pi \in \Pi(\mathbb{P}_X, \mathbb{P}_{\hat{X}})} \int_{[0,1] \times [0,1]}  |x- \hat{x}|^p  \mathrm{d}\pi(x, \hat{x}).                                               
	\end{align}

	Observe that $F_X \in \mathcal{F}_{\epsilon}$ if and only if there exists a coupling $\pi\in\Pi(\mathbb{P}_X, \mathbb{P}_{\hat{X}})$ such that
    $$
    E_{\pi}\left[|Z-\hat{Z}|^{p}\right]=\int_{[0,1] \times [0,1]}|x- \hat{x}|^p  \mathrm{d}\pi(x, \hat{x})\leq\epsilon^p.
    $$
	Indeed, if $F_X\in\mathcal{F}_{\epsilon}$, then by the definition of $\mathcal{F}_{\epsilon},$ $W_p^p(\mathbb{P}_X, \mathbb{P}_{\hat{X}})\leq\epsilon^p$.
	By Theorem 4.1 of Villani (2009) (with cost function $c(x,y):=|x-y|^p$ and $a(x):=b(x):=0$), the infimum in (\ref{2026051401}) is attainable.
	Hence there exists an (optimal) coupling $\pi^*\in\Pi(\mathbb{P}_X, \mathbb{P}_{\hat{X}})$ such that
	\begin{align*}
		E_{\pi^*}\left[|Z-\hat{Z}|^{p}\right]=\int_{[0,1] \times [0,1]}  |x- \hat{x}|^p  \mathrm{d}\pi^*(x, \hat{x}) = W_p^p(\mathbb{P}_X, \mathbb{P}_{\hat{X}}) \leq \epsilon^p.
	\end{align*}
	Conversely, suppose that there is a coupling $\pi\in\Pi(\mathbb{P}_X, \mathbb{P}_{\hat{X}})$ such that $E_{\pi}\left[|Z-\hat{Z}|^{p}\right]\leq\epsilon^p.$ It follows from (\ref{2026051401}) that
	\begin{align*}
		W_p^p(\mathbb{P}_X, \mathbb{P}_{\hat{X}}) \leq \int_{[0,1] \times [0,1]}|x- \hat{x}|^p  \mathrm{d}\pi(x, \hat{x})=E_{\pi}\left[|Z-\hat{Z}|^{p}\right]\leq \epsilon^p,
	\end{align*}
	which yields $F_X\in\mathcal{F}_{\epsilon}$. Therefore, the claimed equivalence is confirmed. 
    
	\vspace{0.2cm}

    In summary, given \(\epsilon>0,\)
	\begin{align}\label{2026051302}
	& \left\{ F_{X}\in\mathcal{F}: W_p^p(\mathbb{P}_X, \mathbb{P}_{\hat{X}}) \leq \epsilon^p \right\} \nonumber\\
	&  \quad =  \{ F_{X}\in\mathcal{F}: \text{ there is a } \pi\in\Pi(\mathbb{P}_X, \mathbb{P}_{\hat{X}})
        \text{ such that } E_{\pi}\left[|Z-\hat{Z}|^{p}\right] \leq \epsilon^p\}.
	\end{align}
	Similarly, given $\epsilon>0$, $p \geq 1$, $0 < c_1 < 1$ and $0 < c_p <\frac{1}{p}$,
	\begin{align}
		\mathcal{F}_{\epsilon}(c_1,c_p) = 
		\{ & F_{X}\in\mathcal{F}: \text{ there is a }  \pi\in\Pi(\mathbb{P}_X, \mathbb{P}_{\hat{X}}) \text{ such that} \nonumber\\
		   & E_{\pi}\left[|Z-\hat{Z}|^{p}\right] \leq \epsilon^p,  \int_{0}^{1}x~\mathrm dF_{X}(x)=c_{1}, \int_{0}^{1}x^{p}~\mathrm dF_{X}(x)=c_{p}\}.
	\end{align}

	\vspace{0.2cm}

	In the present paper, we aim to determine a distribution function $F^*_X$ of $X$ with specific expression
	so that the supremum of the (concave) distortion risk measure of $X$ over distributional uncertainty sets $\mathcal{F}_{\epsilon}$
	and $\mathcal{F}_{\epsilon} (c_{1},c_{p})$ can be attained at $F^*_X$, respectively.
	Precisely, we introduce the optimization problems we will consider in this paper, which are described as follows:
	\begin{equation}\label{supA}
		H_g(F^*_X) = \sup_{F_X \in \mathcal{F}_{\epsilon}} H_g(F_X) \tag{A}
	\end{equation}
	and
	\begin{equation}\label{supB}
		H_g(F^*_X) = \sup_{F_X \in \mathcal{F}_{\epsilon}(c_{1},c_{p})} H_g(F_X), \tag{B}
	\end{equation}
	where $H_g(F_X)$ is defined by (\ref{2001}).
	In both cases, such a $ F^*_X$ is also called an (optimal) solution (i.e. maximizer) to the optimization problem.
 Here, for  simplicity of notations, we have temporarily abused the notation \(F_{X}\) in optimization problems (\ref{supA}) and (\ref{supB}).
On the one hand, \(F_{X}\) always stands for the distribution function of the given random loss \(X\in\mathcal{X}\) under \(\mathbb{P}\). On the other hand,
as a generic element in the uncertainty sets, the notation \(F_{X}\) appearing in the optimization problems   (\ref{supA}) and (\ref{supB}) exactly means that
the distribution function of the given random loss \(X\) is allowed to vary within the uncertainty sets \(\mathcal{F}_{\epsilon}\) and \(\mathcal{F}_{\epsilon}(c_1,c_p)\)
respectively.

	\vspace{0.2cm}
	
	Recall that this paper strives to derive specific expressions for the solution $ F^*_X$ to the optimization problems (\ref{supA}) and (\ref{supB}),
	respectively. In general, it is difficult to derive explicit expression for the maximum values of $H_g(F^*_X)$.
	However, numerical study in Section 4 will demonstrate the computability of the maximum values of $H_g(F^*_X)$ for several distortion functions $g$.
	
	\vspace{0.2cm}
	
	Next, let us comment on the distributional uncertainty sets $\mathcal{F}_{\epsilon}$ and  $\mathcal{F}_{\epsilon}(c_{1},c_{p})$.
	Recall that our aim is to search for certain specific form for the distribution function of $X$.
	Hence, for the optimization problem (\ref{supA}), each element of $\mathcal{F}_{\epsilon}$ represents a plausible  specific form
	(i.e. a specific candidate) for the distribution function of $X$. In other words,
	different elements in  $\mathcal{F}_{\epsilon}$ represent different plausible  specific forms  for the distribution function of $X$.
	Thus, in the formulation of optimization problems (\ref{supA}) and (\ref{supB}) above and also in the sequel,
	we have already used and will continue to use $F_X$ to represent a generic element in $\mathcal{F}_{\epsilon}$ for the distribution function of $X$.
	For instance, the distribution function $F_{\hat{X}}$ of $\hat{X}$ with a specific form,
	which is exactly the empirical distribution function of $X$, represents a specific candidate for the distribution function of $X$.
	Similarly, any solution $F_X^* \in \mathcal{F}_{\epsilon}$ to the optimization problem (\ref{supA}) provides a  specific form
	for the distribution function of $X$. For the sake of terminology, we also call such a $ F_{X}^{*}$ a specific version of the distribution
	function of $X$, or simply, a specific version of $ F_{X}$.
	Analogously, the interpretation above is also applicable to the distributional uncertainty set $\mathcal{F}_{\epsilon}(c_{1},c_{p})$
	and the solution $F_X^* \in \mathcal{F}_{\epsilon}(c_{1},c_{p})$ to the optimization problem (\ref{supB}).
	
	\vspace{0.2cm}
	
	 Note that the optimization problem (\ref{supA}) is well-posed and non-trivial, because $\mathcal{F}_\epsilon$ contains at least two different elements (and hence infinitely many, 
	since $\mathcal{F}_\epsilon$ is convex).
In order to avoid the triviality and ill-posedness of the optimization problem  (\ref{supB}), throughout this paper,
	we assume that the following assumption holds:
	
	\vspace{0.2cm}

	\noindent\textbf{Assumption A} \quad For an appropriately given $\epsilon>0$, $\mathcal{F}_\epsilon(c_1,c_p)$
	contains at least two different elements (and hence infinitely many, since $\mathcal{F}_\epsilon(c_1, c_p)$ is convex),
	so that the optimization problem (\ref{supB}) is well-posed and non-trivial.

	\vspace{0.2cm}
	
	Assumption A is indispensable. For instance, when $p = 2$, then $\epsilon>0$ could not be too small. Indeed, we need that $\epsilon>(\mu-\frac{1}{N}\sum_{i=1}^{N}x_i)^2+(\sigma-\frac{1}{N}(x_i-\frac{1}{N}\sum_{i=1}^{N}x_i))^2$, where $\mu$ and $\sigma$ stand for the mean
	and standard deviation of $X$ respectively.
	For more relevant discussions, we refer to Bernard et al. (2024). Similar assumption was also made in Cornilly et al. (2018, Section 1).
	On the other hand, for general $p \geq 1$, the numerical studies shown in Tables 1 and 2 below also indicate that $\epsilon>0$ should be appropriately chosen.

	\section{Main results}
	
	In this section, we derive closed-form  expressions for the solutions $F^*_X$ to the optimization problems (\ref{supA}) and (\ref{supB}), respectively.
	
	\vspace{0.2cm}
	
	Clearly, a distribution function is a maximizer to optimization problems (A) and (B), respectively,
	if and only if it is a minimizer to the following minimization problems (\ref{inf1A}) and (\ref{inf1B}), respectively:
	\begin{equation}\label{inf1A}
		\min_{F_X \in \mathcal{F_\epsilon}} \int_{0}^{1} \left(1 - g(1 - F_{X}(x))\right)\mathrm dx \tag{A1}
	\end{equation}
	and
	\begin{equation}\label{inf1B}
		\min_{F_X \in\mathcal{F}_{\epsilon}(c_{1},c_{p})} \int_{0}^{1} \left(1 - g(1 - F_{X}(x))\right)\mathrm dx. \tag{B1}
	\end{equation}
	
	\vspace{0.2cm}
	
	Define a function $ \varphi : [0, 1] \to [0, 1]$ by
	\begin{equation*}
		\varphi(y) := 1 - g(1 - y), \quad y \in [0,1].
	\end{equation*}
	Then, $ \varphi$ is strictly convex and twice differentiable, since the distortion function $g$ is strictly concave
	and twice differentiable. Denote by $\varphi'$ and $\varphi''$ the first and second order derivative of $\varphi$, respectively.
	Hence, optimization problems (\ref{inf1A}) and (\ref{inf1B}) can be respectively rewritten as
	\begin{equation}\label{inf2A}
		\inf_{F_{X} \in \mathcal{F}_{\epsilon}} \int_{0}^{1}\varphi(F_{X}(x)) \mathrm dx   \tag{A2}
	\end{equation}
	and
	\begin{equation}\label{inf2B}
		\inf_{F_{X} \in\mathcal{F}_{\epsilon}(c_{1},c_{p})} \int_{0}^{1}\varphi(F_{X}(x)) \mathrm dx.  \tag{B2}
	\end{equation}
	
	\vspace{0.2cm}
	
	Next lemma discusses the existence and uniqueness of optimal solutions to optimization problems (\ref{inf2A}) and (\ref{inf2B}).
	
	\begin{Lemma}\label{lem301}\ \
		There uniquely exist optimal solutions to optimization problems (\ref{inf2A}) and (\ref{inf2B}), respectively.
	\end{Lemma}
	
	\vspace{0.2cm}
	
	\begin{Remark}\label{rmkadd040601}\ \
		In above optimization problems (\ref{inf2A}) and (\ref{inf2B}), if the function $\varphi$ is only supposed to be convex on $[0,1]$,
		then we can replace the first order derivative $\varphi'$ of $\varphi$ on [0,1) with its first order right-derivative $\varphi'_+(x)$, $x\in[0,1)$,
		and assume that $\lim_{x\uparrow1}\varphi'_+(x)\leq\varphi'_-(1)$, where $\varphi'_-(1)$ stands for the first order left-derivative of $\varphi(x)$
		at $x=1$. In this situation, we can find that there still exist optimal solutions to optimization problems (\ref{inf2A}) and (\ref{inf2B}), respectively.
		However, the corresponding optimal solutions might not be unique. Nevertheless, our main results Theorems \ref{thm301} and \ref{thm302} below are still true;
		for details, see Remarks \ref{re302} and \ref{re303} below. Notice also that the above \emph{new} assumptions on $\varphi$ can be equivalently converted to
		the assumptions on the distortion function $g$ as follows: $g$ is supposed to be concave on $[0,1]$; replacing first order derivative $g'$ of $g$ on $(0,1]$
		with its first order left-derivative $g_{-}'$;
		and assuming
		that $\lim_{x\downarrow0}g_{-}'(x) \geq g'_+(0)$, where $g'_+(0)$ stands for the first order right-derivative of $g(x)$ at $x=0$.
	\end{Remark}
	
	\subsection{Solution to optimization problem (\ref{inf2A})}

	First, we rewrite the optimization problem (\ref{inf2A}) as
	\begin{equation}\label{inf3A}
		\inf_{F_{X} \in \mathcal{F}} \int_{0}^{1}\varphi\left(F_{X}(x)\right) \mathrm dx  \tag{A3}
	\end{equation}
	subject to
	\begin{equation}\label{add040201}
		W_{p}^{p}(F_{X},F_{\hat{X}})\leq \epsilon^p.
	\end{equation}
	
	\vspace{0.2cm}

	Next, we transform the set which the optimization problem (A3) is over to the set $\Pi(\mathbb{P}_X, \mathbb{P}_{\hat{X}})$.
	Given any joint distribution $\pi\in\Pi(\mathbb{P}_X, \mathbb{P}_{\hat{X}})$, note that for each  \(1\leq i\leq N\), 
    \[
    \pi(\hat{Z}=x_i)=\pi([0,1]\times\{x_i\})=\mathbb{P}_{\hat{X}}(\{x_i\})=\mathbb{P}(\hat{X}=x_i)=\frac{1}{N},
    \]
    hence by the law of total probability, we have that
	\begin{align}\label{2026060501}
		F_{X}(x) & := \mathbb{P}(X\leq x)=\mathbb{P}_{X}((-\infty,x)) \nonumber\\
                 & = \pi((-\infty,x)\times[0,1]) \nonumber\\
				 & = \sum_{i=1}^{N} \pi((-\infty,x)\times[0,1]|\hat{Z}=x_i) \pi(\hat{Z}=x_i)\nonumber \\
				 & = \frac{1}{N} \sum_{i=1}^{N} \pi((-\infty,x)\times[0,1]~|~\hat{Z}=x_i) \nonumber \\
				 & := \frac{1}{N} \sum_{i=1}^{N} F_{\pi,Z}^{i}(x),
	\end{align} 
    where we have denoted by $F_{\pi,Z}^{i}$, \(1\leq i \leq N\), the conditional distribution function of $Z$ 
	under \(\pi\) conditioned on the event \(\{\hat{Z}=x_i\}\), that is,
	\begin{align}\label{2026052301}
		F_{\pi,Z}^{i}(x):=\pi((-\infty,x)\times[0,1]~|~\hat{Z}=x_i), \quad x\in\mathbb{R}.
	\end{align}
	Clearly, \(F_{\pi,Z}^{i} \in \mathcal{F}\), \(1\leq i \leq N\).
    
    \vspace{0.2cm}

	We proceed to  equivalently convert the constraint (\ref{add040201}) to a more tractable 
	form. Taking (\ref{2026051302}) into account, for any given \(\pi\in\Pi(\mathbb{P}_{X},\mathbb{P}_{\hat{X}})\), by (\ref{2026052301}) and the law of total expectation, we have that
	\begin{equation}\label{E}
		E_{\pi}\left[|Z-\hat{Z}|^{p}\right]=\int_{[0,1] \times [0,1]}|x- \hat{x}|^p  \mathrm{d}\pi(x, \hat{x}) 
		= \frac{1}{N} \sum_{i=1}^{N} \int_{0}^{1}|x-x_i|^{p}F_{\pi,Z}^{i}(\mathrm dx),
	\end{equation}
    for instance, see Tang and Yang (2023, (A5)),
	and
	\begin{equation}\label{add040302}
		\int_{0}^{1}\varphi(F_{X}(x))\mathrm dx=\int_{0}^{1}\varphi\left(\frac{1}{N}\sum_{i=1}^{N}F_{\pi,Z}^{i}(x)\right) \mathrm dx,
	\end{equation}
	where (\ref{2026060501}) has been used, and the conditional distribution functions $F^{i}_{\pi,Z}$ are as in (\ref{2026052301}), $1\leq i \leq N$.
	
    \vspace{0.2cm}
    
    Note that the construction of conditional distribution functions as in $(\ref{2026052301})$ actually establish a mapping between a 
	joint distribution $\pi\in\Pi(\mathbb{P}_X, \mathbb{P}_{\hat{X}})$ and a collection of conditional distributions $\{F_{\pi,Z}^1,...,F_{\pi,Z}^N\}$;
	for more details, see Lemma 3.2 and its proof below. Moreover, by integration by parts,
	\begin{align*}
		\frac{1}{N} & \sum_{i=1}^{N} \int_{0}^{1} \left| x - x_i \right|^p F_{\pi,Z}^{i} (\mathrm dx)
		= \frac{1}{N} \sum_{i=1}^{N}\left[(1-x_{i})^{p}-p\int_{0}^{1}|x-x_{i}|^{p-1}\mbox{sign}(x-x_{i})F_{\pi,Z}^{i}(x) \mathrm dx\right].
	\end{align*}
	Thus, constraint (\ref{add040201}) becomes that
	\begin{equation}\label{add040406}
		-\frac{1}{N} \sum_{i=1}^{N} \int_{0}^{1} |x-x_{i}|^{p-1} \mbox{sign}(x-x_{i}) F_{\pi,Z}^{i}(x)\mathrm dx
		\leq \frac{1}{p}\left(\epsilon^{p}-\frac{1}{N} \sum_{i=1}^{N}(1-x_{i})^{p}\right).
	\end{equation}
	
	\begin{Lemma}\label{2026051301}
		 $F_{X}^*$ is an optimal
		solution to (\ref{inf3A}) if and only if there is a \(\pi^*\in\Pi(\mathbb{P}_{X},\mathbb{P}_{\hat{X}})\) such that
         the corresponding conditional distribution
		functions $F_{\pi^*,Z}^{1}, \ldots, F_{\pi^*,Z}^{N}$ as in (\ref{2026052301}) constitute an optimal solution to the following
		optimization problem (A4):
		\begin{align}\label{inf4A}
		\inf_{\substack{F^1_{\pi,Z}, \ldots, F^N_{\pi,Z} \in \mathcal{F}}}
		\int_{0}^{1}\varphi\Bigl(\frac{1}{N} \sum_{i=1}^{N}F_{\pi,Z}^{i}(x)\Bigr) \mathrm dx \tag{A4}
		\end{align}
		subject to
		\begin{equation}\label{add040202}
			-\frac{1}{N} \sum_{i=1}^{N} \int_{0}^{1} |x-x_{i}|^{p-1} \mbox{sign}(x-x_{i}) F_{\pi,Z}^{i}(x)\mathrm dx
			\leq \frac{1}{p} \left(\epsilon^{p}-\frac{1}{N} \sum_{i=1}^{N}(1-x_{i})^{p}\right),
		\end{equation}
		where the conditional distribution functions $F_{\pi,Z}^{1}, \ldots, F_{\pi,Z}^{N}$ are as in (\ref{2026052301}).
		 In this case, the relation between \(F_{X}^*\) and the collection of \(\{F_{\pi^*,Z}^1,...,F_{\pi^*,Z}^N\}\)  
		 is given by \(F_{X}^*=\frac{1}{N}\sum_{i=1}^{N}F_{\pi^*,Z}^i\).
	\end{Lemma}
	
    \vspace{0.2cm}
    
	Notice also that the optimal solution to optimization problem (\ref{inf4A}) is also unique, because the mapping
    \begin{align*}
    (F_{\pi,Z}^{1}, \ldots ,F_{\pi,Z}^{N}) \rightarrow  \int_{0}^{1}\varphi\left(\frac{1}{N} \sum_{i=1}^{N}F_{\pi,Z}^{i}(x)\right) \mathrm dx
    \end{align*}
    is strictly convex due to the strict convexity of $\varphi$.
    In summary, we have already equivalently converted optimization problem (\ref{inf3A}) to optimization problem (\ref{inf4A})
    via the relation $F_X^* = \frac{1}{N}\sum_{i=1}^{N} F_{\pi^*,Z}^{i}$. It turns out that the collection
     \(\{F_{\pi^*,Z}^1,...,F_{\pi^*,Z}^N\}\)  constituting the optimal solution to the optimization problem (\ref{inf4A}) satisfies
    the following property: for each \(1\leq i\leq N-1\), 
	\(\sup S(F_{\pi^*,Z}^i)\leq \inf S(F_{\pi^*,Z}^{i+1})\), where \(S(F)\) stands for the support
   of \(F\in\mathcal{F}\) which is defined by \(S(F) := \{x \in [0, 1] : 0 < F(x) < 1\}\);
     for more details, see the explainations after Theorem 3.1 below.

	\vspace{0.2cm}
	
	Next, inspired by Rustagi (1975), we provides sufficient and necessary condition for
	certain conditional distribution functions 
	$F_{\tilde{\pi},Z}^{1}, \ldots ,{F_{\tilde{\pi},Z}^{N}} \in \mathcal{F}$ for some \(\tilde{\pi}\in\Pi(\mathbb{P}_{X},\mathbb{P}_{\hat{X}})\) to constitute
	the unique optimal solution to optimization problem (\ref{inf4A}), 
	which is stated in Lemma \ref{lem302} below.
	
	\vspace{0.2cm}
	
	\begin{Lemma}\label{lem302}\ \
		Fix arbitrarily conditional distribution functions $F_{\pi^*,Z}^{1}, \ldots, F_{\pi^*,Z}^{N} \in \mathcal{F}$ as in (\ref{2026052301}) for some \(\pi^*\in\Pi(\mathbb{P}_{X},\mathbb{P}_{\hat{X}})\)
		such that $F_{\pi^*,Z}^{1}, \ldots ,F_{\pi^*,Z}^{N}$ satisfy the constraint (\ref{add040202}) of optimization problem (\ref{inf4A}).
		Then  $F_{\pi^*,Z}^{1},  \ldots , F_{\pi^*,Z}^{N}$ constitute the unique optimal solution to optimization problem (\ref{inf4A})
		if and only if it holds that
		\begin{align}{\label{(2)}}
			& \int_{0}^{1} \varphi'\left(\frac{1}{N}\sum_{i=1}^{N} F_{\pi^*,Z}^{i}(x)\right) \left(\frac{1}{N}\sum_{j=1}^{N}F_{\pi,Z}^{j}(x)\right) \mathrm dx \nonumber \\
			& \geq \int_{0}^{1} \varphi'\left(\frac{1}{N}\sum_{i=1}^{N}F_{\pi^*,Z}^{i}(x)\right)
			\left(\frac{1}{N}\sum_{j=1}^{N}F_{\pi^*,Z}^{j}(x)\right) \mathrm dx
		\end{align}
		for all conditional distribution functions $F_{\pi,Z}^{1}, \ldots ,F_{\pi,Z}^{N} \in \mathcal{F}$
		with $\frac{1}{N}\sum_{i=1}^{N}F_{\pi,Z}^{i} = F_X \in \mathcal{F}_\epsilon$.
		In addition, $F_X^* := \frac{1}{N}\sum_{i=1}^{N} F_{\pi^*,Z}^{i} \in \mathcal{F}_\epsilon$.
	\end{Lemma}	
	\vspace{0.2cm}
	
	By Lemma \ref{lem302}, we know that if conditional distribution functions 
	$F_{\pi^*,Z}^{1}, \ldots ,F_{\pi^*,Z}^{N}$ for some \(\pi^*\in\Pi(\mathbb{P}_{X},\mathbb{P}_{\hat{X}})\) constitute the unique optimal solution
	to optimization problem (\ref{inf4A}),
	then $F_{\pi^*,Z}^{1}, \ldots ,F_{\pi^*,Z}^{N}$ also constitute an optimal solution to the following auxiliary optimization problem:
	\begin{equation}\label{infC}
		\inf_{\substack{F^1_{\pi,Z}, \ldots, F^N_{\pi,Z} \in \mathcal{F}}}
		\frac{1}{N}\sum_{j=1}^{N} \int_{0}^{1}\varphi'\left(\frac{1}{N}\sum_{i=1}^{N}F_{\pi^*,Z}^{i}(x)\right)F_{\pi,Z}^{j}(x)\mathrm dx \tag{C}
	\end{equation}
	subject to
	\begin{equation}\label{add040407}
		-\frac{1}{N} \sum_{j=1}^{N} \int_{0}^{1} |x-x_{j}|^{p-1} \mbox{sign}(x-x_{j}) F_{\pi,Z}^{j}(x)\mathrm dx
		\leq\frac{1}{p}\left(\epsilon^{p}-\frac{1}{N} \sum_{i=1}^{N}(1-x_{i})^{p}\right).
	\end{equation}
	
	\vspace{0.2cm}
	
	From now on, we think of the optimization problem (\ref{infC}) as a stand-alone optimization problem,
	and let the conditional distribution functions $F_{\pi^*,Z}^{1}, \ldots ,F_{\pi^*,Z}^{N}$ in the optimization problem (\ref{infC}) be such that
	they constitute the unique optimal solution to the optimization problem (\ref{inf4A}).
	
	\vspace{0.2cm}
	
	Our approach is as follows.  By analyzing optimization problem (\ref{infC}),
	we first derive a closed-form expression for each conditional distribution function 
	$F_{\pi^*,Z}^{i}$, $1 \leq i \leq N$, respectively,
	and then aggregate them into a single distribution function via
	 $F_X^* := \frac{1}{N}\sum_{i=1}^{N} F_{\pi^*,Z}^{i}$
	which is exactly the desired unique optimal solution to optimization problem (\ref{inf3A}).
	
	\vspace{0.2cm}
	
	Notice that the constraint condition (\ref{add040407}) of optimization problem (\ref{infC}) has two possibilities:
	\begin{enumerate}
		\item[(i)] the constraint condition (\ref{add040407}) does not make contribution to the optimization problem (\ref{infC}).
		\item[(ii)] the constraint condition (\ref{add040407}) does make contribution to the optimization problem (\ref{infC}).
	\end{enumerate}
	
	\vspace{0.2cm}
	
	First, we consider possibility (i).
	In this case, since $\varphi'>0$, we straightforwardly know that the unique optimal solution $F_{\pi^*,Z}^{1}, \ldots ,F_{\pi^*,Z}^{N}$ to
	the optimization problem (\ref{infC}) are given by
	\begin{align}\label{add070501}
		F_{\pi^*,Z}^{i}(x)=
		\begin{cases}
			0, \quad &0\leq x<1,\\
			1, \quad &x=1,
		\end{cases}
	\end{align}
	$1\leq i\leq N.$ Therefore, the unique optimal solution $F_{X}^{*}:=\frac{1}{N}\sum_{i=1}^{N}F_{\pi^*,Z}^{i}$ to optimization problem (\ref{inf2A}) is given by
	\begin{align}\label{add070301}
		F_{X}^{*}(x) =
		\begin{cases}
			0, \quad &0\leq x<1,\\
			1, \quad &x=1.
		\end{cases}
	\end{align}
	
	\vspace{0.2cm}
	
	Next, we consider possibility (ii).
	In this case, we need one more auxiliary optimization problem (\ref{infD}) below,
	since it is more convenient for us to analyze than the optimization problem (\ref{infC}).
	We will show that the given $F_{\pi^*,Z}^{1}, \ldots, F_{\pi^*,Z}^{N}$ as in the optimization problem (\ref{infC}) also constitute an optimal solution to
	the optimization problem  (\ref{infD}) ; see Lemma \ref{auxiliaryD} below.
	Finally, by examining the properties of $F_{\pi^*,Z}^{1}, \ldots, F_{\pi^*,Z}^{N}$ via analyzing the auxiliary optimization problem  (\ref{infD}),
	we can derive closed-form expressions for $F_{\pi^*,Z}^{i}$, $1 \leq i \leq N$.
	
	\vspace{0.2cm}
	
	\begin{Lemma}\label{auxiliaryD} \ \
		Let $F_{\pi^*,Z}^{1}, \ldots, F_{\pi^*,Z}^{N}$ be as in the optimization problem (\ref{infC}).
		Then $F_{\pi^*,Z}^{1}, \ldots, F_{\pi^*,Z}^{N}$ constitute an optimal solution to the following optimization problem (\ref{infD}):
		\begin{align}\label{infD}
			\inf_{\substack{F^1_{\pi,Z}, \ldots, F^N_{\pi,Z} \in\mathcal{F}}}
			\frac{1}{N}\sum_{i=1}^{N}\int_{0}^{1}    \varphi' \left(  \frac{1}{N}   \left(F_{\pi^*,Z}^{i}(x) + (i-1)\right)  \right)   F_{\pi,Z}^{i}(x)\mathrm dx  \tag{D}
		\end{align}
		subject to
		\begin{equation*}
			-\frac{1}{N} \sum_{i=1}^{N} \int_{0}^{1} |x-x_{i}|^{p-1} \mbox{sign}(x-x_{i}) F_{\pi,Z}^{i}(x)\mathrm dx
			\leq \frac{1}{p} \left(\epsilon^{p}-\frac{1}{N} \sum_{i=1}^{N}(1-x_{i})^{p}\right).
		\end{equation*}
	\end{Lemma}
	
	\vspace{0.2cm}
	
	From now on, we also think of the optimization problem (\ref{infD}) as a stand-alone optimization problem,
	and let the conditional distribution functions $F_{\pi^*,Z}^{1}, \ldots ,F_{\pi^*,Z}^{N}$ in the optimization problem (\ref{infD})
	be as in the optimization problem (\ref{infC}),
	that is, $F_{\pi^*,Z}^{1}, \ldots ,F_{\pi^*,Z}^{N}$ constitute the unique optimal solution to the optimization problem (\ref{inf4A}).
	
	\vspace{0.2cm}
	
	By the Lagrangian multiplier method, we will see that $F_{\pi^*,Z}^{1}, \ldots, F_{\pi^*,Z}^{N}$ as in the optimization problem (\ref{infD})
	constitute a solution to the optimization problem (\ref{infD}) if and only if they, together with a Lagrangian multiplier, constitute a solution
	to the equivalent unconstrained version (\ref{add040610}) of the optimization problem (\ref{infD});
	see the (unconstrained) optimization problem (\ref{add040610}) below.
	Moreover,  let $F_{\pi^*,Z}^{1}, \ldots, F_{\pi^*,Z}^{N}$ be as in the optimization problem (\ref{infD}),
	then from the proof of Lemma \ref{auxiliaryD} below, it follows that for each $1 \leq i \leq N$, $F_{\pi^*,Z}^{i}$ must minimize
	$\int_{0}^{1} A_i(x)F_{\pi,Z}^{i}(x) \mathrm dx$ over $\mathcal{F}$, that is,
	\begin{align*}
		\int_{0}^{1} A_i(x)F_{\pi^*,Z}^{i}(x) \mathrm dx = \inf_{\substack{ F^{i}_{\pi,Z} \in \mathcal{F} } } \int_{0}^{1} A_i(x)F_{\pi,Z}^{i}(x) \mathrm dx,
	\end{align*}
	where we have used the notation $A_i(x)$ as in (\ref{add25073102}) in the proof of Lemma \ref{auxiliaryD}, which is exactly defined by
	\begin{align*}
		A_i(x) :=  \varphi'\left(  \frac{1}{N}   \left(F_{\pi^*,Z}^{i}(x)+ (i-1)\right)  \right)  - \lambda |x-x_{i}|^{p-1} \mbox{sign}(x-x_{i}), \quad x \in [0,1],
	\end{align*}
	where $\lambda > 0$ is the Lagrange multiplier. Inspired by Rustagi (1975), we suggest that $F_{\pi^*,Z}^{1}, \ldots, F_{\pi^*,Z}^{N}$ should posses
	special properties and expressions.
	The following Lemmas \ref{lem304}-\ref{lem306} discuss some special properties of $F_{\pi^*,Z}^{1}, \ldots, F_{\pi^*,Z}^{N}$, which are crucial for
	us to figure out closed-form expression for $F_{\pi^*,Z}^{i}$, $ 1\leq i \leq N $.
	
	\vspace{0.2cm}
	
	\begin{Lemma}\label{lem304}\ \
		For each $1\leq i \leq N$, since \( F_{\pi^*,Z}^{i} \) minimizes $\int_{0}^{1}A_{i}(x)F_{\pi,Z}^{i}(x)\mathrm dx$ over $\mathcal{F}$, the set $ S_{i} := \{x \in (0,1) : \ A_{i}(x)\neq 0\}$ has \( F_{\pi^*,Z}^{i} \)-measure zero.
	\end{Lemma}
	
	\vspace{0.2cm}
	
	If $S_i$ as in Lemma \ref{lem304} is not empty, then $S_i$  must be countable union of either open sub-interval or sub-interval containing left-end point of $(0,1)$ due to the right-continuity of $A_i(x)$;  see the proof of Lemma \ref{lem304} below.
	In this case, Lemma \ref{lem304} says that $F_{\pi^*,Z}^{i}$ remains constant on $S_i$.
	
	\vspace{0.2cm}
	
	\begin{Lemma}\label{lem305}\ \
		For each $1\leq i \leq N$, suppose that $F_{\pi^*,Z}^{i}$ minimizing $\int_{0}^{1}A_{i}(x)F_{\pi,Z}^{i}(x)\mathrm dx$ over $\mathcal{F}$
		remains some constant on certain interval, that is,
		\begin{align*}
			F_{\pi^*,Z}^{i}(x)
			\begin{cases}
				< c, & \text{if} \ x<a, \\
				= c, & \text{if} \ a\leq x<b, \\
				> c, & \text{if} \ x\geq b,
			\end{cases}
		\end{align*}
		with some $0<c<1$ and $0\leq a<b\leq1$. Then \(\int_{a}^{b}A_{i}(x)\mathrm dx=0\).
	\end{Lemma}
	
	\vspace{0.2cm}
	
	If $F_{\pi^*,Z}^{i}$ remains some constant $c\in(0,1)$ on certain interval, then Lemma \ref{lem305} says that
	the constant $c$ and the Lagrange multiplier $\lambda$ have certain relationship between each other. Notice that the value of the constant $c$
	does not influence the integral $\int_{0}^{1}A_{i}(x)F_{\pi,Z}^{i}(x)\mathrm dx.$
	
	\vspace{0.2cm}
	
	\begin{Lemma}\label{lem306}\ \
		For each $1\leq i\leq N$, since \( F_{\pi^*,Z}^{i}\) minimizes $\int_{0}^{1}A_{i}(x)F_{\pi,Z}^{i}(x)\mathrm dx$ over $\mathcal{F}$,
		$F_{\pi^*,Z}^{i}$  has no jumps on the open interval $(0,1)$.
	\end{Lemma}
	
	\vspace{0.2cm}
	
	In summary,  preceding three lemmas show that $F_{\pi^*,Z}^{i}$ remains constant on the set
	$\{ x \in (0,1) : \ A_i(x) \neq 0 \}$, $1\leq i\leq N$. For $x \in (0,1)$ with $A_i(x)=0$,  $F_{\pi^*,Z}^{i}(x)$
	is determined by the equation $A_i(x)=0$. Additionally, $F_{\pi^*,Z}^{i}$ is continuous over the interval (0,1).
	These properties collectively can yield a closed-form expression for $F_{\pi^*,Z}^{i}$, which is stated in the Theorem \ref{thm301} below.
	
	\vspace{0.2cm}
	
	Now, we are ready to state the first main result of this paper, which provides the closed-form expression for the unique optimal solution
	to the optimization problem (\ref{inf2A}). Recall that the conditional distribution functions $F_{\pi^*,Z}^{1}, \ldots ,F_{\pi^*,Z}^{N}$ in the optimization
	problem (\ref{infD}) are already fixed such that they constitute the unique optimal solution to the optimization problem (\ref{inf4A}).
	
	\vspace{0.2cm}
	
	\begin{Theorem}\label{thm301}\ \
		Suppose that $p\geq1$. Let $F_{\pi^*,Z}^{1}, \ldots, F_{\pi^*,Z}^{N}$ be as in the optimization problem (\ref{infD}).
		\begin{enumerate}
			\item[(1)] If the constrain condition (\ref{add040407}) does not make contribution to the optimization problem (\ref{infC}), then the unique optimal solution $F_{X}^{*}$ to the optimization problem (\ref{inf2A}) is given
			by \begin{align*}
				F_{X}^{*}(x) =
				\begin{cases}
					0, \quad &0\leq x<1,\\
					1, \quad &x=1.
				\end{cases}
			\end{align*}
			\item[(2)] If the constrain condition (\ref{add040407}) does make contribution to the optimization problem (\ref{infC}), then $F_{\pi^*,Z}^{1}, \ldots, F_{\pi^*,Z}^{N}$ are given by
			\begin{align*}
				{F}_{\pi^*,Z}^{i}(x)=
				\begin{cases}
					0,& \text{if } x < \min\{a_i,1\}\\
					N-N(g')^{-1}(\lambda^* |x-x_{i}|^{p-1}\mbox{sign}(x-x_{i}))-(i-1),   & \text{otherwise},\\
					1,&\text{if } x\geq \mbox{min} \{b_i, 1\},
				\end{cases}
			\end{align*}
			$1 \leq i \leq N$, where $(g')^{-1}$ is the inverse function of $g'$, $\lambda^* >0$ is determined by
			\begin{align}\label{add070303}
				-\frac{1}{N} \sum_{i=1}^{N} \int_{0}^{1} |x-x_{i}|^{p-1} \mbox{sign}(x-x_{i}) F_{\pi^*,Z}^{i}(x)\mathrm dx
				= \frac{1}{p}\left(\epsilon^{p}-\frac{1}{N} \sum_{i=1}^{N}(1-x_{i})^{p}\right),
			\end{align}
			and for each $1\leq i\leq N$, $a_i>0$ is the unique solution to
			\begin{align}\label{2026060701}
			\lambda^* |x-x_{i}|^{p-1}\mbox{sign}(x-x_{i})=g'\left(\frac{N+1-i}{N}\right)
			\end{align}
			with respect to variable $x$ on $(0,+\infty)$, and $b_i>0$ is the unique solution to
			\begin{align}\label{2026060702}
			\lambda^* |x-x_{i}|^{p-1}\mbox{sign}(x-x_{i})=g'\left(\frac{N-i}{N}\right)
			\end{align}
			with respect to variable $x$ on $(0, +\infty)$.
			
			\quad Furthermore,
			the distribution function $F_{X}^{*} := \frac{1}{N} \sum_{i=1}^{N}F_{\pi^*,Z}^{i}$ is the unique optimal solution (minimizer) to
			the optimization problem (\ref{inf2A}). Precisely, $F_{X}^*$ is given by
			\begin{align*}
				{F}_{X}^{*}(x)=
				\begin{cases}
					\frac{i-1}{N},& \text{if } x < a_i\\
					1-(g')^{-1}(\lambda^* |x-x_{i}|^{p-1}\mbox{sign}(x-x_{i}))   & \text{otherwise},\\
					\frac{i}{N},&\text{if } x\geq  b_i,
				\end{cases}
			\end{align*}
			on $\left(F_{X}^{*-1}(\frac{i-1}{N}),\quad F_{X}^{*-1}(\frac{i}{N})\right],$ $1\leq i\leq N.$
		\end{enumerate}
	\end{Theorem}
	
	\vspace{0.2cm}
		
      	Notice that $a_1\leq b_1\leq a_2 \leq b_2 \leq...\leq a_N\leq b_N$, 
		where \(a_i\) and \(b_i\), \(1\leq i \leq N\), are as in Theorem 3.1(2). 
        Moreover, \(\sup S(F_{\pi^*,Z}^i)\leq \inf S(F_{\pi^*,Z}^{i+1})\), \(1\leq i\leq N-1\).
        In fact, \(g'\geq0\), since \(g\) is increasing.
		Clearly, \(\lambda^*|x-x_i|^{p-1}\mbox{sign}(x-x_i)\) is non-negative whenever \(x\geq x_i\), 
		and negative whenever \(x<x_i\).
		Hence, form (\ref{2026060701}) and (\ref{2026060702}) it follows that
		\begin{align*}
        a_i=\left[\frac{g'(\frac{N+1-i}{N})}{\lambda^*}\right]^{\frac{1}{p-1}}+x_i, \quad
		b_i=\left[\frac{g'(\frac{N-i}{N})}{\lambda^*}\right]^{\frac{1}{p-1}}+x_i. 
        \end{align*}
		Since \(g\) is strictly concave, \(g'\) is non-increasing. Thus \(a_i\leq b_{i}\).
		Meanwhile, \(a_{i+1}=\left[\frac{g'(\frac{N-i}{N})}{\lambda^*}\right]^{\frac{1}{p-1}}+x_{i+1}\). Therefore \(b_{i}\leq a_{i+1}\), because \(x_{i}\leq x_{i+1}\).
        Clearly, from the expression of \(F_{\pi,Z}^{i}\) it follows that 
		\(S(F_{\pi,Z}^{i})\subset[a_i,b_i)\), \(1\leq i \leq N\). 
		Thus, \(\sup S(F_{\pi^*,Z}^i)\leq \inf S(F_{\pi^*,Z}^{i+1})\), \(1\leq i\leq N-1\).

	\vspace{0.2cm}

	\begin{Remark}\label{re302}\ \
		Suppose that  \( \varphi \) is a convex function as in Remark 3.1. Then by checking the procedure of solving the optimization problem (\ref{inf2A}),
		we can find that Theorem \ref{thm301} still remains true.
		Precisely, what we need to modify is only to replace the first order derivative $g'$ of $g$ with its first order left-derivative $g'_-$ in Theorem 3.1.
	\end{Remark}
	
	\vspace{0.2cm}

	\begin{Remark}
		Suppose that $p = 2,$ and $\varphi$ is a convex function. When considering the distribution function on $[0,1]$, the expression of $F_{X}^*$ is influenced
		by the endpoints $0$ and $1$, which can be seen from Theorem \ref{thm301}.
		In fact, we can extend Theorem \ref{thm301} to general unbounded random loss, that is, the loss distribution function is supported on the whole real line.
		Recall that in this situation, the samples $x_1 \leq  \ldots  \leq x_N$ belong to real numbers.
		Precisely, we first consider a distribution function supported on $[-b,b]$, $b>0$, and then take the limit of $b$ to infinity. Then by Theorem \ref{thm301}
		and Remark \ref{re302}, we know that the unique solution $F_{X}^{*}$ to optimization problem (\ref{inf2A}) is given by $F_{X}^{*}:=\frac{1}{N}\sum_{i=1}^{N}F_{\pi^*,Z}^{i},$ where
		\begin{align*}
			{F}_{\pi^*,Z}^{i}(x)=
			\begin{cases}
				0,& \text{if } \lambda^*(x-x_i) < g'_-\left(\frac{N+1-i}{N}\right),\\
				N-N(g'_-)^{-1}(\lambda^* (x-x_{i}))-(i-1),   & \text{otherwise},\\
				1,&\text{if } \lambda^*(x-x_i) \geq g'_-\left(\frac{N-i}{N}\right).
			\end{cases}
		\end{align*}
		
		In addition, the left-continuous inverse function ${F}_{X}^{*-1}$ of the distribution function ${F}_{X}^{*}$ is given by
		\begin{align}\label{add042301}
			\lambda^*F_{X}^{*-1}(q)=g'_-(1-q)+\lambda^*x_i, \quad \frac{i-1}{N}<q\leq \frac{i}{N},~1\leq i\leq N.
		\end{align}
		For  more detailed arguments, we refer to the proof of Corollary 3.1 below by setting $\tilde{\eta}_1,\ \tilde{\eta}_2$ there being zero.
	\end{Remark}
	
	\vspace{0.2cm}
	
	Next, we discuss the connection of (\ref{add042301}) with Theorem 4.8 of Bernard et al. (2024). Suppose that the reference distribution function $F$
	in Theorem 4.8 of Bernard et al. (2024) is given by the empirical distribution function $F_{\hat{X}}$, that is,
	$$
	F(x):=\frac{1}{N}\sum_{i=1}^{N}I_{(-\infty,x]}(x_i),\quad x\in\mathbb{R},
	$$
	where $x_1 \leq  \ldots  \leq x_N$, $x_i\in\mathbb{R}$, $1\leq i\leq N$, are samples.  Then, from Theorem 4.8 of Bernard et al. (2024), it follows that
	the inverse function $G^{-1}$ of the corresponding optimal solution $G$ is given by
	\begin{align}\label{add042302}
		G^{-1}(u)=\mu^*+\sigma^*\left(\frac{g'_-(1-u)+\lambda x_i-a_\lambda}{b_\lambda}\right),\quad \frac{i-1}{N}<u\leq\frac{i}{N},~1\leq i\leq N,
	\end{align}
	where $\lambda$, $a_\lambda$, $b_{\lambda}$, $\mu^*$  and $\sigma^*$ are as in Theorem 4.8 of Bernard et al. (2024). In addition, we can find that
	$\lambda =\frac{b_{\lambda}}{\sigma^*}$ and $\mu^*=\frac{\sigma^*}{b_{\lambda}}a_\lambda$.

	On the other hand, by (\ref{add070303}), we can know that $\lambda^*=\lambda$, where $\lambda$ is as in  Theorem 4.8 of Bernard et al. (2024).
	Substituting $\lambda^*$ into $F_{X}^{*-1}$ as in (\ref{add042301}),
	we find that $F_{X}^{*-1}$ as in (\ref{add042301}) is consistent with  $G^{-1}$ as in (\ref{add042302}).
	
	\vspace{0.2cm}
	
	Notice that unlike the method used by Bernard et al. (2024, Theorem 4.8), in this paper, we employ Lagrangian multiplier method to deal with the case of more general Wasserstein distance of order $p\geq1$.
	Taking the above considerations into account, the present study of the optimization problem (\ref{inf2A}) can be viewed as a meaningful complement
	to the study of Bernard et al. (2024, Theorem 4.8).

	{\subsection{Solution to optimization problem (\ref{inf2B})}}

	In this subsection, we study the optimization problem (\ref{inf2B}), that is, to find $F_X^* \in \mathcal{F}_{\epsilon}(c_1,c_p)$ with
	specific expression such that
	\begin{align*}
		\int_{0}^{1}\varphi(F_{X}^*(x)) \mathrm dx =  \inf_{F_{X} \in \mathcal{F}_{\epsilon}(c_1,c_p)} \int_{0}^{1}\varphi(F_{X}(x)) \mathrm dx.  \tag{B2}
	\end{align*}
	
	\vspace{0.2cm}
	
	Recall that by Lemma \ref{lem301}, the optimal solution to the optimization problem (\ref{inf2B}) is unique. From now on, denote by  $F_X^*$
	the unique solution to the optimization problem (\ref{inf2B}).
	
	\vspace{0.2cm}
	
	\begin{Lemma}\label{lem307}\ \
		There exist Lagrange multipliers
		$\eta_1^*, \eta_p^* \in \mathbb{R}$
		such that $F_X^*$ minimizes the functional
		$$
		F_X \rightarrow \int_{0}^{1}\left[\varphi'(F_{X}^*(x))+\eta_1^*+\eta_p^*x^{p-1}\right]F_{X}(x) \mathrm dx
		$$
		over $\mathcal{F}_{\epsilon}(c_1, c_p)$, that is, $F_X^*$ is an optimal solution to the following auxiliary optimization problem (\ref{infO}):
		\begin{align}\label{infO}
			\inf_{F_{X} \in \mathcal{F}_{\epsilon}}
			\int_{0}^{1}\left[\varphi'(F_{X}^*(x))+\eta_1^*+\eta_p^*x^{p-1}\right]F_{X}(x) \mathrm dx,   \tag{O}
		\end{align}
		subject to
		\begin{align*}
			\int_{0}^{1}x ~\mathrm{d}F_{X}(x)=c_1 \quad \text{and} \quad \int_{0}^{1}x^p ~\mathrm{d}F_{X}(x)=c_p.
		\end{align*}
	\end{Lemma}
	
	\vspace{0.2cm}
	
	From now on, we think of the optimization problem (\ref{infO}) as a stand-alone optimization problem, in which $F_X^*$ is the unique solution
	to optimization problem (\ref{inf2B}).
	Similar to (\ref{2026052301}),
	for any $F_X \in \mathcal{F}_{\epsilon}(c_1,\ c_p)$, we denote by $F^{i}_{\pi,Z}$ the conditional distribution function, $1\leq i \leq N$,
	and by  $F^{i}_{\pi^*,Z}$ the conditional distribution function corresponding to $F_X^*$, $1\leq i \leq N$. Then,
	\begin{equation*}
		F_X = \frac{1}{N} \sum_{i=1}^{N}F_{\pi,Z}^{i} \quad \mbox{and} \quad F_X^* =\frac{1}{N} \sum_{i=1}^{N}F_{\pi^*,Z}^{i}.
	\end{equation*}
	Hence, by (\ref{add040406}) and Lemma \ref{lem307}, we know that $F^{1}_{\pi^*,Z}, \ldots, F^{N}_{\pi^*,Z}$ constitute an optimal solution
	to the following auxiliary optimization problem (\ref{infP}):
	\begin{equation}\label{infP}
		\inf_{F_{\pi,Z}^{1}, \ldots, F_{\pi,Z}^{N} \in \mathcal{F}}
		\frac{1}{N}\sum_{j=1}^{N} \int_{0}^{1}\left(\varphi'\left(\frac{1}{N}\sum_{i=1}^{N}F_{\pi^*,Z}^{i}(x)\right)+\eta_{1}^*+\eta_{p}^*x^{p-1}\right)
		F_{\pi,Z}^{j}(x)\mathrm dx     \tag{P}
	\end{equation}
	subject to
	\begin{align*}
		\int_{0}^{1}x ~\mathrm{d}\left(\frac{1}{N}\sum_{j=1}^{N} F_{\pi,Z}^{j}(x)\right)=c_1,
	\end{align*}
	\begin{align*}
		\int_{0}^{1}x^p ~\mathrm{d}\left(\frac{1}{N}\sum_{j=1}^{N} F_{\pi,Z}^{j}(x)\right)=c_p
	\end{align*}
	and
	\begin{align}\label{add070305}
		-\frac{1}{N} \sum_{j=1}^{N}\int_{0}^{1} |x-x_{j}|^{p-1}\mbox{sign}(x-x_{j})F_{\pi,Z}^{j}(x)\mathrm dx
		\leq \frac{1}{p}\left(\epsilon^{p}-\frac{1}{N} \sum_{i=1}^{N}(1-x_{i})^{p}\right).
	\end{align}
	
	\vspace{0.2cm}
	
	It turns out that the collection
     \(\{F_{\pi^*,Z}^1,...,F_{\pi^*,Z}^N\}\)  constituting the optimal solution to optimization problem (\ref{infP}) satisfies
    the following property: for each \(1\leq i\leq N-1\), \(\sup S(F_{\pi^*,Z}^i)\leq \inf S(F_{\pi^*,Z}^{i+1})\);
     for more details, see the explainations after Theorem 3.2 below.

   \vspace{0.2cm}

	From now on, we think of the optimization problem (\ref{infP}) as a stand-alone optimization problem,
	and fix the conditional distribution functions $F_{\pi^*,Z}^{1}, \ldots ,F_{\pi^*,Z}^{N}$ as in the optimization problem (\ref{infP}) such that $\frac{1}{N}\sum_{i=1}^{N}F_{\pi^*,Z}^{i}=F_X^*$.
	Recall that in this situation, $F_{\pi^*,Z}^{1}, \ldots ,F_{\pi^*,Z}^{N}$ constitute an optimal solution to the optimization problem (\ref{infP}), and that $F_X^*$ is the unique optimal solution to the optimization problem (\ref{inf2B}), and thus also an optimal solution to the optimization problem (\ref{infO}).
	
	\vspace{0.2cm}
	
	Similar to previous subsetion,
	the constraint condition (\ref{add070305}) of optimization problem (\ref{infP}) has two possibilities:
	\begin{enumerate}
		\item[(i)] the constraint condition (\ref{add070305}) does not make contribution to the optimization problem (\ref{infP}).
		\item[(ii)] the constraint condition (\ref{add070305}) does make contribution to the optimization problem (\ref{infP}).
	\end{enumerate}
	
	\indent First, we consider possibility (i). In this case, we can directly study the optimization problem (\ref{infO}) itself, and can steadily obtain that the optimal solution $F_{X}^{*}$ to the optimization  problem (\ref{infO}) (hence also to the optimization problem (B2)) is given by
	\begin{align*}
		F_{X}^{*}(x)=
		\begin{cases}
			0, \quad &\text{if~}  x<\max\left(\left(\frac{g'(1)+\eta_1^*}{-\eta_p^*}\right)^{\frac{1}{p-1}},0\right),\\
			1-(g')^{-1}(-\eta_1^*-\eta_p^*x^{p-1}), \quad &\text{otherwise},\\
			1, \quad &\text{if~}  x\geq \min\left(\left(\frac{g'(0)+\eta_1^*}{-\eta_p^*}\right)^{\frac{1}{p-1}},1\right),
		\end{cases}
	\end{align*}
	where the parameters $\eta_1^*\in\mathbb{R}$ and $\eta_p^*<0$  satisfy
	$$\int_{0}^{1}x~\mathrm{d} F_{X}^{*}(x)=c_1,\quad \text{and} \quad \int_{0}^{1}x^p~\mathrm{d} F_{X}^{*}(x)=c_p.$$
	\noindent For more details, see the proof of Theorem 3.2(1) below.
	
	\vspace{0.2cm}
	
	Next, we consider possibility (ii). In this case, we transform analyzing the optimization problem (\ref{infP}) to analyzing another auxiliary optimization problem (\ref{infQ}) below. We will show that the given $F_{\pi^*,Z}^{1}, \ldots ,F_{\pi^*,Z}^{N}$ as in the optimization problem (\ref{infP}) also constitute an optimal solution to the optimization problem (\ref{infQ}); see Lemma \ref{lem308} below. Finally, by examining the properties of $F_{\pi^*,Z}^{1}, \ldots ,F_{\pi^*,Z}^{N}$ via analyzing the optimization problem (\ref{infQ}), we can derive closed-form expressions for $F_{\pi^*,Z}^{i},$ $1\leq i\leq N$, and hence obtain the unique solution $F_{X}^{*}:=\frac{1}{N}\sum_{i=1}^{N}F_{\pi^*,Z}^{i}$ to the optimization problem (\ref{inf2B}).
	
	\vspace{0.2cm}
	
	\begin{Lemma}\label{lem308}\ \
		Let $F_{\pi^*,Z}^{1}, \ldots ,F_{\pi^*,Z}^{N},$ $\eta_1^*$ and $\eta_p^*$ be as in the optimization problem (\ref{infP}). Then $F_{\pi^*,Z}^{1}, \ldots ,F_{\pi^*,Z}^{N}$ constitute an optimal solution to the following optimaization problem (\ref{infQ}):
		\begin{equation}\label{infQ}
			\inf_{F_{\pi,Z}^{1}, \ldots, F_{\pi,Z}^{N} \in \mathcal{F}}
			\frac{1}{N}\sum_{i=1}^{N} \int_{0}^{1}\left(\varphi' \left(  \frac{1}{N}   \left(F_{\pi^*,Z}^{i}(x)+(i-1)\right)  \right)  +\eta_{1}^*+\eta_{p}^*x^{p-1}\right)
			F_{\pi,Z}^{i}(x)\mathrm dx     \tag{Q}
		\end{equation}
		subject to
		\begin{align*}
			&\int_{0}^{1}x ~\mathrm{d}\left(\frac{1}{N}\sum_{i=1}^{N} F_{\pi,Z}^{i}(x)\right)=c_1,
		\end{align*}
		\begin{align*}
			&\int_{0}^{1}x^p ~\mathrm{d}\left(\frac{1}{N}\sum_{i=1}^{N} F_{\pi,Z}^{i}(x)\right)=c_p
		\end{align*}
		and
		\begin{align*}
			&-\frac{1}{N} \sum_{i=1}^{N}\int_{0}^{1} |x-x_{i}|^{p-1}\mbox{sign}(x-x_{i})F_{\pi,Z}^{i}(x)\mathrm dx
			\leq \frac{1}{p}\left(\epsilon^{p}-\frac{1}{N} \sum_{i=1}^{N}(1-x_{i})^{p}\right).
		\end{align*}
	\end{Lemma}
	
	\vspace{0.2cm}

	Now, we are ready to state the second main result of this paper, which provides the closed-form expression for the unique optimal solution $F_X^*$
	to the optimization problem (\ref{inf2B}).
	
	\vspace{0.2cm}
	
	\begin{Theorem}\label{thm302}\ \
		Suppose that $p\geq 2$. Let $F_{\pi^*,Z}^{1}, \ldots, F_{\pi^*,Z}^{N}$ be as in the optimization problem (\ref{infQ}).
		\begin{enumerate}
			\item[(1)]If the constraint condition (\ref{add070305}) does not make contribution to the optimization problem (\ref{infP}), then, the unique optimal solution $F_{X}^{*}$ to the optimization  problem (\ref{inf2B}) is given by
			\begin{align}\label{add070503}
				F_{X}^{*}(x)=
				\begin{cases}
					0, \quad &\text{if~}  x<\max\left(\left(\frac{g'(1)+\eta_1^*}{-\eta_p^*}\right)^{\frac{1}{p-1}},0\right),\\
					1-(g')^{-1}(-\eta_1^*-\eta_p^*x^{p-1}), \quad &\text{otherwise},\\
					1, \quad &\text{if~}  x\geq \min\left(\left(\frac{g'(0)+\eta_1^*}{-\eta_p^*}\right)^{\frac{1}{p-1}},1\right),
				\end{cases}
			\end{align}
			where the parameters $\eta_1^*\in\mathbb{R}$ and $\eta_p^*<0$  are uniquely determined by
			$$\int_{0}^{1}x~\mathrm{d} F_{X}^{*}(x)=c_1 \quad \text{and} \quad \int_{0}^{1}x^p~\mathrm{d} F_{X}^{*}(x)=c_p.$$
			
			\item[(2)]
			If the constraint condition (\ref{add070305}) does make contribution to the optimization problem (\ref{infP}), then the unique optimal solution $F_{X}^{*}$ to the optimization problem (\ref{inf2B}) is given by
			$F_{X}^{*}(x):= \frac{1}{N}\sum_{i=1}^{N}{F}_{\pi^*,Z}^{i}(x),$ where $F_{\pi^*,Z}^{i}$ is continuous on (0,1), and on intervals where it is not constant, $F_{\pi^*,Z}^{i}$ coincides with some  $F_{\bm\eta}^{i}$ of the form
			\begin{align*}
				{F}_{\bm\eta}^{i}(x):=
				\begin{cases}
					0,& \text{if }a_{i}(x)<g'\left(\frac{N+1-i}{N}\right), \\
					N-N(g')^{-1}(a_i(x) )-(i-1),  & \text{otherwise},\\
					1,&\text{if }a_{i}(x)\geq g'\left(\frac{N-i}{N}\right),
				\end{cases}
			\end{align*}
			where for each $1\leq i\leq N$,
			\begin{align}\label{20260513}
				a_{i}(x) := -\eta_{1}^*-\eta_{p}^*x^{p-1}+\lambda^{*} |x-x_{i}|^{p-1}\mbox{sign}(x-x_{i}), \quad x \in \mathbb{R},
			\end{align}
			
			$\bm\eta:=(\eta_1^*,\eta_p^*)\in\mathbb{R}^2,$ and the parameters $\eta_{1}^*$, $\eta_{p}^*$ and $\lambda^*>0$ are determined by the following equations:
			\begin{align}\label{add041005}
				\int_{0}^{1}x\mathrm d\left(\frac{1}{N}\sum_{i=1}^{N}F_{\bm\eta}^{i}\right)(x) = c_{1},
			\end{align}
			\begin{align}\label{add041006}
				\int_{0}^{1}x^{p}\mathrm d\left(\frac{1}{N}\sum_{i=1}^{N}F_{\bm\eta}^{i}\right)(x) = c_{p}
			\end{align}
			and
			\begin{align}\label{add041007}
				-\frac{1}{N} \sum_{i=1}^{N}\int_{0}^{1}|x-x_{i}|^{p-1}\mbox{sign}(x-x_{i})F_{\bm\eta}^{i}(x)\mathrm dx-\frac{1}{p}\left(\epsilon^{p}
				-\frac{1}{N} \sum_{i=1}^{N}(1-x_{i})^{p}\right) = 0.
			\end{align}
			
		\end{enumerate}
	\end{Theorem}
	
	\vspace{0.2cm}
	
		We would like to claim that the conditional distribution functions \(F_{\pi^*,Z}^{1},...,F_{\pi^*,Z}^{N}\)
		as in Theorem 3.2(2) satisfy
		\[\sup S(F_{\pi^*,Z}^{i})\leq\inf S(F_{\pi^*,Z}^{i+1}), \quad 1\leq i\leq N-1.\]
		To show the claim, we need a little more preparations. Let \(h\) be a continuous function on (0.1].
		We say that \(h\) is strictly increasing at some \(x_0\in(0,1]\) if there is a \(\delta =\delta (x_0)>0\)
		such that for any \(x_0-\delta <x<y<x_0,\) \(h(x_0-\delta )<h(x)<h(y)<h(x_0).\) Denote 
		\[ \text{SI}(h):=\{x\in(0,1]:h \text{ is strictly increasing at } x\}.\]
		Denote by cl(SI\((h)\)) the closure of set SI(\(h\)).
		By  Theorem 3.2(2) we know that if \(F_{\pi^*,Z}^{i}\) is strictly increasing at 
		some \(x_0\in(0,1)\),  then \(F_{\bm\eta}^{i}\) is also strictly increasing at \(x_0\),
		and hence \(a_i(\cdot)\) must be strictly increasing at \(x_0\). For each \(1\leq i \leq N\),
		Denote by \(\underline{x}_i\) the smallest solution to 
		\[a_i(x)=g'\left(\frac{N+1-i}{N}\right)\]
		with respect to variable \(x\) on \((0,+\infty)\cap \text{cl(SI}(a_i(\cdot))\)), and by \(\overline{x}_i\) the 
		smallest solution to 
		\[a_i(x)=g'\left(\frac{N-i}{N}\right)\]
		with respect to variable \(x\) on \((0,+\infty)\cap \text{cl(SI}(a_i(\cdot))\)).
		Clearly, \(\underline{x}_i\leq\overline{x}_i\), \(1\leq i\leq N\). Moreover, it is not hard to
        verify that \(S(F_{\pi^*,Z}^{i})\subset [\underline{x}_i,\overline{x}_i]\), \(1\leq i \leq N.\)
		Apparently, for each \(1\leq i\leq N-1,\) \(a_{i+1}(x)\leq a_i(x)\) for every \(x\in[0,+\infty)\).
		Hence it is not hard to verify that \(\overline{x}_i\leq\underline{x}_{i+1}\) due to the definitions 
		of \(\overline{x}_i\) and \(\underline{x}_{i+1}\), \(1\leq i \leq N-1\). Consequently, for each 
		 \(1\leq i \leq N-1\), \(\sup S(F_{\pi^*,Z}^{i})\leq\inf S(F_{\pi^*,Z}^{i+1}).\)

	\vspace{0.2cm}

	Next, let us briefly comment on the possible sign of $\eta_p^*$ as in Theorem \ref{thm302}(2), because it is particularly interesting.
	Let the parameters $\lambda^*$, $\eta_1^*$ and $\eta_p^*$ be as in Theorem \ref{thm302}(2), that is, they satisfy equations (\ref{add041005})-(\ref{add041007}).
	Then, for some suitable $\epsilon$, $c_1$ and $c_p$, $\eta_p^*$ can be non-positive.
	For instance, fix arbitrarily $\epsilon >0$ and $0 < c_1 < 1$ first, and let $\eta_p^*=0$.
	Then, by solving equations (\ref{add041005}) and (\ref{add041007}),
	we can determine the values of $\lambda^*$ and $\eta_1^*$. Hence, by plugging the obtained values of $\lambda^*$ and $\eta_1^*$ into
	the left-hand side of (\ref{add041006}), we can further determine the value of $c_p$.
	Such arguments are also applicable to the case of $\eta_p^*<0$. Next remark studies the unique solution to the optimization problem (\ref{inf2B})
	in the case of $\eta_p^*\leq0$.
	
	\vspace{0.2cm}
	
	\begin{Remark}\ \ \label{rem041701}
		Let the parameters $\eta_{1}^*$, $\eta_{p}^*$,  $\lambda^*\geq0$, and functions $a_i(x)$, $1\leq i\leq N$, be as in Theorem \ref{thm302}(2).
		If $\eta_{p}^*\leq0$, then $a_{i}(x)$ is strictly increasing with respect to variable $x$ on $(0, +\infty)$.
		Hence, the unique solution $F_{X}^{*}(x)$ to the optimization problem (\ref{inf2B}) is given by $F_{X}^{*}(x)=\frac{1}{N}\sum_{i=1}^{N}F_{\pi^*,Z}^{i}(x),$ where $F_{\pi^*,Z}^{i}(x),$ $1\leq i\leq N$, becomes
		\[
		{F}_{\pi,Z}^{i*}(x)=
		\begin{cases}
			0,         & \text{if } a_i(x)<g'\left(\frac{N+1-i}{N}\right),\\
			N-N(g')^{-1}(a_i(x) )-(i-1), & \text{otherwise},\\
			1,&\text{if } a_i(x)\geq g'\left(\frac{N-i}{N}\right).
		\end{cases}
		\]
	\end{Remark}
	
	\vspace{0.2cm}
	
	Remark 3.4 contains a particular case of $\eta_p^*=0$. Next remark further discusses the unique solution to the optimization problem (\ref{inf2B})
	in the case of $\eta_p^*=0$.
	
	\vspace{0.2cm}
	
	\begin{Remark}\ \
		Consider the following optimization problem (R):
		\begin{align}
			\inf_{F_{X}\in\mathcal{F}_{\epsilon}(c_1)}\int_{0}^{1}\varphi(F_{X}(x))\mathrm dx,\tag{R}
		\end{align}
		where $$
		\mathcal{F}_{\epsilon}(c_1):=~\{F_{X}\in\mathcal{F}_\epsilon :\ \int_{0}^{1}x~\mathrm dF_{X}(x)=c_{1}\}.
		$$
		
		\vspace{0.2cm}
		
		Compared with the optimization problem (\ref{inf2B}), the above optimization problem (R) corresponds to the case where the constraint of $p$-order moment
		in the optimization problem (\ref{inf2B}) disappears, that is the case of $\eta_p^* =0$ in the process of solving the optimization problem (\ref{inf2B}).
		By checking the procedure of solving the optimization problem (\ref{inf2B}), we can steadily know that the unique solution to the optimization problem (R)
		is given by $F_{X}^{*}(x)=\frac{1}{N}\sum_{i=1}^{N}F_{\pi^*,Z}^{i}(x),$ where $F_{\pi^*,Z}^{i}(x),$ $1\leq i\leq N$, is given by
		\[
		{F}_{\pi,Z}^{i*}(x)=
		\begin{cases}
			0,         & \text{if } b_i(x)< g'\left(\frac{N+1-i}{N}\right),\\
			N-N(g')^{-1}(b_i(x) )- (i-1), & \text{otherwise},\\
			1,&\text{if } b_i(x)\geq g'\left(\frac{N-i}{N}\right),
		\end{cases}
		\]
		where for each $1\leq i\leq N$,
		$$b_i(x):=-\eta_{1}^*+\lambda^{*} |x-x_{i}|^{p-1}\mbox{sign}(x-x_{i}), \quad x\in\mathbb{R}.$$
	\end{Remark}
	
	\vspace{0.2cm}
	
	Similar to Remark 3.2, we can relax the strict convexity of $\varphi$ to convexity in the optimization problem (\ref{inf2B}),
	which is summarized in the following remark.
	
	\vspace{0.2cm}

	\begin{Remark}\label{re303}\ \
		Suppose that $\varphi$ is a convex function as in Remark 3.1. Then by checking the procedure of solving the optimization problem (\ref{inf2B}),
		we can find that Theorem \ref{thm302} still remains true. Precisely, what we need to modify is only to replace the first order derivative $g'$ of $g$ with its first order left-derivative $g'_-$, and  replace $g'(0)$ with $g'_+(0)$ in Theorem 3.2.
	\end{Remark}

	\vspace{0.2cm}
	
	Suppose that $p = 2$ in Theorem \ref{thm302}. When we only consider distribution functions supported on the interval $[0,1]$,
	then the parameters $\eta_1^*$ and $\eta_2^*$ cannot be well-expressed due to the influence of the endpoints 0 and 1, which
	can be seen from Theorem \ref{thm302}. In fact, we can extend Theorem \ref{thm302} to general unbounded random loss, that is,
	the (loss) distribution function is supported on the whole real line. Specifically, we first consider
	a (loss) distribution function supported on $[-b,b]$, $b>0$, and then take the limit of $b$ to infinity,
	yielding concise expressions for $\eta_1^*$ and $\eta_2^*$; for instance, see Cornilly et al. (2018, Corollary 2.3).
	In summary, Theorem \ref{thm302} implies the following corollary, which holds for general unbounded random losses.
	Recall that in this situation, the samples $x_1 \leq  \ldots  \leq x_N$ belong to real numbers.
	
	\vspace{0.2cm}
	
	\begin{Corollary}\ \
		Suppose that $p=2$, that is, $\mathcal{F}_\epsilon(c_{1},c_{p})=\mathcal{F}_\epsilon(c_{1},c_{2})$. Then,
		the unique solution $F_{X}^{*}$ to the optimization problem (\ref{inf2B}) is given by
		\begin{align*}
			{F}_{X}^{*}(x) =
			\begin{cases}
				\frac{i-1}{N}, & \text{if } x < \frac{g'\left(\frac{N+1-i}{N}\right)+\lambda^* x_{i}+\eta_{1}^*}{ -\eta_{2}^*+\lambda^*},\\
				1-(g')^{-1}(-\eta_{1}^*+(-\eta_{2}^*+\lambda^*)x-\lambda^* x_{i}),  & \text{otherwise},\\
				\frac{i}{N}, & \text{if }x \geq \frac{g'\left(\frac{N-i}{N}\right)+\lambda^* x_{i}+\eta_{1}^*}{ -\eta_{2}^*+\lambda^*},
			\end{cases}
		\end{align*}
		on $\left(F_{X}^{*-1}\left(\frac{i-1}{N}\right),F_{X}^{*-1}\left(\frac{i-1}{N}\right)\right],$ $1\leq i\leq N,$
		where the constants $\eta_{1}^*$ and $\eta_{2}^*$ are given by
		$$
		\eta_{1}^* = ( - \eta_{2}^* + \lambda^*)c_{1}-\frac{\lambda^*}{N}\sum_{i=1}^{N}x_{i}-1,
		$$
		
		$$
		\eta_{2}^*=\lambda^*-\frac{\sqrt{\sum_{i=1}^{N}\int_{\frac{i-1}{N}}^{\frac{i}{N}}\left[g'(1-p)+\lambda^* x_{i}-1
				-\frac{\lambda^*}{N}\sum_{j=1}^{N}x_{j}\right]^{2} \mathrm dp}} {\sqrt{c_{2}-c_{1}^{2}}}
		$$
		and $\lambda^*\geq0$ satisfies
		\begin{align*}
			\sum_{i=1}^{N}\int_{\frac{i-1}{N}}^{\frac{i}{N}}|F_X^{*-1}(p)-x_i|^2\mathrm dp=\epsilon^2.
		\end{align*}
		
		\vspace{0.2cm}
		
		Furthermore, the left-continuous inverse function $F_{X}^{*-1}$ of $F_{X}^{*}$ is given by
		\begin{align}\label{062907}
			F_X^{*-1}(p) =\frac{1}{-\eta_2^*+\lambda^{*}} \left(g'(1 - p) + \eta_1^* + \lambda^* x_{i}\right) \quad \text{if } \frac{i-1}{N}<p\leq \frac{i}{N},\quad i=1, \ldots ,N.
		\end{align}
	\end{Corollary}
	
	\vspace{0.2cm}
	
	\begin{Remark}\ \ In (\ref{062907}), if $\lambda^{*}=0$, then Corollary 3.1 recovers Corollary 2.3 of Cornilly et al. (2018).
		In fact, in this case, the constraint condition (\ref{add070305}) does not make contribution.
	\end{Remark}
	
	\vspace{0.2cm}
	
	Next, we discuss the connection of Corollary 3.1 with Theorem 3.1 of Bernard et al. (2024), which is summarized in the following remark.
	
	\vspace{0.2cm}
	
	\begin{Remark}\ \
		Suppose that the reference distribution function $F$ in Theorem 3.1 of Bernard et al. (2024) is given by the empirical distribution function $F_{\hat{X}}$,
		that is,
		$$
		F(x):=\frac{1}{N}\sum_{i=1}^{N}I_{(-\infty,x]}(x_{i}), \quad x \in \mathbb{R},
		$$
		where $x_1\leq  \ldots  \leq x_N$, $x_i\in\mathbb{R}$, $1\leq i\leq N$, are samples. Then, from Theorem 3.1 of Bernard et al. (2024), it follows that the left-continuous inverse function $G^{-1}$ of the corresponding optimal solution $G$ is given by
		\begin{equation}\label{add041201}
			G^{-1}(u) = \mu+\sigma\left(\frac{g'_-(1-u)+\lambda x_i-a_\lambda}{b_\lambda}\right), \quad \frac{i-1}{N}<u\leq\frac{i}{N},\ 1\leq i\leq N,
		\end{equation}
		where $ \mu,\ \sigma$ and $g$ are as in Theorem 3.1 of Bernard et al. (2024),
		$\lambda>0$ is the unique positive solution to $W_{2}^{2}(G,F)=\epsilon^2,$  $a_\lambda := E(g'(1-U)+\lambda F^{-1}(U))$
		and $b_\lambda := \mbox{std}(g'(1-U)+\lambda ^{-1}(U)).$ Indeed, by an elementary calculation, we  know that
		$$
		a_\lambda = 1+\frac{\lambda}{N}\sum_{i=1}^{N}x_i
		$$
		and
		$$
		b_\lambda = \sqrt{\sum_{i=1}^{N}\int_{\frac{i-1}{N}}^{\frac{i}{N}}\left[g'_-(1-p)+\lambda x_{i}-1-\frac{\lambda}{N}\sum_{i=1}^{N}x_{i}\right]^{2}\mathrm dp}.
		$$
		
		On the other hand, we determine all parameters in Corollary 3.1. In fact, we can steadily verify that $\lambda^*=\lambda$, $\eta_1^*=\frac{b_\lambda}{\sigma}\mu-a_\lambda$, and $-\eta_2^*+\lambda^*=\frac{b_\lambda}{\sigma}$. By Corollary 3.1 and Remark \ref{re303},
		we know that
		when $g$ is a concave function, then the left-continuous inverse function  $F_{X}^{*-1}$ as in (\ref{062907}) of $F_{X}^{*}$  is given by
		\begin{align}\label{add041702}
			F_X^{*-1}(p) =\frac{1}{-\eta_2^*+\lambda^{*}} \left(g'_-(1 - p) + \eta_1^* + \lambda^* x_{i}\right) \quad \text{if } ~\frac{i-1}{N}<p\leq \frac{i}{N},\quad i=1, \ldots ,N.
		\end{align}
		Substituting $\lambda^*$, $\eta_1^*$ and $\eta_2^*$ into $F_{X}^{*-1}$ as in (\ref{add041702}), we find that $F_{X}^{*-1}$ as in (\ref{add041702}) is
		consistent with $G^{-1}$ as in (\ref{add041201}).
		
		\vspace{0.2cm}
		
		Notice that unlike the method used by Bernard et al. (2024, Theorem 3.1), in this paper, we employ Lagrangian multiplier method to deal with the case of more general Wasserstein distance of order $p\geq2$.
		Taking the above considerations into account, the present study of the optimization problem (\ref{inf2B}) can be viewed as a meaningful complement
		to the study of Bernard et al. (2024, Theorem 3.1).
	\end{Remark}

	\section{Numerical studies}

	In this section, in order to illustrate the computability of the proposed models and results,
	first, we numerically calculate the maximum values of the optimization problems (A) and (B) for several distortion functions.
	Second, we also numerically determine the corresponding maximizing distribution functions for certain distortion functions.
	Final, we further numerically investigate the impacts of the parameters $\epsilon$ and $p$ on the maximum values
	and the corresponding maximizing distribution functions, respectively.

	\subsection{Numerical study for optimization problem (\ref{supA})}

	In this subsection, first, we numerically discuss the maximum values for the optimization problem (A) under different values of $\epsilon$ and
	the order $p$ of the Wasserstein distance with  $p = 1,~ 2$ and $3.$  Specifically, we examine the power distortion function $g(x) = x^\alpha$
	with $\alpha = 0.5,$ $0.2,$ $0.1,$ and the dual power distortion function $g(x) = 1 - (1 - x)^\beta$ with $\beta = 2, ~3, ~5$. For simplicity,
	we consider a dataset of size $N = 200$ generated from the uniform distribution function on [0, 1]. By Theorem 3.1, we list  corresponding
	maximum values of the DRMs in Table 1 below.
	
	\vspace{0.2cm}
	
	\begin{table}[H]
		\centering
		\caption{
			Maximum value $H_{g}(F_X^{*})$ for several distortion functions.}
		\begin{tabular}{|ll|l|l|l|l|l|}
			\hline
			\multicolumn{2}{|l|}{  $\epsilon$  }                          & 0.1 & 0.2 & 0.3 & 0.4 &0.5  \\ \hline
			\multicolumn{1}{|l|}{\multirow{3}{*}{$g(x)=x^{0.1}$}}
			& p=1  &0.94005  &1         &1        &1        &1       \\ \cline{2-7}
			\multicolumn{1}{|l|}{}
			& p=2  &0.93156  &0.95189	&0.96804  &0.98122  &1        \\ \cline{2-7}
			\multicolumn{1}{|l|}{}
			& p=3  &0.93048	 &0.95035	&0.96645  &0.97948  &0.98975   \\ \hline
			\multicolumn{1}{|l|}{\multirow{3}{*}{$g(x)=x^{0.2}$}}
			& p=1  &0.88472  &0.91998   &0.94784  &0.97183  &0.99335    \\ \cline{2-7}
			\multicolumn{1}{|l|}{}
			&p=2   &0.87039  &0.90753   &0.93787  &0.96317  &0.98396    \\ \cline{2-7}
			\multicolumn{1}{|l|}{}
			&p=3   &0.86986	 &0.90502   &0.93506  &0.95986  &0.97980     \\ \hline
			\multicolumn{1}{|l|}{\multirow{3}{*}{$g(x)=x^{0.5}$}}
			&p=1   &0.73943  &0.81362   &0.87511  &0.93113  &0.98346    \\ \cline{2-7}
			\multicolumn{1}{|l|}{}
			&p=2   &0.72317	 &0.79459	&0.85754  &0.91325  &0.96133   \\ \cline{2-7}
			\multicolumn{1}{|l|}{}
			&p=3   &0.72155  &0.79145	&0.85275  &0.90644  &0.95175   \\ \hline
			\multicolumn{1}{|l|}{\multirow{3}{*}{$g(x)=1-(1-x)^{2}$}}
			&p=1   &0.76753  &0.87433	&0.93837  &0.98166  &1         \\ \cline{2-7}
			\multicolumn{1}{|l|}{}
			&p=2   &0.73364  &0.82655   &0.90236  &0.95718  &0.98946  \\ \cline{2-7}
			\multicolumn{1}{|l|}{}
			&p=3   &0.73187  &0.81567   &0.88914  &0.94540  &0.98132   \\ \hline
			\multicolumn{1}{|l|}{\multirow{2}{*}{$g(x)=1-(1-x)^{3}$}}
			&p=1   &0.87248  &0.95309   &0.98649  &0.99752  &1         \\ \cline{2-7}
			\multicolumn{1}{|l|}{}
			&p=2   &0.82372  &0.90801   &0.96193  &0.98911  &0.99869    \\ \cline{2-7}
			\multicolumn{1}{|l|}{}
			&p=3   &0.81950  &0.89327   &0.94969  &0.98259  &0.99645    \\ \hline
			\multicolumn{1}{|l|}{\multirow{2}{*}{$g(x)=1-(1-x)^{5}$}}
			&p=1   &0.87248  &0.95309   &0.98649  &0.99752  &1          \\ \cline{2-7}
			\multicolumn{1}{|l|}{}
			&p=2   &0.82372  &0.90801	&0.96193  &0.98911  &0.99869   \\ \cline{2-7}
			\multicolumn{1}{|l|}{}
			&p=3   &0.81950  &0.89327	&0.94969  &0.98259	&0.99645   \\ \hline
		\end{tabular}
	\end{table}
	
	\vspace{0.2cm}
	
	From each row in Table 1, we can see that the maximum values increase when the radius $\epsilon$ of the Wasserstein ball becomes larger, which is
	consistent with intuition. From each column in Table 1, for a fixed power or dual power distortion function, we can find that the maximum value decrease
	when the order $p$ of Wasserstein distance becomes larger, which is also consistent with intuition. For the power and dual power distortion functions
	$g(x)=x^{0.2}$ and $g(x)=1-(1-x)^2$, from Figures 1 and 2, we can visually see the influence of different values of $\epsilon$ and $p$ on the maximum value $H_g(F_{X}^*)$, respectively.
	
	\vspace{0.2cm}
	
	\begin{figure}[H]
		\centering
		\begin{minipage}{0.45\textwidth}
			\centering
			\includegraphics[width=\linewidth]{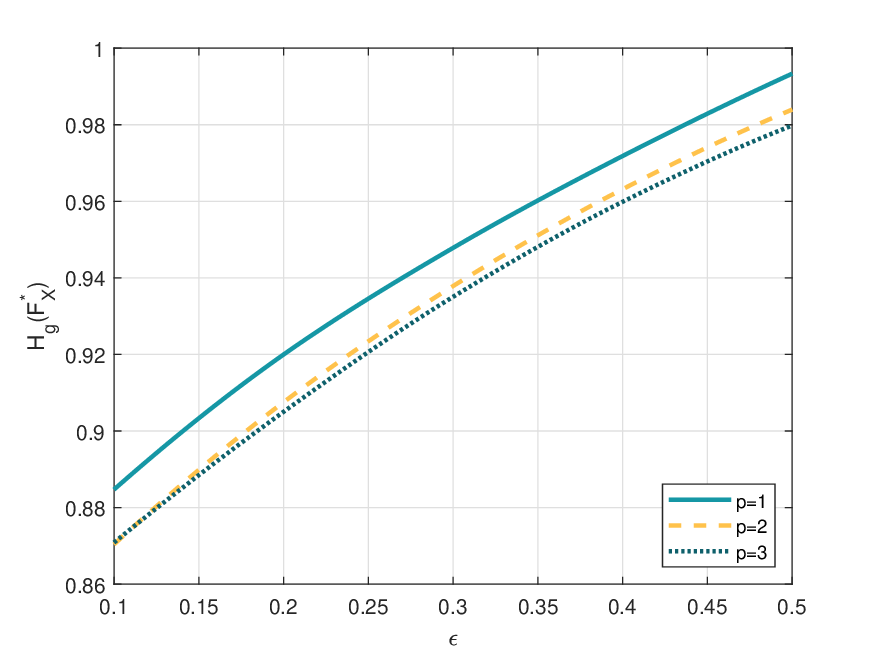}
			\caption{Maximum value for $g(x)=x^{0.2}.$}
		\end{minipage}
		\hfill
		\begin{minipage}{0.45\textwidth}
			\centering
			\includegraphics[width=\linewidth]{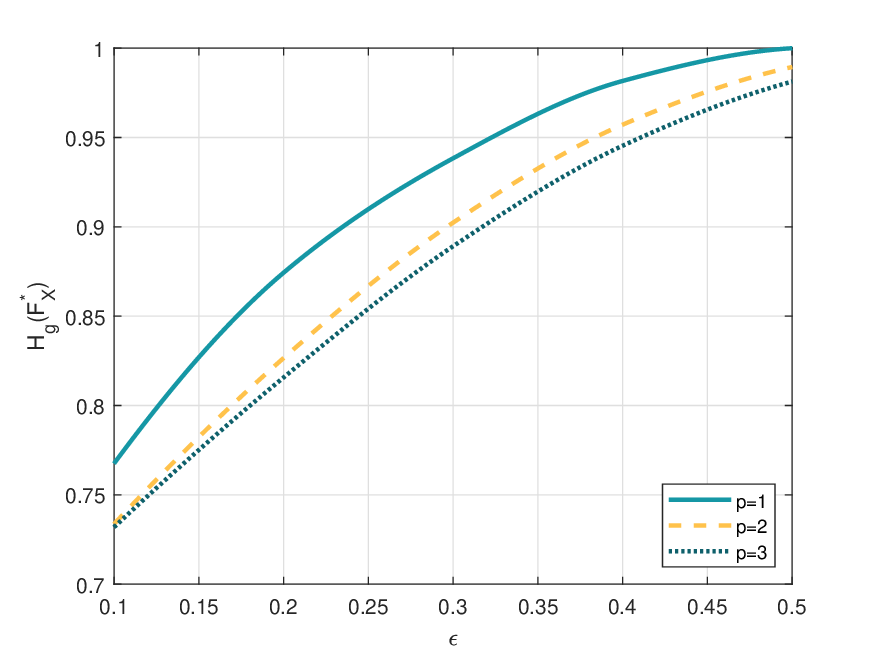}
			\caption{Maximum value for $g(x)=1-(1-x)^2.$}
		\end{minipage}
	\end{figure}
	
	\vspace{0.2cm}
	
	In order to have a visual look at the relationship between the distortion function and the corresponding maximizing distribution function,
	we continue to focus on the distortion functions  $g(x)=x^{0.2}$ and $g(x)=1-(1-x)^2$ under $\epsilon := 0.1$, as displayed in Figures 3 and 4 below.
	We can find that the maximizing distribution functions are either continuous or a mixture of continuous and discrete ones.
	\begin{figure}[H]
		\centering
		\begin{minipage}{0.45\textwidth}
			\centering
			\includegraphics[width=\linewidth]{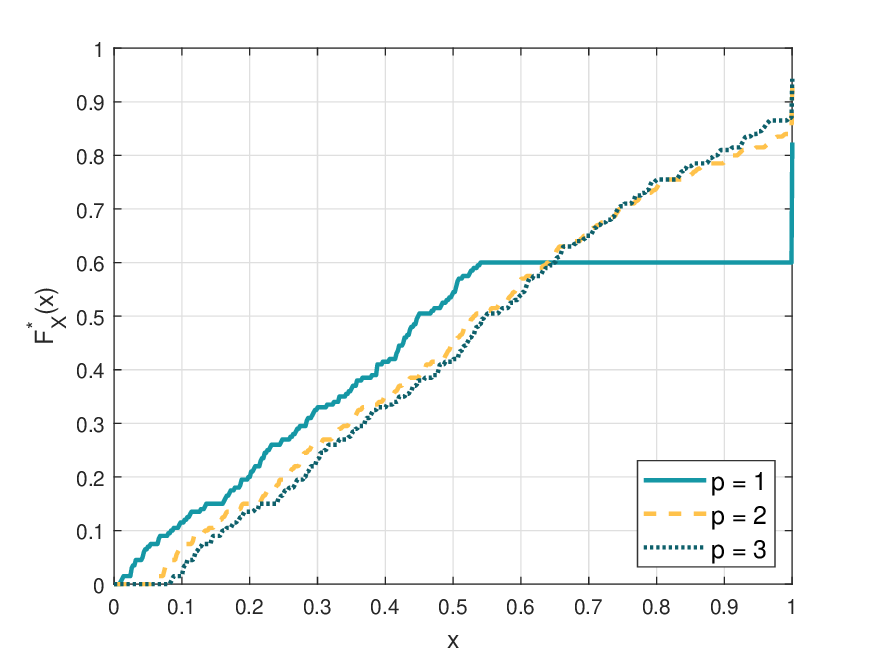}
			\caption{Maximizing distribution functions for $g(x)=x^{0.2}.$}
		\end{minipage}
		\hfill
		\begin{minipage}{0.45\textwidth}
			\centering
			\includegraphics[width=\linewidth]{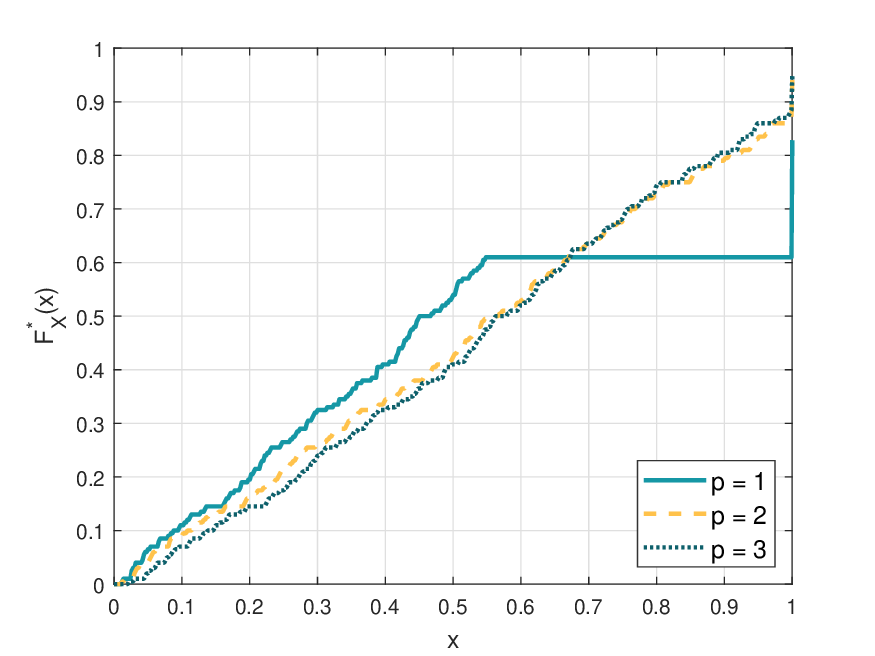}
			\caption{Maximizing distribution functions for $g(x)=1-(1-x)^2.$}
		\end{minipage}
	\end{figure}

	\subsection{Numerical study for optimization problem (\ref{supB})}
	
	In this subsection, by Corollary 3.1, we numerically examine the worst-case DRMs for optimization problem (B) under different values
	of $\epsilon$ for several dual power distortion functions.
	In Table 2 below, we set  $c_1 := \frac{1}{N}\sum_{i=1}^{N}x_{i}$ and $c_2:=\frac{1}{N}\sum_{i=1}^{N}(x_{i}-c_{1})^{2}+c_1^2$. For simplicity,
	we consider a dataset of size $N = 200$ generated from the uniform distribution function on [0, 10], and the results are displayed in Table 2 below.
	Table 2 indicates the difference of the maximum values of the DRMs between cases of having and not having Wasserstein ball constraint.
	Precisely, the maximum values of
	the DRMs with Wasserstein ball constraint are smaller than those without Wasserstein ball constraint (i.e. $\epsilon = \infty$).
	This phenomenon indicates that the Wasserstein ball constraint has notable influence on the maximum value of the distortion risk measure.
	\begin{table}[h!]
		\caption{
			Maximum value $H_{g}(F_X^{*})$ for several dual power distortion functions.}
		\begin{tabular}{|c|c|c|c|c|c|c|}
			\hline
			$\epsilon$ & 0.1 & 0.2 & 0.3 & 0.4 & 0.5 &$\infty$\\
			\hline
			$g(x)=1-(1-x)^2$ &9.047508	&9.053153  &9.053153   &9.053153  &9.053153  &9.053153 \\ \hline
			$g(x)=1-(1-x)^3$ &11.334653 &11.377295 &11.411109  &11.436039 &11.452029 &11.459253\\ \hline
			$g(x)=1-(1-x)^5$ &13.822679 &13.973414 &14.111965 &14.238159 &14.351848 &14.789888\\ \hline
		\end{tabular}
	\end{table}
	
	\vspace{0.2cm}
	
	Next, we briefly discuss the influence of the first two moments constraints on the maximum values of the distortion risk measure. For this purpose,
	we use the same dataset used in Table 1, which has a sample mean 0.467202 and sample variance 0.075883. We set the radius $\epsilon$ of the Wasserstein
	ball be 0.1, and choose the dual power distortion function $g(x)=1-(1-x)^3$ for simplicity. Moreover, we set $c_1:=0.5$ and $c_2:=0.35$.
	By Remark 3.3 and Corollary 3.1, we display the maximum values of the distortion risk measure in Table 3 below. In Table 3, the first five maximum values correspond to the cases of having Wasserstein ball constraint only, while the last one corresponds to the case of having both Wasserstein ball and
	the first two moments constraints. Table 3 indicates that the maximum value (i.e. 0.53354) of the distortion risk measure with the first two moments
	constraints is smaller than that (i.e. 0.88377) of the distortion risk measure without the first two moments constraints.
	This phenomenon indicates that the moment constraints have notable influence on the maximum value of the distortion risk measure.
	
	\begin{table}[h!]
		\centering
		\caption{Maximum value $H_g(F_X^*)$ under $\epsilon=0.1.$}
		\begin{tabular}{|c|c|c|c|c|c|c|}
			\hline
			&$p=1$ &$p=2$ & $p=3$& $p=4$ &$p=5$ & $(c_1,c_2)=(0.5,0.35)$ \\
			\hline
			$g(x)=1-(1-x)^3$&1.65104 &0.88377 & 0.85213& 0.85108& 0.84629 &0.53354 \\
			\hline
		\end{tabular}
	\end{table}

	\section{Concluding remarks}

	When the underlying loss distribution function lies within a Wasserstein ball around an empirical distribution function, or when it has a given mean
	and any other higher order moment while being within a Wasserstein ball around an empirical distribution function, we derive closed-form expression
	for the distribution function maximizing the DRMs with strictly concave and twice differentiable distortion functions,
	and, furthermore, extend to the case of concave distortion functions.
	Our findings show that the maximizing distribution function is either continuous or a mixture of continuous and discrete distribution functions.
	Moreover, it also turns out that the worst-case value of DRM  in the second scenario is smaller than that in the first scenario.
	This finding is indeed consistent with the standard in the sense that less reliable information leads to a more conservative risk assessment.
	In other words, if there is less information available, then we would tend to adopt a more conservative approach to risk assessment.
	
	\vspace{0.2cm}
	
 We would like to mention that this study does not consider additional characteristics of the underlying loss distribution function (or the corresponding random loss);
for instance, neither the uni- or multimodality and symmetry of the underlying loss distribution function nor the unboundedness of the random loss for general Wasserstein distance
of \(p\) order. Moreover, a more general case of the underlying distortion function could be a general distortion function except for the concave one. Hence, it would be interesting
to see relevant study worked out in the future, in which some of above characteristics are in use.

   \section*{Acknowledgements}	The authors are very grateful to the Editors   and the anonymous referees for their constructive and valuable comments and suggestions, which led to the present
greatly improved version of the manuscript. In particular, the possible topics for future study mentioned in the concluding remarks
were motivated by the anonymous referees. The authors are also very grateful to Professor  Steven
Vanduffel for his very helpful
discussions and comments on an earlier version of the manuscript.
	
	\section*{Appendix}
	\setcounter{equation}{0}
	\setcounter{subsection}{0}
	\renewcommand{\theequation}{A.\arabic{equation}}
	\renewcommand{\thesubsection}{A.\arabic{subsection}}

	In this appendix, we provide proofs of all main results of this paper, including the proofs of all lemmas.
	
	\vspace{0.2cm}
	
	\noindent{\bf Proof of Lemma \ref{lem301}}\ \
	
	\vspace{0.2cm}
	
	We only show the existence and uniqueness of solution to the optimization problem (\ref{inf2B}), because the existence and uniqueness of solution to
	the optimization problem (\ref{inf2A}) is a special case. Our argumentation is inspired by Rustagi (1957) and Cornilly (2018).
	
	\vspace{0.2cm}
	
	We begin with showing the existence of solution to optimization problem (\ref{inf2B}).
	Observe that $\int_{0}^{1} x^{p} \mathrm dF(x) = c_{p}$ if and only if $\int_{0}^{1} x^{p-1} F (x) \mathrm d x = d_{p}$ for certain
	appropriate $d_{p}$. Consider the mapping $T: \mathcal{F}_\epsilon(c_{1}, c_{p}) \rightarrow$ Ran($T$) defined by
	\begin{align*}
		T(F) := \left( \int_{0}^{1} \varphi(F(x))\mathrm d x, \int_{0}^{1} F(x) \mathrm d x, \int_{0}^{1} x^{p-1} F(x) \mathrm d x \right),
	\end{align*}
	where Ran($T$) stands for the range of $T$, which is a subset of $\mathbb{R}^{3}$.  Then, the mapping $T$ is continuous and convex. Moreover, $T$ maps $\mathcal{F}_\epsilon(c_{1}, c_{p})$ into a convex and compact set, since $\mathcal{F}_\epsilon(c_{1}, c_{p})$ is convex and compact in the topology of convergence in distribution.  Note that the convexity and compactness of $\mathcal{F}_\epsilon(c_{1}, c_{p})$ under the topology of convergence
	in distribution follow from
	Villani (2009, Remark 6.19), Villani (2003, page 220) and Rustagi (1957, Lemma 3.1).
	Hence, the imposed moment conditions on $F$ imply that Ran($T$) is non-empty, bounded and closed, and thus there is a solution to
	the optimization problem (\ref{inf2B}).

	\vspace{0.2cm}
	
	Next, we turn to show the uniqueness of solution to the optimization problem (\ref{inf2B}). Suppose that $F_{0}$ is a solution to optimization problem (\ref{inf2B}),
	and $F_{1}$ is another admissible solution. Denote
	\begin{align*}
		M := \int_{0}^{1} \varphi(F_{0}(x))\mathrm d x = \int_{0}^{1} \varphi(F_{1}(x)) \mathrm d x.
	\end{align*}
	Since $\varphi$ is strictly convex, we know that for any $\lambda \in (0, 1)$,
	\begin{align*}
		\varphi( \lambda F_{0}(x) + (1-\lambda) F_{1}(x))<\lambda\varphi(F_{0}(x)) + (1-\lambda)\varphi(F_{1}(x)), \quad x \in [0,1].
	\end{align*}
	Thus, we have that
	\begin{align*}
		\int_{0}^{1} \varphi( \lambda F_{0}(x) + (1-\lambda) F_{1}(x))\mathrm d x
		< \lambda \int_{0}^{1} \varphi(F_{0}(x))\mathrm d x + (1-\lambda) \int_{0}^{1} \varphi(F_{1}(x))\mathrm d x = M,
	\end{align*}
	which contradict the fact that $F_0$ and $F_1$ are solutions to the optimization problem (\ref{inf2B}). Lemma \ref{lem301} is proved.
	
	\vspace{0.2cm}

	\noindent\textbf{Proof of Lemma~\ref{2026051301}}

	First, we  conclude that
	\begin{align}{\label{2026052401}}
		\inf_{F_{X}\in \mathcal{F}_{\epsilon}} \int_{0}^{1} \varphi(F_{X}(x))\mathrm dx
	  = &\inf_{\pi\in\Pi(\mathbb{P}_{X},\mathbb{P}_{\hat{X}})} \int_{0}^{1} 
	    \varphi\left(\frac{1}{N}\sum_{i=1}^{N}F_{\pi,Z}^{i}(x)\right)\mathrm dx. \nonumber \\
        &\text{subject to }  E_{\pi}\left[|Z-\hat{Z}|^{p}\right]\leq \epsilon^{p}
	\end{align}
    In fact, for any \(F_X \in \mathcal{F}_\epsilon\), by (\ref{2026051302}) and (\ref{2026060501}),
	there exists \(\tilde{\pi}\in\Pi(\mathbb{P}_{X},\mathbb{P}_{\hat{X}})\) with
	\begin{align*}
		 F_{X}(x)=\frac{1}{N}\sum_{i=1}^{N}F_{\tilde{\pi},Z}^{i}(x), \quad x \in \mathbb{R}, 
	\end{align*}
  such that \(E_{\tilde{\pi}}\left[|Z-\hat{Z}|^{p}\right] \leq \epsilon^{p}\). Hence, 
	\begin{align*}
		  \int_{0}^{1} \varphi(F_{X}(x))\mathrm dx
	  = & \int_{0}^{1} \varphi\left(\frac{1}{N}\sum_{i=1}^{N}F_{\tilde{\pi},Z}^{i}(x)\right)\mathrm dx\\
	 \geq &  \inf_{\pi\in\Pi(\mathbb{P}_{X},\mathbb{P}_{\hat{X}})} \int_{0}^{1} 
	    \varphi\left(\frac{1}{N}\sum_{i=1}^{N}F_{\pi,Z}^{i}(x)\right)\mathrm dx, \\
		& \text{ subject to } E_{\pi}\left[|Z-\hat{Z}|^{p}\right]\leq \epsilon^{p}
	\end{align*}
	and thus
	\begin{align}\label{2026060502}
		\inf_{F_{X}\in \mathcal{F}_{\epsilon}} \int_{0}^{1} \varphi(F_{X}(x))\mathrm dx
		 \geq & \inf_{\pi\in\Pi(\mathbb{P}_{X},\mathbb{P}_{\hat{X}})} \int_{0}^{1} 
	    \varphi\left(\frac{1}{N}\sum_{i=1}^{N}F_{\pi,Z}^{i}(x)\right)\mathrm dx. \nonumber\\
		& \text{ subject to } E_{\pi}\left[|Z-\hat{Z}|^{p}\right]\leq \epsilon^{p}
	\end{align}  
    
    Conversely, let some \(\pi\in\Pi(\mathbb{P}_{X},\mathbb{P}_{\hat{X}})\) 
	with \(E_{\pi}\left[|Z-\hat{Z}|^{p}\right]\leq \epsilon^{p}\). 
	From (\ref{2026051302}) and (\ref{2026060501}), we know that
	\begin{align*}
		F_{X}(x)=\frac{1}{N}\sum_{i=1}^{N}F_{\pi,Z}^{i}(x), \quad x\in\mathbb{R},
	\end{align*}
    and \(F_{X}\in\mathcal{F}_{\epsilon}\). Hence,
	\begin{align*}
		\int_{0}^{1} \varphi\left(\frac{1}{N}\sum_{i=1}^{N}F_{\tilde{\pi},Z}^{i}(x)\right)\mathrm dx
	& = \int_{0}^{1} \varphi(F_{X}(x))\mathrm dx \geq \inf_{F_{X}\in \mathcal{F}_{\epsilon}} \int_{0}^{1} \varphi(F_{X}(x))\mathrm dx.
	\end{align*}
    Thus,
	\begin{align}\label{2026060503}
		& \inf_{\pi\in\Pi(\mathbb{P}_{X},\mathbb{P}_{\hat{X}})} \int_{0}^{1} 
	    \varphi\left(\frac{1}{N}\sum_{i=1}^{N}F_{\pi,Z}^{i}(x)\right)\mathrm dx  \geq \inf_{F_{X}\in \mathcal{F}_{\epsilon}} \int_{0}^{1} \varphi(F_{X}(x))\mathrm dx.\\
		& \text{ subject to } E_{\pi}\left[|Z-\hat{Z}|^{p}\right]\leq \epsilon^{p} \nonumber
	\end{align}
	Therefore, (\ref{2026052401}) follows immediately from (\ref{2026060502}) and (\ref{2026060503}).
	
	Second, we conclude that 
    \begin{align}\label{2026060504}
    &\inf_{\pi\in\Pi(\mathbb{P}_{X},\mathbb{P}_{\hat{X}})} \int_{0}^{1}
     \varphi\left(\frac{1}{N}\sum_{i=1}^{N}F_{\pi,Z}^{i}(x)\right)\mathrm dx \nonumber \\
     &\text{subject to }  E_{\pi}\left[|Z-\hat{Z}|^{p}\right]\leq \epsilon^{p} \nonumber\\
     & \quad =\inf_{F_{\pi,Z}^{1},...,F_{\pi,Z}^{N}\in\mathcal{F}}\int_{0}^{1} 
     \varphi\left(\frac{1}{N}\sum_{i=1}^{N}F_{\pi,Z}^{i}(x)\right)\mathrm dx. \nonumber\\
     & \quad \quad \text{ subject to } \frac{1}{N}\sum_{i=1}^{N}\int_{0}^{1}|x-x_i|^p F_{\pi,Z}^{i}(\mathrm dx) \leq \epsilon^{p}
    \end{align}
	
    Indeed, given conditional distribution functions \(F_{\pi,Z}^{1},...,F_{\pi,Z}^{N}\in\mathcal{F}\) as in (\ref{2026052301}) for some 
	\(\pi\in\Pi(\mathbb{P}_X, \mathbb{P}_{\hat{X}})\) such that 
	\begin{align*}
		\frac{1}{N}\sum_{i=1}^{N}\int_{0}^{1} |x-x_i|^p F_{\pi,Z}^{i}(\mathrm dx) \leq \epsilon^{p},
	\end{align*}
	then by (\ref{E}),
	\begin{align*}
		E_{\pi}\left[|Z-\hat{Z}|^p\right]=\frac{1}{N}\sum_{i=1}^{N}\int_{0}^{1} |x-x_i|^p F_{\pi,Z}^{i}(\mathrm dx) \leq \epsilon^{p},
	\end{align*}
    and hence \(F_X\in\mathcal{F}_{\epsilon}\) due to (\ref{2026051302}), and 
	\begin{align*}
		F_X(x)=\frac{1}{N}\sum_{i=1}^{N} F_{\pi,Z}^{i}(x), \quad x\in\mathbb{R},
	\end{align*}
    due to (\ref{2026060501}), where \(F_{\pi,Z}^{i}\), \(1\leq i\leq N\), is as in (\ref{2026052301}).
	Thus, by (\ref{2026052401}),
	\begin{align*}
		  \int_{0}^{1} \varphi \left(\frac{1}{N}\sum_{i=1}^{N}F_{\pi,Z}^{i}(x)\right) \mathrm dx
		&=\int_{0}^{1} \varphi \left( F_{X}(x) \right) \mathrm dx\\
        &\geq \inf_{F_X\in\mathcal{F}_{\epsilon}} \int_{0}^{1} \varphi \left( F_{X}(x) \right) \mathrm dx\\
		&=\inf_{\pi\in\Pi(\mathbb{P}_{X},\mathbb{P}_{\hat{X}})} \int_{0}^{1} 
	    \varphi\left(\frac{1}{N}\sum_{i=1}^{N}F_{\pi,Z}^{i}(x)\right)\mathrm dx. \nonumber \\
        &\quad  \text{subject to }  E_{\pi}\left[|Z-\hat{Z}|^{p}\right]\leq \epsilon^{p}
	\end{align*}
	Therefore,
	\begin{align}\label{2026060505}
		& \inf_{F_{\pi,Z}^{1},...,F_{\pi,Z}^{N}\in\mathcal{F}}\int_{0}^{1} 
	    \varphi\left(\frac{1}{N}\sum_{i=1}^{N}F_{\pi,Z}^{i}(x)\right)\mathrm dx \nonumber\\
        & \text{ subject to } \frac{1}{N}\sum_{i=1}^{N}\int_{0}^{1}|x-x_i|^p F_{\pi,Z}^{i}(\mathrm dx) \leq \epsilon^{p}\nonumber\\
		&\quad \geq \inf_{\pi\in\Pi(\mathbb{P}_{X},\mathbb{P}_{\hat{X}})} \int_{0}^{1} 
	    \varphi\left(\frac{1}{N}\sum_{i=1}^{N}F_{\pi,Z}^{i}(x)\right)\mathrm dx. \nonumber \\
        &\quad \quad \text{subject to }  E_{\pi}\left[|Z-\hat{Z}|^{p}\right]\leq \epsilon^{p}
	\end{align}
    
    Conversely, taking (\ref{2026052301}) and (\ref{E}) into account, it is not hard to verify that
    \begin{align*}
		& \inf_{F_{\pi,Z}^{1},...,F_{\pi,Z}^{N}\in\mathcal{F}}\int_{0}^{1} 
	    \varphi\left(\frac{1}{N}\sum_{i=1}^{N}F_{\pi,Z}^{i}(x)\right)\mathrm dx \nonumber\\
        & \text{ subject to } \frac{1}{N}\sum_{i=1}^{N}\int_{0}^{1}|x-x_i|^p F_{\pi,Z}^{i}(\mathrm dx) \leq \epsilon^{p}\nonumber\\
		&\quad \leq \inf_{\pi\in\Pi(\mathbb{P}_{X},\mathbb{P}_{\hat{X}})} \int_{0}^{1} 
	    \varphi\left(\frac{1}{N}\sum_{i=1}^{N}F_{\pi,Z}^{i}(x)\right)\mathrm dx, \nonumber \\
        &\quad \quad\text{subject to }  E_{\pi}\left[|Z-\hat{Z}|^{p}\right]\leq \epsilon^{p}
	\end{align*}
    which together with (\ref{2026060505}), implies (\ref{2026060504}). Consequently, the sufficiency and necessity  follow immediately from (\ref{2026052401}) and (\ref{2026060504}). 
    The relation that \(F_{X}^*=\frac{1}{N}\sum_{i=1}^{N}F_{\pi^*,Z}^i\) is clear in the course of proceeding proofs. 
    Lemma 3.2 is proved.

	\vspace{0.2cm}
	
	\noindent{\bf Proof of Lemma \ref{lem302}}\ \
	
	\vspace{0.2cm}
	
	Let $F_{\pi^*,Z}^{1}, \ldots ,F_{\pi^*,Z}^{N} \in \mathcal{F}$ satisfying the constraint (\ref{add040202}) of the optimization problem (\ref{inf4A})
	be arbitrarily given.
	For any $F_{\pi,Z}^{1}, \ldots ,F_{\pi,Z}^{N} \in \mathcal{F}$ such that $\frac{1}{N}\sum_{i=1}^{N}F_{\pi,Z}^{i}(x) \in \mathcal{F}_\epsilon$,
	define a function $I : \ [0,1] \rightarrow \mathbb{R}$ by
	\begin{align*}
		I(\lambda) := \int_{0}^{1} \varphi\left(\lambda\left(\frac{1}{N}\sum_{i=1}^{N}F_{\pi^*,Z}^{i}(x)\right)
		+ (1 - \lambda)\left(\frac{1}{N}\sum_{i=1}^{N}F_{\pi,Z}^{i}(x)\right)\right) \mathrm dx, \quad 0 \leq \lambda \leq 1.
	\end{align*}
	Since \(\varphi\) is strictly convex, it is not hard to verify that \(I(\lambda)\) is a strictly convex function of \(\lambda\).
	
	\vspace{0.2cm}
	
	For the sake of notational simplicity,  denote $F_X^* := \frac{1}{N}\sum_{i=1}^{N}F_{\pi^*,Z}^{i}$ and $F_X := \frac{1}{N}\sum_{i=1}^{N}F_{\pi,Z}^{i}$.
	Then, $F_{X}^{*}, F_{X} \in \mathcal{F}_\epsilon$.
	Since \(\varphi\) is twice differentiable, the first order derivative \(\varphi'\) of $\varphi$ exists and is continuous.
	Hence, \(I\) is differentiable, and its first order derivative $I'$ is given by
	\begin{align*}
		I'(\lambda)= \int_{0}^{1} \varphi'\left(\lambda F_{X}^{*}(x) + (1 - \lambda)F_{X}(x)\right)
		\left(F_{X}^{*}(x)-F_{X}(x)\right) \mathrm dx, \quad 0 \leq \lambda \leq 1.
	\end{align*}
	
	\vspace{0.2cm}
	
	For any \(0<\lambda_{1}<\lambda_{2}<1\), there exists some $\alpha \in (0,1)$ such that \(\lambda_{2}=\alpha\lambda_{1}+(1-\alpha)\cdot1\) and \(1-\lambda_{2}=\alpha(1-\lambda_{1})\). Hence, for any $x \in [0,1]$,
	\begin{align}\label{add040501}
		\lambda_{2}F_{X}^{*}(x)+(1-\lambda_{2})F_{X}(x)=\alpha[\lambda_{1}F_{X}^{*}(x)+(1-\lambda_{1})F_{X}(x)]+(1-\alpha)F_{X}^{*}(x).
	\end{align}
	By (\ref{add040501}) and the strict convexity of $\varphi$, we have that
	\begin{align}\label{add040502}
		\int_{0}^{1} & \varphi(\lambda_{2}F_{X}^{*}(x)+(1-\lambda_{2})F_{X}(x))\mathrm dx \nonumber \\
		& < \alpha\int_{0}^{1}\varphi(\lambda_{1}F_{X}^{*}(x)+(1-\lambda_{1})F_{X}(x))\mathrm dx
		+(1-\alpha)\int_{0}^{1}\varphi(F_{X}^{*}(x))\mathrm dx.
	\end{align}
	
	\vspace{0.2cm}
	
	To show necessity. Suppose that $F_{\pi^*,Z}^{1}, \ldots ,F_{\pi^*,Z}^{N}$ constitute a solution to optimization problem (\ref{inf4A}).
	Then \(F_{X}^{*}\) minimizes $\int_{0}^{1}\varphi(F_{X}(x))\mathrm dx$. Hence, we know that
	\begin{align}\label{add040503}
		\int_{0}^{1}\varphi(\lambda_{2}F_{X}^{*}(x)+(1-\lambda_{2})F_{X}(x))\mathrm dx\geq \int_{0}^{1}\varphi(F_{X}^{*}(x))\mathrm dx.
	\end{align}
	From (\ref{add040502}) and (\ref{add040503}), it follows that
	\begin{align*}
		I(\lambda_{2}) = \int_{0}^{1}\varphi(\lambda_{2}F_{X}^{*}(x)+(1-\lambda_{2})F_{X}(x))\mathrm dx
		< \int_{0}^{1}\varphi(\lambda_{1}F_{X}^{*}(x)+(1-\lambda_{1})F_{X}(x))\mathrm dx = I(\lambda_{1}),
	\end{align*}
	which yields that $I$ is strictly decreasing on $(0,1)$, and hence $I(\lambda)$ attains its minimum over $[0,1]$ at \(\lambda = 1\).
	Since $I$ is strictly convex on $[0,1]$,
	$I(\lambda)$ attains its minimum over $[0,1]$ at \(\lambda = 1\) if and only if $ I'(1) \leq 0$, that is,
	\begin{align*}
		\int_{0}^{1} \varphi'(F_{X}^{*}(x))\left(F_{X}^{*}(x)-F_{X}(x)\right)\mathrm dx \leq 0,
	\end{align*}
	which is exactly equivalent to
	\begin{align*}
		\int_{0}^{1} \varphi'(F_{X}^{*}(x))F_{X}^{*}(x)\mathrm dx \leq \int_{0}^{1} \varphi'(F_{X}^{*}(x))F_{X}(x)\mathrm dx,
	\end{align*}
	which implies that (\ref{(2)}) holds. Necessity is proved.
	
	\vspace{0.2cm}
	
	To show sufficiency. Assume that $(\ref{(2)})$ holds true. For any $F_{\pi,Z}^{1}, \ldots ,F_{\pi,Z}^{N}$ with
	$F_X := \frac{1}{N}\sum_{i=1}^{N}F_{\pi,Z}^{i} \neq F_X^*$, we know that
	\begin{align*}
		I'(1) =  \int_{0}^{1} \varphi'(F_{X}^{*}(x))\left(F_{X}^{*}(x)-F_{X}(x)\right)\mathrm dx \leq 0.
	\end{align*}
	Hence, by the strict convexity of $I$, we have that $I(1) < I(0)$, which exactly yields that
	\begin{align*}
		\int_{0}^{1} \varphi(F_{X}^{*}(x))\mathrm dx < \int_{0}^{1} \varphi(F_{X}(x))\mathrm dx,
	\end{align*}
	which means that $F_{X}^{*}$ is the unique minimizer minimizing $\int_{0}^{1}\varphi(F_{X}(x))\mathrm dx$,
	and thus $F_{\pi^*,Z}^{1}, \ldots ,F_{\pi^*,Z}^{N}$ constitute a solution to optimization problem (\ref{inf4A}).
	Sufficiency is proved. Lemma \ref{lem302} is proved.
	
	\vspace{0.2cm}

	\noindent\textbf{Proof of Lemma $\ref{auxiliaryD}$}
	
	\vspace{0.2cm}
	
	We first explore the properties of $F_{\pi^*,Z}^{1}, \ldots, F_{\pi^*,Z}^{N}$ by making use of its optimality in the auxiliary optimization problem (\ref{infC}).
	By introducing the Lagrange multiplier $\beta$ to the optimization problem (\ref{infC}), we have that
	\begin{align}\label{add25073101}
		& \inf_{F_{\pi,Z}^{1}, \ldots, F_{\pi,Z}^{N}\in\mathcal{F}} \sup_{\beta\geq0}
		\left\{ \frac{1}{N}\sum_{j=1}^{N} \int_{0}^{1} \varphi' \left( \frac{1}{N}\sum_{i=1}^{N}F_{\pi^*,Z}^{i}(x) \right) F_{\pi,Z}^{j}(x) \mathrm dx \right. \nonumber\\
		& \quad \quad  \quad \quad \quad \quad \quad
		\left. -\beta \left[ \frac{1}{N} \sum_{j=1}^{N} \int_{0}^{1} |x-x_{j}|^{p-1}\mbox{sign}(x-x_{j}) F_{\pi,Z}^{j}(x) \mathrm dx +\kappa \right] \right\}
		\nonumber\\
		& \qquad  = \sup_{\beta\geq 0}\inf_{F_{\pi,Z}^{1}, \ldots, F_{\pi,Z}^{N}\in\mathcal{F}}
		\left\{ \frac{1}{N}\sum_{j=1}^{N} \int_{0}^{1} \left[ \varphi' \left( \frac{1}{N}\sum_{i=1}^{N}F_{\pi^*,Z}^{i}(x) \right) \right. \right. \nonumber\\
		& \quad \quad  \quad \quad \quad \quad \quad \quad \quad \quad
		\left. \left. -\beta |x-x_{j}|^{p-1}\mbox{sign}(x-x_{j}) \right] F_{\pi,Z}^{j}(x) \mathrm dx - \beta \kappa \right\} \nonumber\\
		& \qquad = \sup_{\beta\geq 0} \left\{ \frac{1}{N}\sum_{j=1}^{N} \inf_{F_{\pi,Z}^{j}\in\mathcal{F}} \int_{0}^{1}
		H_j(x; \beta) F_{\pi,Z}^{j}(x) \mathrm dx - \beta \kappa \right\},
	\end{align}
	where for each $1\leq j\leq N,$
	\begin{align*}
		H_{j}(x;\beta) := \varphi' \left( \frac{1}{N}\sum_{i=1}^{N}F_{\pi^*,Z}^{i}(x) \right) -\beta |x-x_{j}|^{p-1}\mbox{sign}(x-x_{j}), \quad x\in[0,1],
	\end{align*}
	\begin{align}\label{add25073103}
		\kappa := \frac{1}{p} \left( \epsilon^{p}-\frac{1}{N} \sum_{i=1}^{N}(1-x_{i})^{p} \right),
	\end{align}
	and we have interchanged the order of $\inf$ and $\sup$ which is guaranteed by a strong duality in worst-case estimation problems; for instance,
	see Gao and Kleywegt (2023, Theorem 1 and Proposition 2), or Tang and Yang (2023, page 463).
	
	\vspace{0.2cm}
	
	By (\ref{add25073101}), we know that if some $F_{\hat{\pi},Z}^1, \ldots, F_{\hat{\pi},Z}^N$ constitute an optimal solution to the optimization
	problem (\ref{infC}), then for each $ 1 \leq j \leq N,$  $F_{\hat{\pi},Z}^j$ must minimize the inner infimum $\inf_{F_{\pi,Z}^{j}\in\mathcal{F}}$
	in (\ref{add25073101}).
	
	\vspace{0.2cm}
	
	Now, suppose that $F_{\hat{\pi},Z}^1, \ldots, F_{\hat{\pi},Z}^N$ constitute an optimal 
	solution to the  optimization problem (\ref{infC}).
	Then there exists a Lagrange multiplier $\hat\beta\geq0$ such that 
	 $F_{\hat{\pi},Z}^1, \ldots, F_{\hat{\pi},Z}^N$ and $\hat{\beta}$ jointly satisfy
	the generalized Kuhn-Tucker conditions for the optimization problem (\ref{infC}).
	In order to apply the generalized Kuhn-Tucker conditions for infinite-dimensional spaces to our case,
	we naturally extend the objective function from domain $\mathcal{F}$ to the Banach space \(\textbf{B}[0,1]:=\{\text{all bounded functions on}\ [0,1]\}\)
	endowed with the maximum norm $ ||f||:=\sup|f|$. In other words, we regard $\mathcal{F}$ as a convex subset of \(\textbf{B}[0,1]\).
	Hence, the generalized Kuhn-Tucker conditions for infinite-dimensional spaces imply that
	\begin{enumerate}
		\item[(1)]
		\begin{equation*}\label{062903}
			\hat\beta\geq0;
		\end{equation*}
		\item[(2)]
		\begin{align}\label{062904}
			-\frac{1}{N}\sum_{j=1}^{N}\int_{0}^{1}|x-x_j|^{p-1}\mbox{sign}(x-x_j)F_{\hat{\pi},Z}^{j}(x)\mathrm dx - \kappa \leq 0;
		\end{align}
		\item[(3)]
		\begin{align}\label{062905}
			\hat\beta \left( -\frac{1}{N}\sum_{j=1}^{N}\int_{0}^{1}|x-x_j|^{p-1}\mbox{sign}(x-x_j)F_{\hat{\pi},Z}^{j}(x)\mathrm dx - \kappa \right) = 0;
		\end{align}
		\item[(4)]
		\begin{align}\label{062906}
			&\nabla f \left(\left(F^{1}_{\hat{\pi},Z}, \ldots ,F^{N}_{\hat{\pi},Z}\right);\left(h_1, \ldots ,h_N\right)\right) \nonumber\\
			& \quad = \frac{1}{N}\sum_{j=1}^{N}\left\langle \varphi'\left(\frac{1}{N}\sum_{i=1}^{N}F_{\pi^*,Z}^{i}(x)\right)
			-\hat\beta |x-x_{j}|^{p-1}\mbox{sign}(x-x_{j}),\ h_j\right\rangle \nonumber\\
			& \quad = \frac{1}{N}\sum_{j=1}^{N} \left\langle H_j(x;\hat\beta), \ h_j \right\rangle \nonumber\\
			& \quad = 0;
		\end{align}
	\end{enumerate}
	where $f : \mathbf{B}[0,1]^{N} \rightarrow \mathbb{R}$ is defined by
	\begin{align*}
		f(F_{\pi,Z}^{1}, & \ldots ,F_{\pi,Z}^{N}) \\
		& :=  \frac{1}{N}\sum_{j=1}^{N}\int_{0}^{1}\left[\varphi'\left(\frac{1}{N}\sum_{i=1}^{N}F_{\pi^*,Z}^{i}(x)\right)
		-\hat\beta|x-x_j|^{p-1}\mbox{sign}(x-x_j)\right]F_{\pi,Z}^{j}(x)\mathrm dx - \hat\beta \kappa,
	\end{align*}
	$(F_{\pi,Z}^{1}, \ldots ,F_{\pi,Z}^{N})\in \mathbf{B}[0,1]^{N},$
	$\nabla f \left(\left(F^{1}_{\hat{\pi},Z}, \ldots ,F^{N}_{\hat{\pi},Z}\right);\left(h_1, \ldots ,h_N\right)\right)$
	stands for the Fr\'{e}chet derivative of $f$ at $\left(F^{1}_{\hat{\pi},Z}, \ldots ,F^{N}_{\hat{\pi},Z}\right)$ with increment $\left(h_1, \ldots ,h_N\right)$,
	and the inner product $\left\langle u,\ v \right\rangle := \int_{0}^{1} u(x)v(x) \mathrm dx,$ provided that the integral exists,
	for $u, v \in \mathbf{B}[0,1].$
	Note that since the constraint condition (\ref{add040407}) does make contribution to the optimization problem (\ref{infC}),
	hence $\hat{\beta}>0$, and thus equality in (\ref{062904}) holds due to (\ref{062905}). In (\ref{062906}), we have used the fact that
	\begin{align}\label{062901}
		&\nabla f \left(\left(F^{1}_{\hat{\pi},Z}, \ldots ,F^{N}_{\hat{\pi},Z}\right);\left(h_1, \ldots ,h_N\right)\right)\nonumber\\
		& \quad= \frac{1}{N}\sum_{j=1}^{N}\left\langle \varphi'\left(\frac{1}{N}\sum_{i=1}^{N}F_{\pi^*,Z}^{i}(x)\right)-\hat\beta |x-x_{j}|^{p-1}\mbox{sign}(x-x_{j}), h_j\right\rangle.
	\end{align}
	Indeed,
	\begin{align}\label{062902}
		\lim_{\parallel h\parallel\rightarrow0}\frac{\left|f\left(F_{\pi,Z}^{1}+h_{1}, \ldots ,F_{\pi,Z}^{N}+h_{N}\right)-f(F_{\pi,Z}^{1}, \ldots ,F_{\pi,Z}^{N})-\frac{1}{N}\sum_{j=1}^{N}\left\langle H_j(x;\hat\beta),h_j\right\rangle\right|}{\parallel h\parallel}=0,
	\end{align}
	where \(||h||:= || h_1||+\cdots+||h_N||,\) which yields (\ref{062901}). For more details, we refer to pages 247-250 of Luenberger (1969). For Lagrangian multiplier method in finite-dimensional setting, we refer to Boyd and Vandenberghe (2004).
	
	\vspace{0.2cm}
	
	Similarly, by introducing the Lagrange multiplier $\lambda$ to the optimization problem (\ref{infD}), we also know that
	\begin{align}\label{add040610}
		&   \inf_{F_{\pi,Z}^{1}, \ldots, F_{\pi,Z}^{N}\in\mathcal{F}}  \sup_{\lambda\geq 0}
		\left  \{ \frac{1}{N}\sum_{i=1}^{N} \int_{0}^{1} \varphi' \left( \frac{1}{N} \left( F_{\pi^*,Z}^{i}(x)+(i-1) \right) \right) F_{\pi,Z}^{i}(x) \mathrm dx
		\right.  \nonumber\\
		&  \quad \quad \quad \quad \quad \quad \quad
		\left. -\lambda \left[ \frac{1}{N} \sum_{i=1}^{N}\int_{0}^{1} |x-x_{i}|^{p-1}\mbox{sign}(x-x_{i}) F_{\pi,Z}^{i}(x) \mathrm dx + \kappa \right] \right\}
		\nonumber\\
		& \qquad = \sup_{\lambda\geq 0} \inf_{F_{\pi,Z}^{1}, \ldots, F_{\pi,Z}^{N}\in\mathcal{F}}
		\left\{ \frac{1}{N}\sum_{i=1}^{N}  \int_{0}^{1}   \left[ \varphi' \left( \frac{1}{N} \left( F_{\pi^*,Z}^{i}(x)+(i-1) \right) \right)
		\right. \right.   \nonumber\\
		& \quad \quad \quad \quad \quad \quad \quad \quad \quad \quad \quad \quad \quad \quad \quad
		\left. \left.  -\lambda |x-x_{i}|^{p-1}\mbox{sign}(x-x_{i}) \right] F_{\pi,Z}^{i}(x) \mathrm dx -\lambda \kappa \right\} \nonumber\\
		& \qquad = \sup_{\lambda\geq 0} \left\{ \frac{1}{N}\sum_{i=1}^{N}  \inf_{F_{\pi,Z}^{i}\in\mathcal{F}}
		\int_{0}^{1}  A_i(x) F_{\pi,Z}^{i}(x) \mathrm dx - \lambda \kappa \right\},
	\end{align}
	where $\kappa$ is as in (\ref{add25073103}), for each $1\leq i\leq N$,
	\begin{align}\label{add25073102}
		A_i(x) := \varphi'\left(  \frac{1}{N}  \left(F_{\pi^*,Z}^{i}(x)+(i-1)\right)  \right)-\lambda |x-x_{i}|^{p-1}\mbox{sign}(x-x_{i}), \quad x\in[0,1],
	\end{align}
	and we have interchanged the order of $\inf$ and $\sup$ in (\ref{add040610}) again.
	
	\vspace{0.2cm}
	
	By (\ref{add040610}), we know that if some $\tilde{F}_{\pi,Z}^{1}, \ldots, \tilde{F}_{\pi,Z}^{N}$ constitute an optimal solution to the optimization
	problem (\ref{infD}), then for each $ 1 \leq i \leq N,$  $\tilde{F}_{\pi,Z}^{i}$ must minimize the inner infimum $\inf_{F_{\pi,Z}^{i}\in\mathcal{F}}$
	in (\ref{add040610}).
	
	\vspace{0.2cm}
	
	Now, suppose that some $F_{\tilde{\pi},Z}^{1}, \ldots, F_{\tilde{\pi},Z}^{N}$ 
	constitute an optimal solution to the optimization problem (\ref{infD}).
	Then there is a Lagrange multiplier $\tilde{\lambda} \geq 0$ such that  
	$F_{\tilde{\pi},Z}^{1}, \ldots, F_{\tilde{\pi},Z}^{N}$ and
	$\tilde{\lambda}$ jointly satisfy the generalized Kuhn-Tucker conditions for the optimization problem (\ref{infD}), that is,
	\begin{enumerate}
		\item[($1'$)]
		\begin{align}\label{add041001}
			\tilde\lambda\geq0;
		\end{align}
		\item[($2'$)]
		\begin{align}\label{add041002}
			-\frac{1}{N}\sum_{i=1}^{N}\int_{0}^{1}|x-x_i|^{p-1}\mbox{sign}(x-x_i)F_{\tilde{\pi},Z}^{i}(x)\mathrm dx - \kappa \leq 0;
		\end{align}
		\item[($3'$)]
		\begin{align}\label{add041003}
			\tilde\lambda \left( -\frac{1}{N}\sum_{i=1}^{N}\int_{0}^{1}|x-x_i|^{p-1}\mbox{sign}(x-x_i)F_{\tilde{\pi},Z}^{i}(x)\mathrm dx
			- \kappa \right) = 0;
		\end{align}
		\item[($4'$)]
		\begin{align}\label{add041004}
			&\nabla G\left(\left(F^{1}_{\tilde{\pi},Z}, \ldots ,F^{N}_{\tilde{\pi},Z}\right);(h_1, \ldots ,h_N)\right)\nonumber\\
			& \quad   =\frac{1}{N}\sum_{i=1}^{N}\left\langle\varphi'\left(  \frac{1}{N}   \left(F_{\pi^*,Z}^{i}(x)+(i-1)\right)  \right)-\tilde\lambda|x-x_i|^{p-1}\mbox{sign}(x-x_i),h_i\right\rangle\nonumber\\
			& \quad =0;
		\end{align}
	\end{enumerate}
	where $G : \mathbf{B}[0,1]^{N} \rightarrow \mathbb{R}$ is defined by
	\begin{align*}
		G\left(F^{1}_{\pi,Z}, \ldots ,F^{N}_{\pi,Z}\right)
		& := \frac{1}{N}\sum_{i=1}^{N} \int_{0}^{1} \left[ \varphi' \left( \frac{1}{N} \left( F_{\pi^*,Z}^{i}(x)+(i-1) \right) \right) \right. \\
		& \quad \quad \quad \quad \quad \quad \quad
		\left.  -\tilde\lambda |x-x_{i}|^{p-1}\mbox{sign}(x-x_{i}) \right] F_{\pi,Z}^{i}(x) \mathrm dx - \tilde\lambda \kappa,
	\end{align*}
	$\left( F^{1}_{\pi,Z}, \ldots ,F^{N}_{\pi,Z} \right) \in \mathbf{B}[0,1]^{N}.$ In (\ref{add041004}), we have used the fact that the  Fr\'{e}chet derivative
    \(\nabla G\left(\left(F^{1}_{\tilde{\pi},Z}, \ldots ,F^{N}_{\tilde{\pi},Z}\right);(h_1, \ldots ,h_N)\right)\)
	of $G$ at $\left(F^{1}_{\tilde{\pi},Z}, \ldots ,F^{N}_{\tilde{\pi},Z}\right)$ with increment $\left(h_1, \ldots ,h_N\right)$ is given by
	\begin{align}\label{2026060703}
		\frac{1}{N}\sum_{i=1}^{N} \left\langle \varphi' \left( \frac{1}{N}  \left( F_{\pi^*,Z}^{i}(x)+(i-1) \right) \right)
		- \tilde\lambda |x-x_i|^{p-1}\mbox{sign}(x-x_i), \ h_i \right\rangle,
	\end{align}
	which can be verified similarly to (\ref{062902}) by the definition of the Fr\'{e}chet derivative.
	
	\vspace{0.2cm}
	
	We claim that \(F^{1}_{\hat{\pi},Z}, \ldots ,F^{N}_{\hat{\pi},Z}\) and \(\tilde{\lambda}\) also satisfy the generalized Kuhn-Tucker conditions
	(\ref{add041001})-(\ref{add041004}) for the optimization problem (\ref{infD}).
	Indeed, clearly, $F_{\hat{\pi},Z}^1, \ldots, F_{\hat{\pi},Z}^N$ and $\tilde\lambda$  satisfy the generalized Kuhn-Tucker conditions
	(\ref{add041001})-(\ref{add041003}).  Moreover,  similar to (\ref{2026060703}), we know that
	\begin{align*}
		& \nabla G\left(\left(F^{1}_{\hat{\pi},Z}, \ldots ,F^{N}_{\hat{\pi},Z}\right);(h_1, \ldots ,h_N)\right) \\
		& \quad = \frac{1}{N}\sum_{i=1}^{N} \left\langle \varphi' \left(  \frac{1}{N} \left( F_{\pi^*,Z}^{i}(x)+(i-1) \right) \right)
		-\tilde\lambda|x-x_i|^{p-1}\mbox{sign}(x-x_i), \ h_i \right\rangle   \\
		& \quad = 0,
	\end{align*}
	where the second equality holds  because that
	 \[\frac{1}{N}\sum_{i=1}^{N} \left\langle \varphi' \left(  \frac{1}{N} \left( F_{\pi^*,Z}^{i}(x)+(i-1) \right) \right)
		-\tilde\lambda|x-x_i|^{p-1}\mbox{sign}(x-x_i), \ h_i \right\rangle  = 0\]
	due to (\ref{add041004}). The proceeding discussions just  imply that $F_{\hat{\pi},Z}^1, \ldots, F_{\hat{\pi},Z}^N$ also satisfy (\ref{add041004}).
	
	\vspace{0.2cm}
	
	Notice that the optimization problem (\ref{infD}) is a convex one with differentiable objective,
	and that the constraint function satisfies the so-called regularity condition.
	Hence, by Luenberger (1969, page 221), we know that
	$ F^{1}_{\hat{\pi},Z}, \ldots ,F^{N}_{\hat{\pi},Z} $ also constitute an optimal solution to optimization problem (\ref{infD}).
	Recall that $F_{\pi^*,Z}^{1}, \ldots, F_{\pi^*,Z}^{N}$ constitute an optimal solution to the optimization problem (\ref{infC}).
	Therefore, letting $ F^{i}_{\hat{\pi},Z} = F_{\pi^*,Z}^{i}$, $1 \leq i \leq N,$ implies that
	$F_{\pi^*,Z}^{1}, \ldots, F_{\pi^*,Z}^{N}$ constitute an optimal solution to the optimization problem (\ref{infD}).
	Lemma \ref{auxiliaryD} is proved.

	\vspace{0.2cm}
	
	\noindent\textbf{Proof of Lemma $\ref{lem304}$}
	
	\vspace{0.2cm}
	
	For each $1\leq i\leq N,$ by the definition of Lebesgue-Stieltjes measure induced by a distribution function , \(L_{F_{\pi^*,Z}^{i}}((a,b])=L_{F_{\pi^*,Z}^{i}}(b)-L_{F_{\pi^*,Z}^{i}}(a)\), $a<b$. To show the lemma, it is sufficient for us to show that \(S_{i1}:=\{x: A_i(x)>0, 0<x<1\}\) and  \(S_{i2}:=\{x: A_i(x)<0, 0<x<1\}\) have $F_{\pi^*,Z}^{i}$-measure zero, that is \(L_{F_{\pi^*,Z}^{i}}(S_{i1})=L_{F_{\pi^*,Z}^{i}}(S_{i2})=0\).
	
	\vspace{0.2cm}
	
	We begin with showing $L_{F_{\pi^*,Z}^{i}}(S_{i1})=0$. First, we decompose the set \(S_{i1}\).  For any rational number \(q\in S_{i1}\) with \(A_{i}(q)>0\),  let $I_{q}$ be the largest interval containing $q$ and $I_{q}\subseteq S_{i1}$. By  the right continuity of $A_{i}(x)$, there exists a \(\delta=\delta(q)>0\) such that $A_{i}(x)>0$ whenever \(x \in [q, q+\delta) \).  Denote \(a_i(q):=\inf\{a:  a\leq q, A_{i}(a)>0\}\), \(b_i(q):=\sup\{b:  b> q, A_{i}(b)>0\}\). Clearly,  $a_i(q)\leq q<b_i(q)$. Hence, $I_{q}=[a_i(q), b_i(q))$. Apparently,  $q\in I_{q}$ and $I_{q}\subseteq S_{i1}$. Thus, we know that $S_{i1}=\cup_{q\in\mathcal{Q}}I_{q}$, that is, $S_{i1}$ is a denumerable union of intervals $[x_i , x_{i+1})$, where $[x_i , x_{i+1})$ stands for a generic interval of form of $I_{q}$. Therefore, it suffices for us to show that each interval $[x_i , x_{i+1})$ has $F_{\pi^*,Z}^{i}$-measure zero, that is, \(F_{\pi^*,Z}^{i}(x_{i}-)=F_{\pi^*,Z}^{i}(x_{i+1}-)\).  We will show this assertion via proof by contradiction. Suppose that for certain interval $[x_i , x_{i+1})$, \(F_{\pi^*,Z}^{i}(x_{i}-)\neq F_{\pi^*,Z}^{i}(x_{i+1}-)\). Then \(F_{\pi^*,Z}^{i}(x_{i}-)< F_{\pi^*,Z}^{i}(x_{i+1}-)\). Notice that there are two possibilities: (i) \(F_{\pi^*,Z}^{i}(x)\) is constant on the interval \([x_{i}, x_{i+1})\); (ii) \(F_{\pi^*,Z}^{i}(x)\) is not constant on the interval \([x_{i}, x_{i+1})\). Hence, we will show the desired assertion by discussing these two possibilities, respectively.
	
	\vspace{0.2cm}
	
	(i) Suppose that \(F_{\pi^*,Z}^{i}(x)\) is constant on the interval \([x_{i}, x_{i+1})\), that is, there is some $0<c\leq1$ such that  \(F_{\pi^*,Z}^{i}(x)=c,\) for all \(x\in [x_{i},x_{i+1}).\) If \(F_{\pi^*,Z}^{i}(x_{i}-)=c,\) then \(F_{\pi^*,Z}^{i}(x_{i}-)-F_{\pi^*,Z}^{i}(x_{i+1}-)=0,\) which leads to a contradiction.  If \(F_{\pi^*,Z}^{i}(x_{i}-)<c,\) then for any \( x'\in (x_{i},x_{i+1})\), \(F_{\pi^*,Z}^{i}(x_{i}-)<c=F_{\pi^*,Z}^{i}(x').\) Keeping in mind that  \(A_{i}(x)>0\) for all \(x\in [x_{i}, x_{i+1})\), thus,  we have that
	\begin{align}\label{add063001}
		& \int_{x_{i}}^{x_{i+1}}A_{i}(x)F_{\pi^*,Z}^{i}(x_{i}-)\mathrm dx\nonumber\\
		& \quad <\int_{x_{i}}^{x'}A_i(x)F_{\pi^*,Z}^{i}(x_{i}-)\mathrm dx+\int_{x'}^{x_{i+1}}A_{i}(x)F_{\pi^*,Z}^{i}(x')\mathrm dx\nonumber\\
		& \quad \leq\int_{x_{i}}^{x_{i+1}}A_{i}(x)F_{\pi^*,Z}^{i}(x)\mathrm dx.
	\end{align}
	Now, we define a new distribution function $F_{\tilde{\pi},Z}^{i}$ by
	\[
	F_{\tilde{\pi},Z}^i(x):=
	\begin{cases}
		F_{\pi^*,Z}^{i}(x_{i}-),& \text{if } x \in [x_{i},x_{i+1}),\\
		F_{\pi^*,Z}^{i}(x), & \text{otherwise}.
	\end{cases}
	\]
	Consequently, (\ref{add063001}) implies that \(F_{\pi^*,Z}^{i}\) is not the minimizer, which leads to a contradiction.
	
	\vspace{0.2cm}
	
	(ii) Suppose that \(F_{\pi^*,Z}^{i}(x)\) is not constant on the interval \([x_{i}, x_{i+1})\), that is, there exists an \( x'\in [x_{i},x_{i+1}), \) such that \(F_{\pi^*,Z}^{i}(x_{i}-)<F_{X}^{*}(x')\leq F_{\pi^*,Z}^{i}(x_{i+1})\).  Keeping in mind that \(A_i(x)>0\) for all \( x\in [x_{i},x_{i+1})\), hence, we have that
	\begin{align}\label{add063002}
		& \int_{x_{i}}^{x_{i+1}}A_{i}(x)F_{\pi^*,Z}^{i}(x_{i}-)\mathrm dx\nonumber\\
		& \quad <\int_{x_{i}}^{x'}A_i(x)F_{\pi^*,Z}^{i}(x_{i}-)\mathrm dx+\int_{x'}^{x_{i+1}}A_{i}(x)F_{\pi^*,Z}^{i}(x')\mathrm dx\nonumber\\
		& \quad \leq\int_{x_{i}}^{x_{i+1}}A_{i}(x)F_{\pi^*,Z}^{i}(x)\mathrm dx.
	\end{align}
	Now, we define a new distribution function $F_{\tilde{\pi},Z}^{i}$ by
	\[
	F_{\tilde{\pi},Z}^i(x):=
	\begin{cases}
		F_{\pi^*,Z}^{i}(x_{i}-),& \text{if } x \in [x_{i},x_{i+1}),\\
		F_{\pi^*,Z}^{i}(x),& \text{otherwise}.
	\end{cases}
	\]
	Consequently, (\ref{add063002}) implies that \(F_{\pi^*,Z}^{i}\) is not the minimizer, which leads to a contradiction.

	\vspace{0.2cm}
	
	Next, we proceed to show $L_{F_{\pi^*,Z}^{i}}(S_{i2})=0$. First, we claim that the set \(S_{i2}\) is a denumerable union of open intervals. In fact, for any \( x_{0}\in S_{i2}\), by the right continuity of $A_i(x)$, there exists a  \(\delta_{1}>0\) such that \(A_{i}(x)<0\) whenever \(~ x\in[ x_{0},  x_{0}+\delta_{1})\). By the definition of $A_i(x)$,
	\begin{align*}
		\lim_{x\rightarrow x_{0}-}A_{i}(x)&=\varphi'\left(\frac{1}{N}\left(F_{\pi^*,Z}^{i}(x_{0}-)-(i-1)\right)\right)-\lambda k|x_{0}-x_{i}|^{k-1}\mbox{sign}(x_{0}-x_{i})\\
		&\leq \varphi'\left(\frac{1}{N}\left(F_{\pi^*,Z}^{i}(x_{0})-(i-1)\right)\right)-\lambda k|x_{0}-x_{i}|^{k-1}\mbox{sign}(x_{0}-x_{i})<0.
	\end{align*}
	Thus, there exists certain \(\delta_{2}>0\) such that \(A_{i}(x)<0\) whenever \( x\in(x_{0}-\delta_{2}, x_{0}]\).  For any rational number \( q\in S_{i2}\), let \(I_{q}\) be the largest  interval containing $q$ and  \(I_{q}\subseteq S_{i2}\). Denote \(a_i(q):=\inf\{a:a<q, (a,q)\subseteq S_{i2}\}, b_i(q):=\sup\{b:b>q, (q,b)\subseteq S_{q}\}\). Clearly, \(a_i(q)<q<b_i(q)\). Thus \(I_{q}=(a_i(q),b_i(q))\). Obviously, \(q\in I_{q}\) and \(I_{q}\subseteq S_{i2}\). Hence, we know that \(S_{i2}=\cup_{q\in\mathcal{Q}} I_{q}\), that is,  \(S_{i2}\) is a denumerable union of open intervals $(x_i,x_{i+1})$, where $(x_i,x_{i+1})$ stands for a generic interval of form of $I_q$.
	
	\vspace{0.2cm}
	
	Therefore, it is sufficient for us to show that each such interval $(x_i , x_{i+1})$ has $F_{\pi^*,Z}^{i}$-measure zero, that is, \(F_{\pi^*,Z}^{i}(x_{i})=F_{\pi^*,Z}^{i}(x_{i+1})\). We will show this assertion via proof by contradiction. Suppose that for certain interval $(x_i , x_{i+1})$, \(F_{\pi^*,Z}^{i}(x_{i})\neq F_{\pi^*,Z}^{i}(x_{i+1})\). Then \(F_{\pi^*,Z}^{i}(x_{i})< F_{\pi^*,Z}^{i}(x_{i+1})\).  Thus, there exists an \(x'\in(x_{i},x_{i+1})\) such that \(F_{\pi^*,Z}^{i}(x_{i})<F_{\pi^*,Z}^{i}(x')\leq F_{\pi^*,Z}^{i}(x_{i+1})\). Keeping in mind that \(A_{i}(x)<0\) for all \(x\in(x_{i},x_{i+1})\), hence, we have that
	
	\begin{align}\label{add063003}
		\int_{x_{i}}^{x_{i+1}}A_{i}(x)F_{\pi^*,Z}^{i}(x_{i+1})\mathrm dx
		& < \int_{x_{i}}^{x'}A_{i}(x)F_{\pi^*,Z}^{i}(x)\mathrm dx+\int_{x'}^{x_{i+1}}A_{i}(x)F_{\pi^*,Z}^{i}(x_{i+1})\mathrm dx\nonumber\\
		& \leq \int_{x_{i}}^{x_{i+1}}A_{i}(x)F_{\pi^*,Z}^{i}(x)\mathrm dx.
	\end{align}
	Now, we define a new distribution function $F_{\tilde{\pi},Z}^{i}$ by
	\[
	F_{\tilde{\pi},Z}^i(x):=
	\begin{cases}
		F_{\pi^*,Z}^{i}(x_{i+1}),& \text{if } x \in (x_{i},x_{i+1}),\\
		F_{\pi^*,Z}^{i}(x), & \text{otherwise}.
	\end{cases}
	\]
	Consequently, (\ref{add063003}) implies that \(F_{\pi^*,Z}^{i}\) is not the minimizer, which leads to a contradiction. Lemma 3.4 is proved.

	\vspace{0.2cm}

	\noindent\textbf{Proof of Lemma $\ref{lem305}$ }
	
	\vspace{0.2cm}
	
	We will show the lemma via proof by contradiction. Assume that the assertion is not true. Then, there are two cases: Case 1: $\int_{a}^{b} A_{i}(x) \, \mathrm{d}x < 0$; Case 2: $\int_{a}^{b} A_{i}(x) \, \mathrm{d}x > 0$. Next, we proceed to discuss these two cases, respectively.
	
	\vspace{0.2cm}
	
	Case 1: For any small number $\delta > 0$ with $b<1$ and $b+\delta<1$, we replace $F_{\pi^*,Z}^{i}(x)$ on the interval $[a, b + \delta)$ with the constant  $F_{\pi^*,Z}^{i}(b + \delta)$.  Denote by $\nu$  the increase in $\int_{0}^{1}A_{i}(x)F_{\pi,Z}^{i}(x)\mathrm{d}x$ due to this replacement.  Then,
	
	\begin{align*}
		\nu :&= \int_{a}^{b+\delta} A_{i}(x) F_{\pi^*,Z}^{i}(b + \delta) \mathrm  dx - \int_{a}^{b+\delta} A_i(x) F_{\pi^*,Z}^{i}(x) \mathrm dx\\
		&     = (F_{\pi^*,Z}^{i}(b + \delta) - c) \int_{a}^{b+\delta} A_{i}(x) \mathrm  dx - \int_{a}^{b+\delta} A_{i}(x) [F_{\pi^*,Z}^{i}(x) - c]  \mathrm dx\\
		&    \leq  (F_{\pi^*,Z}^{i}(b + \delta) - c) \int_{a}^{b+\delta} A_{i}(x) \mathrm  dx +\int_{a}^{b+\delta}\left|A_{i}(x) [F_{\pi^*,Z}^{i}(x) - c]\right|\mathrm dx\\
		&    \leq [F_{\pi^*,Z}^{i}(b + \delta) - c] \left( \int_{a}^{b} A_{i}(x) \mathrm  dx + 2M\delta \right),
	\end{align*}
	where \( |A_{i}(x)| < M\) whenever $x\in[0,1]$ for certain $M>0$.
	
	\vspace{0.2cm}
	
	Letting $\delta \to 0$, we know that $\nu$ becomes negative. 
	Define a new distribution function $F_{\tilde{\pi},Z}^i$ by
	\[
	F_{\tilde{\pi},Z}^i(x):=
	\begin{cases}
		F_{\pi^*,Z}^{i}(b+\delta),& \text{if } x \in [a,b+\delta),\\
		F_{\pi^*,Z}^{i}(x), & \text{otherwise}.
	\end{cases}
	\]
	Notice that $\nu$ is negative for small enough $\delta>0$. Hence, in this case, \(F_{\pi^*,Z}^{i}\) is not the minimizer, which leads to a  contradiction. The case where $b =1$ is trivial, because we can directly set $\delta=0$ in the previous argument.
	
	\vspace{0.2cm}
	
	Case 2: First, we consider the situation where $a>0$ and $a$ is a point of continuity of  $F_{\pi^*,Z}^{i}$. Then by an argument similar to that as in Case 1, we can get a contradiction. Indeed, for small enough $\delta>0$ with $a-\delta>0$,  by replacing $F_{\pi^*,Z}^{i}(x)$ on $(a - \delta, b)$ with $F_{\pi^*,Z}^{i}(a - \delta)$, we can obtain that
	\begin{align*}
		\tau & := \int_{a-\delta}^{b} A_{i}(x) F_{\pi^*,Z}^{i}(a- \delta) \mathrm  dx - \int_{a-\delta}^{b} A_{i}(x) F_{\pi^*,Z}^{i}(x) \mathrm dx\\
		&= (F_{\pi^*,Z}^{i}(a-\delta) - c) \int_{a-\delta}^{b} A_{i}(x) \mathrm  dx - \int_{a-\delta}^{b} A_{i}(x) [F_{\pi^*,Z}^{i}(x) - c]  \mathrm dx\\
		&\leq(F_{\pi^*,Z}^{i}(a-\delta) - c) \int_{a-\delta}^{b} A_{i}(x) \mathrm  dx +\int_{a-\delta}^{b}\left|A_{i}(x) [F_{\pi^*,Z}^{i}(x) - c]\right|\mathrm dx\\
		&\leq [F_{\pi^*,Z}^{i}(a-\delta) - c] \left(\int_{a}^{b} A_{i}(x)\mathrm dx+2\delta M\right),
	\end{align*}
	where $|A_{i}(x)|<M$ whenever $x\in[0,1]$ for certain $M>0$.
	
	\vspace{0.2cm}
	
	Letting $\delta \to 0$, we know that $\tau$ becomes non-positive. 
	Define a new distribution function  $F_{\tilde{\pi},Z}^i$ by
	\[
	F_{\tilde{\pi},Z}^i(x):=
	\begin{cases}
		F_{\pi^*,Z}^{i}(a-\delta),& \text{if } x \in (a-\delta,b),\\
		F_{\pi^*,Z}^{i}(x), & \text{otherwise}.
	\end{cases}
	\]
	Notice that $\tau$ is negative for small enough $\delta>0$. Hence, in this situation, \(F_{\pi^*,Z}^{i}\) is not the minimizer, which leads to a  contradiction.\\
	
	\vspace{0.2cm}
	
	Next, we consider the situation where either $a=0$ or there is a jump in \(F_{\pi^*,Z}^{i}\) at $a$. Indeed, these situation is trivial, because the argument is totally similar to that of previous situation. Lemma 3.5 is proved.
	
	\vspace{0.2cm}
	
	\noindent\textbf{Proof of Lemma $\ref{lem306}$}
	
	\vspace{0.2cm}
	
	We will show this lemma via proof by contradiction. Suppose that for some $1\leq i\leq N$, $F_{\pi^*,Z}^{i}$ has a jump at $a$, $a\in(0,1)$. By checking the proof of Lemma $\ref{lem304}$, we can get that $A_{i}(a) = 0$. Keeping in mind that $\varphi'$ is strictly increasing, hence, we know that $a$ must be the right-hand end-point of certain open interval $(x_i,x_{i+1})$ with $A_{i}(x) < 0$ whenever $x\in(x_i,x_{i+1})$.
	On the other hand, by checking again the proof of Lemma $\ref{lem304}$, we know that $ F_{\pi^*,Z}^{i}(x) =  F_{\pi^*,Z}^{i}(a)$ on $(x_i,x_{i+1})$. Thus, $ F_{\pi^*,Z}^{i}$ is left continuous at $a$.  Therefore,  $ F_{\pi^*,Z}^{i}$ is continuous at $a$, which leads to a contradiction. Lemma 3.6 is proved.
	
	\vspace{0.2cm}

	\noindent\textbf{Proof of Theorem $\ref{thm301}$}
	
	\vspace{0.2cm}
	
	(1) The assertion immediately follows from (\ref{add070501}) and (\ref{add070301}).
	
	\vspace{0.2cm}
	
	(2) Our argumentation is also partially inspired by Cornilly et al. (2018). Let $F_{\pi^*,Z}^{1}, \ldots ,F_{\pi^*,Z}^{N}$ be as in the optimal problem (\ref{infD}),
	that is, they constitute the unique optimal solution to the optimal problem (\ref{inf4A}). From Lemmas $\ref{lem304}-\ref{lem306}$,
	we know that for each $1\leq i \leq N$, $F_{\pi,Z}^{i}$ must satisfy the following equation
	\begin{align}\label{add070102}
		\varphi'\left(  \frac{1}{N}   \left(F_{\pi^*,Z}^{i}(x)+(i-1)\right)  \right)-\lambda |x-x_{i}|^{p-1}\mbox{sign}(x-x_{i})=0,
	\end{align}
	for $x\in (0,1)$ with $A_i(x)=0$.
	
	\vspace{0.2cm}
	
	Since, $\varphi'$ is strictly increasing, (\ref{add070102}) suggests that we can steadily invert on some set $I$ to derive an expression for $F_{\pi^*,Z}^{i}$ in terms of the original distortion function $g$. More precisely, for any given $1\leq i\leq N$, define a mapping \( F_{\pi,Z}^{i} : I_i\rightarrow[0,1] \) by
	\begin{align*}
		F_{\pi,Z}^{i}(x) := N - N(g')^{-1}\left(\lambda |x-x_{i}|^{p-1}\mbox{sign}(x-x_{i})\right) - (i-1), \quad x\in I_i,
	\end{align*}
	where $\lambda>0$ is the Lagrange multiplier, \(I_{i}\) is defined as
	
	\begin{align*}
		I_i := \left\{ x \in (0, 1) : \lambda |x-x_{i}|^{p-1}\mbox{sign}(x-x_{i}) \in \left[g'\left(\frac{N+1-i}{N}\right), g'\left(\frac{N-i}{N}\right)\right] \right\}.
	\end{align*}
	Hence, we can obtain an expression for $F_{\pi^*,Z}^{i}$ as follows:
	\begin{align}\label{add070103}
		F_{\pi^*,Z}^{i}(x):=
		\begin{cases}
			0,& \text{if }x<\mbox{min}\{a_i,1\},\\
			N- N(g')^{-1}(\lambda^* |x-x_{i}|^{p-1}\mbox{sign}(x-x_{i})) - (i-1),   & \text{otherwise},\\
			1,&\text{if }x\geq \mbox{min}\{b_i,1\},
		\end{cases}
	\end{align}
	where the Lagrange multiplier $\lambda^*>0$  satisfies
	$$-\frac{1}{N} \sum_{i=1}^{N}\int_{0}^{1}|x-x_{i}|^{p-1}\mbox{sign}(x-x_{i}){F}_{\pi^*,Z}^{i}(x)\mathrm dx-\frac{1}{p}\left(\epsilon^{p}-\frac{1}{N} \sum_{i=1}^{N}(1-x_{i})^{p}\right)=0,$$
	and $a_i>0$ is the unique solution to
	$$\lambda^* |x-x_{i}|^{p-1}\mbox{sign}(x-x_{i})=g'\left(\frac{N+1-i}{N}\right)$$
	with respect to variable $x$ on $(0,\infty)$,
	and $b_i>0$ is the unique solution to
	$$\lambda^* |x-x_{i}|^{p-1}\mbox{sign}(x-x_{i})=g'\left(\frac{N-i}{N}\right)$$
	with respect to variable $x$ on $(0,\infty)$.
	
	\vspace{0.2cm}
	
	Furthermore, for each $1\leq i\leq N$, by an elementary calculation, we have that
	\begin{align}\label{add070104}
		F_{\pi^*,Z}^{i}(x):=
		\begin{cases}
			0,& \text{if }x<F_{X}^{*-1}(\frac{i-1}{N}),\\
			NF_{X}^{*}(x) - (i-1),   & \text{otherwise},\\
			1,&\text{if }x\geq F_{X}^{*-1}(\frac{i}{N}),
		\end{cases}
	\end{align}
	where \(F_{X}^{*}(x):=\frac{1}{N}\sum_{i=1}^{N}F_{\pi^*,Z}^{i}(x),\) \(x\in [0,1]\), is the unique optimal solution to optimization problem (\ref{inf2A}). Notice that if $\min\{a_i,1\}=a_i$ and $\min\{b_i,1\}=b_i$, then $F_{X}^{*-1}(\frac{i-1}{N})\leq a_i$ and $b_i= F_{X}^{*-1}(\frac{i}{N}).$ If $\min\{a_i,1\}=1$, then $\min\{b_i,1\}=1$, and thus
	\[
	F_{\pi^*,Z}^{i}(x):=
	\begin{cases}
		0,& \text{if }0\leq x<1,\\
		1,&\text{if }x=1.
	\end{cases}
	\]
	
	\vspace{0.2cm}
	
	In summary, from (\ref{add070103}) and (\ref{add070104}) it follows that $F_{X}^*$ is given by
	\begin{align*}
		F_{X}^*(x)=
		\begin{cases}
			\frac{i-1}{N},                                               & \text{if }  x<a_i,\\
			1-(g')^{-1}(\lambda^* |x-x_{i}|^{p-1}\mbox{sign}(x-x_{i})),  & \text{otherwise},\\
			\frac{i}{N},                                                 & \text{if }  x \geq  b_i,
		\end{cases}
	\end{align*}
	on $(F_{X}^{*-1}(\frac{i-1}{N}),\quad F_{X}^{*-1}(\frac{i}{N})],$ $1\leq i\leq N.$
	Theorem 3.1 is proved.
	
	\vspace{0.2cm}
	
	\noindent\textbf{Proof of Lemma $\ref{lem307}$}
	
	\vspace{0.2cm}
	
	Let $F_{X}^*$ be the minimizer of optimization problem (\ref{inf2B}). By an argument similar to the proof of  Lemma 3.2, we know that $F_{X}^*$ minimizes optimization problem (\ref{inf2B}) if and only if it holds that
	$$\int_0^1\varphi'(F_{X}^*(x))F_{X}(x)\mathrm dx\geq\int_0^1\varphi'(F_{X}^*(x))F_{X}^*(x)\mathrm dx$$
	for all $F_{X}\in\mathcal{F}_{\epsilon}(c_1,c_p).$
	Thus, $F_{X}^*$ minimizes the functional
	$$ F_{X}\rightarrow\int_0^1\varphi'(F_{X}^*(x))F_{X}(x)\mathrm dx$$
	over $\mathcal{F}_{\epsilon}(c_1,c_p)$, that is, $F_{X}^*$ is an optimal solution to the following optimization problem (\ref{add070201}):
	\begin{align}\label{add070201}
		\inf_{F_{X}\in\mathcal{F}_{\epsilon}}\int_0^1\varphi'(F_{X}^*(x))F_{X}(x)\mathrm dx.\tag{S}
	\end{align}
	subject to
	\begin{align*}
		\begin{cases}
			\int_{0}^{1}F_{X}(x)\mathrm dx=d_1,\\
			\int_{0}^{1}x^{p-1}F_{X}(x)\mathrm dx=d_p,
		\end{cases}
	\end{align*}
	where $d_1:=1-c_1$, $d_p:=\frac{1}{p}-c_p$. Introducing the Lagrange multipliers $\eta_1$ and $\eta_p\in\mathbb{R}$ to the optimization problem (\ref{add070201}), we obtain the Lagrange dual problem:
	\begin{align*}
		\sup_{\eta_1,\eta_p\in\mathbb{R}} \inf_{F_{X}\in\mathcal{F}_{\epsilon}} \left\{\int_{0}^{1}\left(\varphi'\left(F_{X}^*(x)\right)+\eta_1+\eta_px^{p-1}\right)F_{X}(x)\mathrm dx-\eta_1d_1-\eta_pd_p\right\}.
	\end{align*}
	
	\vspace{0.2cm}
	
	Since $F_{X}^*$ is an optimal solution to the optimization problem (\ref{add070201}), there exist Lagrange multipliers  $\eta_1^*$, $\eta_p^*\in\mathbb{R}$ such that
	\begin{align}\label{add070502}
		&\inf_{F_{X}\in\mathcal{F}_{\epsilon}}\sup_{\eta_1,\eta_p\in\mathbb{R}} \left\{\int_{0}^{1}\left(\varphi'\left(F_{X}^*(x)\right)+\eta_1+\eta_px^{p-1}\right)F_{X}(x)\mathrm dx-\eta_1d_1-\eta_pd_p.\right\}\nonumber\\
		& \quad =\sup_{\eta_1,\eta_p\in\mathbb{R}} \inf_{F_{X}\in\mathcal{F}_{\epsilon}} \left\{\int_{0}^{1}\left(\varphi'\left(F_{X}^*(x)\right)+\eta_1+\eta_px^{p-1}\right)F_{X}(x)\mathrm dx-\eta_1d_1-\eta_pd_p.\right\}\nonumber\\
		& \quad =\int_{0}^{1}\left(\varphi'\left(F_{X}^*(x)\right)+\eta_1^*+\eta_p^*x^{p-1}\right)F_{X}^*(x)\mathrm dx-\eta_1^*d_1-\eta_p^*d_p,
	\end{align}
	where we can switch the order of $\sup$ and $\inf$ which is guaranteed by an strong duality in worst-case estimation problems; for instance, see Theorem 1 and Proposition 2 of Gao and Kleywegt (2023).
	
	\vspace{0.2cm}
	
	(\ref{add070502}) implies that $F_{X}^*$ is an optimal solution to the optimization problem (\ref{infO}). Lemma \ref{lem307} is proved.
	
	\vspace{0.2cm}
	
	\noindent\textbf{Proof of Lemme \ref{lem308}}
	
	\vspace{0.2cm}
	
	The proof is totally similar to that of Lemma \ref{auxiliaryD}. We just sketch it here. First, we explore the properties of $F_{\pi^*,Z}^{1}, \ldots ,F_{\pi^*,Z}^{N}$
	by making use of its optimality in the optimization problem (\ref{infP}). By introducing the Lagrange multiplier  $\beta$ to the optimization
	problem (\ref{infP}), we can obtain an unconstrained version of the optimization problem (\ref{infP}).
	Suppose that some $F_{\hat{\pi},Z}^{1}, \ldots ,F_{\hat{\pi},Z}^{N}$ constitute an optimal solution to the optimization problem (\ref{infP}).
	Then, there exists a Lagrange multiplier $\hat{\beta}>0$ such that 
	$F_{\hat{\pi},Z}^{1}, \ldots ,F_{\hat{\pi},Z}^{N}$ and $\hat{\beta}$ jointly satisfy
	the generalized  Kuhn-Tucker conditions for the optimization problem (\ref{infP}).
	
	\vspace{0.2cm}
	
	Next, by introducing the Lagrange multiplier  $\lambda$ to the optimization problem (\ref{infQ}), we can obtain an unconstrained version of
	the optimization problem (\ref{infQ}). Suppose that some 
	$F_{\tilde{\pi},Z}^{1}, \ldots ,F_{\tilde{\pi},Z}^{N}$ constitute an optimal solution to
	the optimization problem (\ref{infQ}). 
	Then there exists a Lagrange multiplier 
	$\tilde\lambda\geq0$ such that 
	$F_{\tilde{\pi},Z}^{1}, \ldots ,F_{\tilde{\pi},Z}^{N}$
	and $\tilde\lambda$ jointly satisfy the generalized Kuhn-Tucker conditions for the optimization problem (\ref{infQ}).
	
	\vspace{0.2cm}
	
	Clearly, we know that $F_{\hat{\pi},Z}^{1}, \ldots ,F_{\hat{\pi},Z}^{N}$ and 
	$\tilde\lambda$ also satisfy the dual feasibility, 
	primal feasibility and complementary slackness of the generalized Kuhn-Tucker conditions 
	for the optimization problem (\ref{infQ}). 
	We conclude that $F_{\hat{\pi},Z}^{1}, \ldots ,F_{\hat{\pi},Z}^{N}$
	and $\tilde\lambda$ also satisfy the stationarity of the generalized Kuhn-Tucker condition
	 for the optimization problem (\ref{infQ})
	by a similar argument as in the proof of Lemma 3.3. In summary,
	$F_{\hat{\pi},Z}^{1}, \ldots ,F_{\hat{\pi},Z}^{N}$ and $\tilde\lambda$ 
	satisfy the generalized Kuhn-Tucker conditions for optimization problem (\ref{infQ}).
	
	\vspace{0.2cm}
	
	Notice that the optimization problem (\ref{infQ}) is a convex one with 
	differentiable objective, and that the constraint functions satisfy
	the so-called regularity condition. Hence, by Luenberger (1969, page 221), 
	we know that $F_{\hat{\pi},Z}^{1}, \ldots ,F_{\hat{\pi},Z}^{N}$ also constitute
	an optimal solution to the optimization problem (\ref{infQ}).
	Recall that $F_{\pi^*,Z}^{1}, \ldots, F_{\pi^*,Z}^{N}$ constitute an optimal solution
	 to the optimization problem (\ref{infP}).
	Therefore, letting $ F^{i}_{\hat{\pi},Z} = F_{\pi^*,Z}^{i}$, $1 \leq i \leq N,$ 
	implies that
	$F_{\pi^*,Z}^{1}, \ldots, F_{\pi^*,Z}^{N}$ constitute an optimal solution to 
	the optimization problem (\ref{infQ}).
	Lemma 3.8 is proved.
	
	\vspace{0.2cm}
	
	\noindent\textbf{Proof of Theorem \ref{thm302}}

	\vspace{0.2cm}
	The proof is similar to that of Theorem \ref{thm301}. Our argumentation is also inspired by Cornilly et al. (2018). We sketch it here.

	\vspace{0.2cm}

	(1) In this case, the optimization problem (\ref{inf2B}) reduces to the optimization problem studied by Cornilly et al. (2018), and hence (\ref{add070503}) follows from Cornilly (2018, Theorem 2.2).

	\vspace{0.2cm}

	(2) Let $F_{\pi^*,Z}^{1}, \ldots ,F_{\pi^*,Z}^{N}$ be as in optimal problem (\ref{infQ}), that is, \(F_{X}^{*}:=\frac{1}{N}\sum_{i=1}^{N}F_{\pi^*,Z}^{i}\) is the unique optimal solution to optimal problem (\ref{inf2B}). By an argument similar to Lemmas $\ref{lem304}-\ref{lem306}$, we know that for each $1\leq i \leq N$, $F_{\pi,Z}^{i}$ must satisfy the following equation
	\begin{align}\label{add070202}
		B_{i}(x):=\varphi'\left(  \frac{1}{N}   \left(F_{\pi^*,Z}^{i}(x)+(i-1)\right)  \right)+\eta_{1}^*+\eta_{p}^*x^{p-1}-\lambda|x-x_{i}|^{p-1}\mbox{sign}(x-x_{i})=0,
	\end{align}
	for $x\in(0,1)$.

	\vspace{0.2cm}

	Since, $\varphi'$ is strictly increasing, (\ref{add070202}) suggests that we can 
	steadily invert on some set $I$ to derive an expression for $F_{\pi^*,Z}^{i}$ in terms of the original distortion function $g$. More precisely, for any given $1\leq i\leq N$, define a mapping \( F_{\pi,Z}^{i} : I_i\rightarrow[0,1] \) by
	\begin{align}\label{add070101}
		F_{\pi,Z}^{i}(x) := N - N(g')^{-1}\left(-\eta_1^*-\eta_p^*x^{p-1}+\lambda |x-x_{i}|^{p-1}\mbox{sign}(x-x_{i})\right) - (i-1),
	\end{align}
	$x\in I_i$, where $\eta_1^*$, $\eta_p^*$ and $\lambda>0$ are the Lagrange multipliers, \(I_{i}\) is defined as
	\begin{align*}
		I_i &:=\\
		&  \left\{ x \in (0, 1) :-\eta_1^*-\eta_p^*x^{p-1}+ \lambda |x-x_{i}|^{p-1}\mbox{sign}(x-x_{i}) \in \left[g'\left(\frac{N+1-i}{N}\right), g'\left(\frac{N-i}{N}\right)\right] \right\}.
	\end{align*}

	\vspace{0.2cm}

	Since $-\eta_1^*-\eta_p^*x^{p-1}+\lambda |x-x_{i}|^{p-1}\mbox{sign}(x-x_{i})$ 
	could not be non-decreasing with respect to $x$, $F^{i}_{\pi,Z}$ as in (\ref{add070101})
	 could not be non-decreasing. Therefore, we need to modify  
	 $F^{i}_{\pi,Z}$ as in (\ref{add070101}) so that we can 
	 figure out the possible form of $F^{i}_{\pi^*,Z}$. Precisely, 
	 notice that $F^{i}_{\pi,Z}$ as in (\ref{add070101}) 
	 is continuous on $(0,1)$. Hence, we can replace strictly decreasing 
	 segments of $F^{i}_{\pi,Z}$ with constants, and keep the modified function 
	 being continuous on $(0,1)$ and coinciding with $F^{i}_{\pi,Z}$ on the intervals
	  where the modified function is not constant. Thus, the resulting modified 
	  function is just the desired $F^{i}_{\pi^*,Z}$.

	\vspace{0.2cm}

	In summary, we know that $F^{i}_{\pi^*,Z}$ coincides with some $F_{\bm\eta}^{i}(x)$ of the following form on intervals where is not constant, where
	\begin{align*}
		{F}_{\bm\eta}^{i}(x):=
		\begin{cases}
			0,& \text{if }a_{i}(x)<g'\left(\frac{N+1-i}{N}\right), \\
			N-N(g')^{-1}(a_i(x) )-(i-1),  & \text{otherwise},\\
			1,&\text{if }a_{i}(x)\geq g'\left(\frac{N-i}{N}\right),
		\end{cases}
	\end{align*}
	where for each $1\leq i\leq N$,
	$$
	a_{i}(x) := -\eta_{1}^*-\eta_{p}^*x^{p-1}+\lambda^{*} |x-x_{i}|^{p-1}\mbox{sign}(x-x_{i}), \quad x \in \mathbb{R},
	$$
	$\bm\eta:=(\eta_1^*,\eta_p^*),$ and the parameters $\lambda^*>0$, $\eta_{1}^*$ and $\eta_{p}^*$ are determined by following equations:
	\begin{align*}
		-\frac{1}{N} \sum_{i=1}^{N}\int_{0}^{1}|x-x_{i}|^{p-1}\mbox{sign}(x-x_{i})F_{\bm\eta}^{i}(x)\mathrm dx-\frac{1}{p}\left(\epsilon^{p}
		-\frac{1}{N} \sum_{i=1}^{N}(1-x_{i})^{p}\right) = 0,
	\end{align*}
	\begin{align*}
		\int_{0}^{1}x\mathrm d\left(\frac{1}{N}\sum_{i=1}^{N}F_{\bm\eta}^{i}\right)(x) = c_{1}
	\end{align*}
	and
	\begin{align*}
		\int_{0}^{1}x^{p}\mathrm d\left(\frac{1}{N}\sum_{i=1}^{N}F_{\bm\eta}^{i}\right)(x) = c_{p}.
	\end{align*}
	Theorem \ref{thm302} is proved.
	
	\vspace{0.2cm}
	
	\noindent\textbf{Proof of Corollary 3.1}
	
	\vspace{0.2cm}
	
	Inspired by Cornilly et al. (2018), we consider the random variable \(X\) taking values in \([-b, b]\), $b>0$, and the transformed random
	variable \(Y := (X + b)/2b\).
	Denote by $\mu_{X}$ and $\sigma^2_{X}$ the mean and variance of $X$ respectively. Then the mean $\mu_{Y}$ of $Y$ equals
	$(\mu_{X} + b)/2b$ and the  variance $\sigma_{Y}^{2}$ of $Y$ equals $\sigma_{X}^2/4b^2$. Notice that
	\begin{align}{\label{add071301}}
		F_{X}(x)=F_{Y}\left(\frac{x+b}{2b}\right),\quad x\in[-b,b] \quad \text{and} \quad F_{X}^{-1}(p)=2bF_{Y}^{-1}(p)-b,\quad p\in[0,1].
	\end{align}
	
	\vspace{0.2cm}
	
	Clearly, the distribution function $F_Y$ of $Y$ is supported on [0,1]. Hence, by applying Theorem 3.2(2) to the random variable $Y$, we know that the optimal distribution function $F_Y^*$ for $Y$, maximizing the distortion risk measure of $Y$, is given by
	\begin{align}\label{add070504}
		{F}_{Y}^{*}(y):=
		\begin{cases}
			\frac{i-1}{N}& \text{if } y<\frac{g'(1)+\tilde{\lambda} y_{i}+\tilde{\eta_{1}}}{-\tilde{\eta_{2}}+\tilde{\lambda}},\\
			1-(g')^{-1}(-\tilde{\eta_{1}}+(-\tilde{\eta_{2}}+\tilde{\lambda})y-\tilde{\lambda} y_{i})  & \text{otherwise},\\
			\frac{i}{N}&\text{if }y\geq \frac{g'(0)+\tilde{\lambda} y_{i}+\tilde{\eta_{1}}}{-\tilde{\eta_{2}}+\tilde{\lambda}},
		\end{cases}
	\end{align}
	on $({F}_{Y}^{*-1}(\frac{i-1}{N}),\quad{F}_{Y}^{*-1}(\frac{i}{N})]$, $1\leq i\leq N$, and thus the left-continuous inverse function $F_{Y}^{*-1}$ of $F_{Y}^{*}$ is given by
	
	\begin{align}\label{add071302}
		F_Y^{*-1}(p) =
		\begin{cases}
			0,                                                                                             & \text{if } p \leq F_{Y}^*(0), \\
			\frac{1}{-\tilde{\eta_2}+\tilde{\lambda}} \left(g'(1 - p) +\tilde{ \eta_1}+\tilde{\lambda} y_{1}\right), & \text{if } F_{Y}^*(0)<p\leq \frac{1}{N} , \\
			\frac{1}{-\tilde{\eta_2}+\tilde{\lambda}} \left(g'(1 - p) +\tilde{ \eta_1}+\tilde{\lambda}y_{i}\right), & \text{if } \frac{i-1}{N}<p\leq \frac{i}{N},\quad i=2, \ldots ,N-1, \\
			\frac{1}{-\tilde{\eta_2}+\tilde{\lambda}} \left(g'(1 - p) +\tilde{ \eta_1}+\tilde{\lambda} y_{N}\right), & \text{if } \frac{N-1}{N}<p\leq \lim_{y\uparrow1^-}F_{Y}^*(y),\\
			1,                                                                                             & \text{if } p > \lim_{y\uparrow1}F_{Y}^*(y),
		\end{cases}
	\end{align}
	where \(y_i:=\frac{x_i+b}{2b}\), $1\leq i\leq N$, $\tilde{\eta_1}$, $\tilde{\eta_2}$ and $\tilde{\lambda}$ are determined by following conditions
	\begin{align*}
		\mu_Y=\int_{0}^{1}  F_Y^{*-1}(p)\mathrm dp, \quad  \quad \mu_Y^2+\sigma_{Y}^2=\int_{0}^{1} (F_Y^{*-1}(p))^2\mathrm dp
	\end{align*}
	and
	\begin{align*}
		\sum_{i=1}^{N}\int_{\frac{i-1}{N}}^{\frac{i}{N}}|F_Y^{*-1}(p)-y_i|^2\mathrm d p=\left(\frac{\epsilon}{2b}\right)^2.
	\end{align*}
	Indeed,
	\begin{align*}
		\tilde{\eta_1} =& \frac{1}{\beta-\alpha} \left[ g(1 - \beta) - g(1 - \alpha)+(-\tilde{\eta_2}+\tilde{\lambda})(\mu_Y - (1 - \beta))\right.\\
		& \left.-\tilde{\lambda}\left(\left(\frac{1}{N}-\alpha\right)y_1+\frac{1}{N}\sum_{i=2}^{N-1}y_i+\left(\beta-\frac{N-1}{N}\right)y_N\right) \right],
	\end{align*}
	
	\begin{align}\label{add070505}
		&\tilde{\eta}_2= \tilde{\lambda}-\sqrt{\frac{\sum_{i=2}^{N-1}\int_{\frac{i-1}{N}}^{\frac{i}{N}}h_i(p;\tilde{\lambda},y_1, \ldots ,y_N)^2\mathrm dp+\tau_1(\alpha)+\tau_2(\beta)}{{\sigma_Y^2 + \mu_Y^2 - (1 - \beta) - (\beta - \alpha)^{-1} (\mu_Y - (1 - \beta))^2}}},
	\end{align}
	where $\tau_1(\alpha):=\int_{\alpha}^{\frac{1}{N}}h_1(p;\tilde{\lambda},y_1, \ldots ,y_N)^2\mathrm dp$, $\tau_2(\beta):=\int_{\frac{N-1}{N}}^{\beta}h_N(p;\tilde{\lambda},y_1, \ldots ,y_N)^2\mathrm dp$, $\alpha := F_Y^*(0)$, $\beta := \lim_{y \uparrow 1} F_Y^*(y)$, and for each $1\leq i\leq N$,
	\begin{align*}
		&h_i(p;\tilde{\lambda},y_1, \ldots ,y_N)\\
		&:=g'(1-p)+\tilde{\lambda}y_i+\frac{g(1-\beta)-g(1-\alpha)-\tilde{\lambda}\left((\frac{1}{N}-\alpha)y_1+\frac{1}{N}\sum_{i=2}^{N-1}y_i+(\beta-\frac{N-1}{N})y_N\right)}{\beta-\alpha}.
	\end{align*}
	Notice that the above expressions are possibly implicit, depending  on whether or not \(\alpha = 0\) \text{and} \(\beta = 1\). Moreover, (\ref{add070505}) yields that $-\tilde{\eta}_2+\tilde{\lambda}>0$, which guarantees that (\ref{add070504}) is well-defined.
	
	\vspace{0.2cm}
	
	From (\ref{add071301}), (\ref{add070504}) and (\ref{add071302}), it follows that the optimal solution $F_{X}^*$ to optimization problem (B2) and its left-continuous inverse  function \(F_X^{*-1}\) are respectively  given by
	\begin{align}\label{add071303}
		{F}_{X}^{*}(x) =
		\begin{cases}
			\frac{i-1}{N}, & \text{if } x < \frac{g'\left(\frac{N+1-i}{N}\right)+\lambda^* x_{i}+\eta_{1}^*}{ -\eta_{2}^*+\lambda^*},\\
			1-(g')^{-1}(-\eta_{1}^*+(-\eta_{2}^*+\lambda^*)x-\lambda^* x_{i}),  & \text{otherwise},\\
			\frac{i}{N}, & \text{if }x \geq \frac{g'\left(\frac{N-i}{N}\right)+\lambda^* x_{i}+\eta_{1}^*}{ -\eta_{2}^*+\lambda^*},
		\end{cases}
	\end{align}
	on $\left(F_{X}^{*-1}\left(\frac{i-1}{N}\right),F_{X}^{*-1}\left(\frac{i-1}{N}\right)\right],$ $1\leq i\leq N,$ and
	
	\begin{align}
		F_X^{*-1}(p) =
		\begin{cases}
			-b,                                                                                             & \text{if } p \leq F_{X}^*(-b), \\
			\frac{1}{-\eta_2^*+\lambda^*} \left(g'(1 - p) + \eta_1^* + \lambda^* x_{1}\right), & \text{if } F_{X}^*(-b)<p\leq \frac{1}{N} , \\
			\frac{1}{-\eta_2^*+\lambda^{*}} \left(g'(1 - p) + \eta_1^* + \lambda^* x_{i}\right), & \text{if } \frac{i-1}{N}<p\leq \frac{i}{N},\quad i=2, \ldots ,N-1, \\
			\frac{1}{-\eta_2^*+\lambda^{*}} \left(g'(1 - p) + \eta_1^* + \lambda^* x_{N}\right), & \text{if } \frac{N-1}{N}<p\leq \lim_{x\uparrow b}F_{X}^*(x),\\
			b,                                                                                             & \text{if } p > \lim_{x\uparrow b}F_{X}^*(x),
		\end{cases}
	\end{align}
	where
	\begin{align*}
		\eta_1^* =& \frac{1}{\beta-\alpha} \left[ g(1 - \beta) - g(1 - \alpha)+2b(-\eta_2^*+\lambda^{*})(\mu_Y - (1 - \beta)-\frac{\beta-\alpha}{2})\right.\\
		& \left.-\lambda^{*}\left(\left(\frac{1}{N}-\alpha\right)x_1+\frac{1}{N}\sum_{i=2}^{N-1}x_i+\left(\beta-\frac{N-1}{N}\right)x_N\right) \right],
	\end{align*}
	\begin{align*}
		&\eta_2^*=\lambda^*-\frac{1}{2b}\sqrt{\frac{\sum_{i=2}^{N-1}\int_{\frac{i-1}{N}}^{\frac{i}{N}}h_i(p;\lambda^{*},x_1, \ldots ,x_N)^2\mathrm dp+\theta_1(\alpha)+\theta_2(\beta)}{{\sigma_Y^2 + \mu_Y^2 - (1 - \beta) - (\beta - \alpha)^{-1} (\mu_Y - (1 - \beta))^2}}}
	\end{align*}
	where $\theta_1(\alpha):=\int_{\alpha}^{\frac{1}{N}}h_1(p;\lambda^{*},x_1, \ldots ,x_N)^2\mathrm dp$, $\theta_2(\beta):=\int_{\frac{N-1}{N}}^{\beta}h_N(p;\lambda^{*},x_1, \ldots ,x_N)^2\mathrm dp$
	and $\lambda^*\geq0$ satisfies
	\begin{align*}
		\sum_{i=1}^{N}\int_{\frac{i-1}{N}}^{\frac{i}{N}}|F_X^{*-1}(p)-x_i|^2\mathrm dp=\epsilon^2.
	\end{align*}
	
	\vspace{0.2cm}
	
	Recall that $F_{X}^*$ as in (\ref{add071303})  maximizes the distortion risk measure over all distribution functions on \([-b, b]\) with mean \(\mu_X\) and variance \(\sigma_X^2\).
	
	\vspace{0.2cm}
	
	Apparently, for fixed \( \mu_X \) and \( \sigma_X^2 \), $\lim_{b\rightarrow\infty}\mu_Y=\lim_{b\rightarrow\infty}\frac{\mu_X+b}{2b}=0.5$, $\lim_{b\rightarrow\infty}\sigma_Y^2=\lim_{b\rightarrow\infty}\frac{\sigma_X^2}{4b^2}=0$, $\lim_{b\rightarrow\infty}\alpha=\lim_{b\rightarrow\infty}F_{X}^*(-b)=0$
	and $\lim_{b\rightarrow\infty}\beta=\lim_{b\rightarrow\infty}F_{X}^*(b-)=1$. Hence, $g(1 - \alpha)$ goes to 1  when $\alpha$ decreases to zero, and $g(1 - \beta)$ goes to zero when $\beta$ increases to 1.
	Consequently, the constants \( \eta_1^* \) and \( \eta_2^*\) become
	\[
	\eta_1^* = (-\eta_2^*+\lambda^*) \mu_X -\frac{\lambda^*}{N}\sum_{i=1}^{N}x_{i}-1\quad
	\]
	and
	\[
	\eta_2^* = \lambda^*-\sqrt{\frac{\sum_{i=1}^{N}\int_{\frac{i-1}{N}}^{\frac{i}{N}}\left[g'(1-p)+\lambda^* x_{i}-1
			-\frac{\lambda^*}{N}\sum_{j=1}^{N}x_{j}\right]^{2} \mathrm dp}{{\sigma_X^2}}}.
	\]
	Corollary 3.1 is proved.

	\bibliographystyle{model1-num-names}

\end{document}